\documentclass[12pt,a4paper]{article}
\usepackage[dvips]{graphicx}

\renewcommand{\theequation}{\arabic{section}.\arabic{equation}}
\newcommand{\elimine}[1]{ }
\renewcommand{\log}{{\rm ln}}
\newcommand{\q}{\quad}

\newcommand{\be}{\begin{equation}}
\newcommand{\ee}{\end{equation}}
\newcommand{\ba}{\begin{array}}
\newcommand{\ea}{\end{array}}
\newcommand{\pag}{\left(}
\newcommand{\pad}{\right)}
\newcommand{\acg}{\left\{}
\newcommand{\acd}{\right\}}
\newcommand{\crg}{\left[}
\newcommand{\crd}{\right]}
\newcommand{\pog}{\left.}
\newcommand{\pod}{\right.}
\newcommand{\ud}{{1/2}}
\newcommand{\usd}{{\frac{1}{2}}}
\newcommand{\usq}{{\frac{1}{4}}}
\newcommand{\Tr}[1]{ { \hbox{Tr}\ {#1}  } }
\newcommand{\tru}[1]{ { \hbox{tr}_n \, {#1} \, } }
\newcommand{\trd}[1]{ { \hbox{tr}_{2n} \, {#1} \, } }
\newcommand{\dtt}[1]{ { \hbox{det} ( {#1} ) } }

\newcommand{\ento}[1]{{[#1]}}
\newcommand{\ap}{{\ento{a}}}
\newcommand{\AP}{{\ento{A}}}
\newcommand{\bp}{{\ento{ad}}}
\newcommand{\cpp}{{\ento{da}}}
\newcommand{\ep}{{\ento{d}}}
\newcommand{\ip}{{\ento{i}}}
\newcommand{\fpp}{{\ento{gdg'}}}

\newcommand{\bup}{{\ento{agdg'}}}
\newcommand{\bdp}{{\ento{g'agd}}}
\newcommand{\cvp}{{\ento{gdg'a}}}
\newcommand{\cdp}{{\ento{dg'ag}}}
\newcommand{\gup}{{\ento{g}}}
\newcommand{\gvp}{{\ento{g'}}}
\newcommand{\pbp}{{\ento{a.d.}}}
\newcommand{\pcp}{{\ento{d.a.}}}
\newcommand{\pbq}{{\ento{.a.d}}}
\newcommand{\pcq}{{\ento{.d.a}}}
\newcommand{\cze}{{\ento{0}}}
\newcommand{\cpi}{{\ento{1}}}

\newcommand{\ADh}{{\hat{{A}}}}
\newcommand{\DDh}{{\hat{{D}}}}
\newcommand{\HDh}{{\hat{{H}}}}
\newcommand{\JDh}{{\hat{{J}}}}
\newcommand{\KDh}{{\hat{{K}}}}
\newcommand{\MDh}{{\hat{{M}}}}
\newcommand{\NDh}{{\hat{{N}}}}
\newcommand{\DNh}{{\Delta\NDh}}

\newcommand{\PDh}{{\hat{{P}}}}
\newcommand{\QDh}{{\hat{{Q}}}}
\newcommand{\RDh}{{\hat{{R}}}}
\newcommand{\TDh}{{\hat{{T}}}}
\newcommand{\UDh}{{\hat{{U}}}}
\newcommand{\VDh}{{\hat{{V}}}}
\newcommand{\XDh}{{\hat{{X}}}}
\newcommand{\YDh}{{\hat{{Y}}}}
\newcommand{\Deltah}{{\hat{{\Delta}}}}
\newcommand{\TDha}{{\TDh^\ap}}
\newcommand{\TDhf}{{\TDh^\fpp}}

\newcommand{\TDhd}{{\TDh^\ep}}
\newcommand{\TDhb}{{\TDh^\bp}}
\newcommand{\TDhc}{{\TDh^\cpp}}
\newcommand{\TDhgu}{{\TDh^\gup}}
\newcommand{\TDhgv}{{\TDh^\gvp}}

\newcommand{\DIhz}{{{\hat{\cal D}}_0}}

\newcommand{\DIhu}{{{\hat{\cal D}}_1}}
\newcommand{\BO}{{B}}

\newcommand{\NO}{{N}}

\newcommand{\VO}{{V}}
\newcommand{\AIh}{{\hat{\cal A}}}
\newcommand{\DIh}{{\hat{\cal D}}}

\newcommand{\MIh}{{\hat{\cal M}}}

\newcommand{\EI}{{E}}
\newcommand{\FI}{{F}}
\newcommand{\GI}{{\cal G}}
\newcommand{\HI}{{\cal H}}
\newcommand{\II}{{\cal I}}
\newcommand{\IIs}{{ \II_{\rm s} }}

\newcommand{\LI}{{\cal L}}
\newcommand{\MI}{{\cal M}}
\newcommand{\NI}{{\cal N}}
\newcommand{\QI}{{\cal Q}}
\newcommand{\RI}{{\cal R}}
\newcommand{\SI}{{S}}
\newcommand{\TI}{{\cal T}}
\newcommand{\XI}{{\cal Y}}
\newcommand{\UI}{{\cal U}}

\newcommand{\YI}{{Z}}
\newcommand{\ZI}{{\zeta}}
\newcommand{\RIp}{{\RI'}}
\newcommand{\HIp}{{\HI'}}
\newcommand{\li}{{l}}
\newcommand{\qi}{{q}}

\newcommand{\lia}{{\li^\ap}}
\newcommand{\lid}{{\li^\ep}}
\newcommand{\lib}{{\li^\bp}}
\newcommand{\lic}{{\li^\cpp}}

\newcommand{\ligu}{{\li^\gup}}
\newcommand{\ligv}{{\li^\gvp}}
\newcommand{\LIa}{{\LI^\ap}}
\newcommand{\LId}{{\LI^\ep}}
\newcommand{\LIb}{{\LI^\bp}}
\newcommand{\LIc}{{\LI^\cpp}}

\newcommand{\LIgu}{{\LI^\gup}}

\newcommand{\RIa}{{\RI^\ap}}
\newcommand{\RId}{{\RI^\ep}}
\newcommand{\RIb}{{\RI^\bp}}
\newcommand{\RIc}{{\RI^\cpp}}
\newcommand{\RIf}{{\RI^\fpp}}

\newcommand{\TIa}{{\TI^\ap}}
\newcommand{\TId}{{\TI^\ep}}
\newcommand{\TIb}{{\TI^\bp}}
\newcommand{\TIc}{{\TI^\cpp}}
\newcommand{\TIf}{{\TI^\fpp}}
\newcommand{\TIgu}{{\TI^\gup}}
\newcommand{\TIgv}{{\TI^\gvp}}
\newcommand{\ZIa}{{\ZI^\ap}}
\newcommand{\ZId}{{\ZI^\ep}}
\newcommand{\ZIb}{{\ZI^\bp}}
\newcommand{\ZIc}{{\ZI^\cpp}}

\newcommand{\EIzb}{{{\bar \EI}}}
\newcommand{\SIzb}{{{\check{\SI}}}}
\newcommand{\FIgb}{{{\FI}_{\rm G}}}

\newcommand{\RIbze}{{\RI^\bp_0}}
\newcommand{\RIcze}{{\RI^\cpp_0}}
\newcommand{\RIdze}{{\RI^\ep_0}}

\newcommand{\TIaze}{{\TI^\ap_0}}

\newcommand{\TIdze}{{\TI^\ep_0}}

\newcommand{\RIze}{{\RI_0}}
\newcommand{\RIzeg}{{\RI_0^g}}

\newcommand{\RIig}{{\RI^\ip_\gamma}}

\newcommand{\TIig}{{\TI^\ip_\gamma}}

\newcommand{\RIad}{{\RI^\ap_\delta}}

\newcommand{\RIbd}{{\RI^\bp_\delta}}
\newcommand{\RIcd}{{\RI^\cpp_\delta}}
\newcommand{\RIdd}{{\RI^\ep_\delta}}

\newcommand{\TIad}{{\TI^\ap_\delta}}

\newcommand{\Tru}[1]{ { \hbox{Tr}_1\ {#1}  } }
\newcommand{\HIu}{{\HI_1}}
\newcommand{\Omk}{{\hat{\Omega}_k}}
\newcommand{\QDhg}{ { \hat{Q} _\gamma } }
\newcommand{\QDhd}{ { \hat{Q} _\delta } }
\newcommand{\avOmk}{\langle\Omk\rangle}

\newcommand{\avQDhg}{\langle\QDhg\rangle}

\newcommand{\Cgd}{{C_{\gamma\delta}}}

\newcommand{\RIbup}{{\RI^\bup}}
\newcommand{\RIbdp}{{\RI^\bdp}}
\newcommand{\RIcup}{{\RI^\cvp}}
\newcommand{\RIcdp}{{\RI^\cdp}}
\newcommand{\RIdag}{{\RI^{[dag]}}}
\newcommand{\RIdga}{{\RI^{[dga]}}}
\newcommand{\RIgda}{{\RI^{[gda]}}}
\newcommand{\RIadg}{{\RI^{[adg]}}}
\newcommand{\RIagd}{{\RI^{[agd]}}}
\newcommand{\RIgad}{{\RI^{[gad]}}}
\newcommand{\HIsbu} { {\HI^{\bup} } }
\newcommand{\HIsbd} { {\HI^{\bdp} } }
\newcommand{\HIscu} { {\HI^{\cvp} } }
\newcommand{\HIscd} { {\HI^{\cdp} } }
\newcommand{\ksdgg}{ {k_d^{g'g} } }
\newcommand{\ksagg}{ {k_a^{gg'} } }

\newcommand{\ksu}{ {k^{1} } }

\newcommand{\ksdzu}{ {k_d^{01} } }

\newcommand{\TIm}{{\TI^{-1}}}
\newcommand{\TIam}{{{\TIa}^{-1}}}
\newcommand{\TIdm}{{{\TId}^{-1}}}

\newcommand{\EB}{{{\bar\EI}}}

\newcommand{\IB}{{{\II}}}
\newcommand{\RB}{{{\bar\RI}}}
\newcommand{\YB}{{{\check{\YI}}}}

\newcommand{\EBb}{\EB^{(ad)}}
\newcommand{\EBc}{\EB^{(da)}}

\newcommand{\RBb}{\RB^{(ad)}}
\newcommand{\RBc}{\RB^{(da)}}
\newcommand{\RBbu}{ { \RB^{[a.d.]} } }
\newcommand{\RBbd}{ { \RB^{[.a.d]} } }
\newcommand{\RBcu}{ { \RB^{[d.a.]} } }
\newcommand{\RBcd}{ { \RB^{[.d.a]} } }

\newcommand{\GL}{{ \bar{\GI} }}
\newcommand{\HL}{{ \bar{\HI} }}

\newcommand{\SL}{{{\bf S}}}
\newcommand{\UL}{{{\bf 1}}}
\newcommand{\SIL}{{{\bf \Sigma}}}
\newcommand{\YIg}{{\YI^{\,g}}}
\newcommand{\YIgg}{{\YI^{\,g\,g'}}}

\newcommand{\TLa}{{\bf T}^\ap}
\newcommand{\TLd}{{\bf T}^\ep}
\newcommand{\TLam}{{{\TLa}^{-1}}}
\newcommand{\TLdm}{{{\TLd}^{-1}}}
\newcommand{\GLbu}{ { \GL^\pbp } }
\newcommand{\GLbd}{ { \GL^\pbq } }
\newcommand{\GLcu}{ { \GL^\pcp } }
\newcommand{\GLcd}{ { \GL^\pcq } }
\newcommand{\HLbu}{ { \HL^\pbp } }
\newcommand{\HLbd}{ { \HL^\pbq } }
\newcommand{\HLcu}{ { \HL^\pcp } }
\newcommand{\HLcd}{ { \HL^\pcq } }
\newcommand{\CLl}{{\bf C}}
\newcommand{\ELl}{{\bf E}}
\newcommand{\FLl}{{\bf F}}
\newcommand{\HLl}{ {\HI} }
\newcommand{\KLl}{{\bf K}}
\newcommand{\LLl}{ {\LI} }
\newcommand{\SLl}{{\bf S}}
\newcommand{\RLl}{ {\RI} }
\newcommand{\dRLl} {{\delta\RLl}}
\newcommand{\dLLl} {{\delta\LLl}}

\newcommand{\MLbu}{{{\bar{\bf M}}^\pbp}}
\newcommand{\MLcu}{{{\bar{\bf M}}^\pcp}}
\newcommand{\MLbd}{{{\bar{\bf M}}^\pbq}}
\newcommand{\MLcd}{{{\bar{\bf M}}^\pcq}}
\newcommand{\MLbum}{{{\bar{\bf M}}^{\pbp -1}}}

\newcommand{\MLcdm}{{{\bar{\bf M}}^{\pcq -1}}}
\newcommand{\oo}{\omega}
\newcommand{\bb}{\beta}
\newcommand{\ggg}{\gamma}
\newcommand{\doo}{{\frac{\hbox{d}\ }{\hbox{d}\oo} }}
\newcommand{\du}{{\frac{\hbox{d}\ }{\hbox{d}u} }}
\newcommand{\db}{{\frac{\hbox{d}\ }{\hbox{d}\bb}}}
\newcommand{\duh}[1]{{ \frac{\displaystyle  \hbox{d}{#1}}
{\displaystyle \hbox{d}u} }}

\newcommand{\intzb}{{ \int_0^\bb\, \hbox{d}u\  }}
\newcommand{\la}{{\lambda}}
\newcommand{\ttt}{{\tau}}
\newcommand{\tta}{{\ttt^\ap}}
\newcommand{\ttd}{{\ttt^\ep}}
\newcommand{\ttb}{{\ttt^\bp}}
\newcommand{\ttc}{{\ttt^\cpp}}
\newcommand{\ttf}{{\ttt^\fpp}}

\newcommand{\ttgu}{{\ttt^\gup}}
\newcommand{\ttgv}{{\ttt^\gvp}}

\newcommand{\zz}{{(0)}}
\newcommand{\bbb}{{(\bb)}}
\newcommand{\uu}{{(u)}}

\newcommand{\ssf}{{\int_g}}
\newcommand{\sff}{{\int_{\,g\,g'}}}
\newcommand{\fiu}{{g}}

\newcommand{\intfi}{{{\frac{1}{2\pi}}\,{\int_0^{2\pi}}
\,\hbox{d}\theta\,}}
\newcommand{\Nz}{{\NO_0}}

\newcommand{\exup}{{\TIgu}}
\newcommand{\exum}{{\TIgu^{-1}}}
\newcommand{\exdp}{{\TIgv}}
\newcommand{\exdm}{{\TIgv^{-1}}}
\newcommand{\gb}{\alpha}
\newcommand{\gbd}{\mbox{\boldmath$\gb$}}
\newcommand{\gbt}{\mbox{\boldmath$\gb$}^\dagger}
\newcommand{\si}{\sigma}
\newcommand{\ab}{a}
\newcommand{\abd}{{\ab}}
\newcommand{\abt}{{\ab^\dagger}}
\newcommand{\uni}{ {\hat {\bf 1} }  }
\newcommand{\den}{\rho}
\newcommand{\kap}{\kappa}
\newcommand{\hhh}{h}
\newcommand{\del}{\Delta}
\newcommand{\kapp}{{\kap_+}}
\newcommand{\kapm}{{\kap_-}}
\newcommand{\delp}{{\del_+}}
\newcommand{\delm}{{\del_-}}

\newcommand{\ebcstz}{{{\tilde e}_0}}

\newcommand{\epbcsp}{{\epsilon_p}}
\newcommand{\ebcsp}{{e_p}}
\newcommand{\ebcstp}{{{\tilde e}_p}}
\newcommand{\adbcsp}{{a_p}}
\newcommand{\acbcsp}{{a^\dagger_p}}
\newcommand{\adbcsb}{{a_{\bar p}}}
\newcommand{\acbcsb}{{a^\dagger_{\bar p}}}
\newcommand{\bdbcsp}{{b_p}}
\newcommand{\bcbcsp}{{b^\dagger_p}}
\newcommand{\bdbcsb}{{b_{\bar p}}}
\newcommand{\bcbcsb}{{b^\dagger_{\bar p}}}
\newcommand{\ubcsp}{{u_p}}
\newcommand{\vbcsp}{{v_p}}

\newcommand{\tbcsp}{{t_p}}
\newcommand{\qbcsp}{{t_p^{-1}}}

\newcommand{\Gbcspq}{{G_{pq}}}

\newcommand{\ebcsq}{{e_q}}
\newcommand{\ebcstq}{{{\tilde e}_q}}
\newcommand{\ubcsq}{{u_q}}
\newcommand{\vbcsq}{{v_q}}
\newcommand{\tbcsq}{{t_q}}
\newcommand{\qbcsq}{{t_q^{-1}}}
\newcommand{\adbcsq}{{a_q}}

\newcommand{\adbcsc}{{a_{\bar q}}}

\begin{document}
\title{Extended BCS Theories}

\author{R. BALIAN\\
CEA/Saclay, Service de Physique Th\'eorique\\
91191 Gif-sur-Yvette Cedex, France
\and
H.FLOCARD and M. V\'EN\'ERONI\\
Division de Physique Th\'eorique\thanks{
Unit\'e de recherches des Universit\'es Paris XI et Paris VI
associ\'ee au CNRS. E-mail: flocard@ipno.in2p3.fr
}, Institut de Physique Nucl\'eaire\\
91406 Orsay Cedex, France
}
\date{June 11 1997}

\maketitle
\vspace{1cm}


\vspace{1cm}

\vspace{1cm}

IPNO/TH  97-13

\newpage 
\begin{abstract}
Extensions of the Hartree-Fock-Bogoliubov theory are worked out
which are tailored for, (i) the consistent  evaluation of fluctuations
and correlations  and (ii) the restoration through projection of
broken symmetries. For both purposes we rely on a single
variational principle which optimizes the characteristic function.
The Bloch equation is used as a constraint to define the equilibrium
state, and the trial quantities are a density operator and a
Lagrangian multiplier matrix which is
akin to an observable. The conditions of stationarity
are respectively a Schr\"odinger--like equation and a
Heisenberg--like equation with an imaginary time running backwards.
General conditions for the trial spaces are stated that
warrant the preservation of thermodynamic relations.
The connection with the standard minimum principle for thermodynamic
functions is discussed.
When the trial spaces are chosen to be of the
independent--quasi--particle type, the ensuing coupled equations
provide an extension of the Hartree--Fock--Bogoliubov approximation,
which optimizes the characteristic function.
An expansion of the latter in powers of its sources yields
for the fluctuations and correlations
compact formulae in which the RPA kernel emerges
variationally. Variational expressions for thermodynamic
quantities or characteristic functions are also obtained with projected
trial states,
whether an invariance symmetry is broken or not. In particular, the
projection on even or odd particle number is worked out for a pairing
Hamiltonian, which leads to new equations replacing the BCS ones.
Qualitative differences between even and odd systems,
depending on the temperature $T$, the level density and
the strength of the pairing force,
are investigated analytically and numerically.
When the single-particle level spacing is small compared to the
BCS gap $\Delta$ at zero temperature,
pairing correlations are effective, for both even and odd projected cases,
at all temperatures below the BCS critical temperature
$T_{\rm c}$. There exists a crossover temperature $T_{\times}$
such that odd-even effects disappear for $T_{\times}<T<T_{\rm c}$.
Below $T_{\times}$, the free-energy difference between odd and even
systems decreases quasi-linearly with $T$. The low temperature
entropy for odd systems has the Sakur-Tetrode form.
When the level spacing is comparable with $\Delta$, pairing in odd systems
takes place only between two critical temperatures,
exhibiting thus a reentrance effect.

\end{abstract}

\newpage
\begin{center}
Contents
\end{center}

\noindent 1. {\it Introduction}

\noindent 2. {\it The Variational Principle.}
\newline
2.1 {The Action-like Functional and the Stationarity Conditions.}
\newline\noindent
2.2 {General Properties.}
\newline\noindent
2.3 {Cumulants of Conserved Observables.}
\newline\noindent
2.4 {Connection with the Standard Variational Principle.}
\newline\noindent
2.5 {Variational Principle for $\DIh\uu$ Alone.}

\noindent 3. {\it Extended Finite-Temperature HFB
Approximation for Characteristic Functions.}
\newline\noindent
3.1 {Generalities and Notations.}
\newline\noindent
3.2 {The Variational Ans\"atze and the Associated Functional.}
\newline\noindent
3.3 {The Coupled Equations.}

\noindent 4. {\it Expansion of the Extended HFB Approximation 
: Fluctuations and Correlations.}
\newline\noindent
4.1 {The HFB Approximation Recovered.}
4.1.1 {Thermodynamic Quantities.}
4.1.2 {Expectation Values of Observables.}
\newline\noindent
4.2 {Fluctuations and Correlations of Conserved Observables.}
4.2.1 {The Particle Number Operator.}
4.2.2 {Characteristic Function for Conserved 
Single-Quasi-Particle Observables.}
4.2.3 {Broken Invariances.}
\newline\noindent
4.3 {Fluctuations and Correlations of Non-Conserved Observables.}
4.3.1 {Kubo Correlations.}
4.3.2 {Standard Correlations.}
4.3.3 {Diagrammatic Interpretation.}

\noindent 5. {\it Projected Extension of the Thermal
HFB Approximation.}
\newline\noindent
5.1 {Generalities and Notations.}
\newline\noindent
5.2 {The Variational Ans\"atze and the Associated Functional.}
\newline\noindent
5.3 {The Coupled Equations.}
\newline\noindent
5.4 {Unbroken $\PDh$-Invariance.}
5.4.1 {Notation and General Formalism.}
5.4.2 {Partition Function, Entropy.}

\noindent 6. {\it Projection on Even or Odd Particle Number.}
\newline\noindent
6.1 {Characteristic Function in the Particle-Number Parity 
Projected HFB Approximation.}
\newline\noindent
6.2 {The Partition Function in the Particle-Number-Parity
Projected HFB Approximation.}
\newline\noindent
6.3 {The Projected BCS Model: Generalities.}
6.3.1 {The Variational Space.}
6.3.2 {The Variational Equations.}
6.3.3 {Comments on the Equations.}
\newline\noindent
6.4 {The Projected BCS Model: Limiting Cases.}
6.4.1 {Low Temperatures.}
6.4.2 {High Temperatures.}
6.4.3 {Large Systems.}
\newline\noindent
6.5 {The Projected BCS Model: A Numerical Illustration.}
6.5.1 {Superdeformed Heavy Nuclei.}
6.5.2 {Mesoscopic Metallic Islands.}
6.5.3 {The Free-Energy Difference.}
6.5.4 {Ultrasmall Metallic Grains.}
6.5.5 {Comments.}

\noindent 7. {\it Summary and Perspectives.}

\noindent Appendix A {\it Geometric Features of the HFB Theory.}
\newline\noindent
A.1 {The Reduced Liouville Space.}
\newline\noindent
A.2 {Expansion of the HFB Energy, Entropy and Grand Potential.}
\newline\noindent
A.3 {The HFB Grand Potential and the RPA Equation.}
\newline\noindent
A.4 {Riemannian Structure of the HFB Theory.}
\newline\noindent
A.5 {Lie-Poisson Structure of the Time-Dependent HFB Theory.}

\noindent Appendix B {\it Liouville Formulation of the Projected
Finite-Temperature HFB Equations.}

\newpage

\setcounter{equation}{0}
\section[S1]{Introduction}

Let us consider a finite physical system in thermodynamic equilibrium.
We want to evaluate {\it variationally}
the thermodynamic functions of this system
along with the expectation
values, fluctuations and correlations of some set of observables
$\QDh_\ggg$.
The equilibrium state of the system
is defined from our knowledge of conserved
quantities such as the number of particles,
the energy, the total angular
momentum, the parity.
However, because our system of interest is finite, these
equilibrium data should be handled with care.
Indeed, depending on the circumstances, 
their values can be given in two
different ways, either with certainty or statistically,
and this allows us to classify
the conserved quantities into
two types.
Take for instance a finite, closed
system thermalized
with a heat bath.
Among the variables that determine its state,
the particle number $\NO$ is given with certainty ; we shall call $\NO$
a conserved quantity of type~I.
On the other hand, the energy is known only statistically as
only its expectation 
value is fixed; let us call it a conserved quantity of
type~II.
The Gibbs argument, which identifies a heat bath
with a large set of copies of the system, entails
that this system is in a canonical Boltzmann-Gibbs equilibrium
characterized by a Lagrangian
multiplier $\beta$ defining the inverse temperature.
Apart from $\NO$, our information is specified by $\beta$
or equivalently by the expectation value of the Hamiltonian.
Take now an open system which communicates
with heat and particle baths.
In this case, both the energy and the particle number are
conserved quantities of type~II, and the Gibbs argument leads
to the grand canonical distribution. 
For other equilibrium data one has likewise to recognize, according
to the circumstances, whether they are of type~I or type~II.
Note that, aside from the truly conserved quantities, the type~II
equilibrium data may also include nearly 
conserved quantities (or order parameters) which are here
assumed to have all been identified beforehand.

For macroscopic systems, this distinction between types~I and II
is immaterial
because the relative fluctuations that occur for
observables of type~II are
negligible in the thermodynamic limit.
Canonical and grand canonical distributions then become
equivalent.
When dealing with {\it finite systems}, however, it is important
to specify
carefully which statistical ensemble is relevant to the
physical situation under consideration.

This problem is encountered in various systems of current interest.
In nuclear physics, depending on the
experimental conditions
and/or the accuracy of the description, 
we may have to describe nuclei having either a
well-defined or a fluctuating particle number, angular momentum or 
energy.
Similar situations may occur for atoms or molecules.
Another example is that of metallic or atomic clusters
which can be prepared with a given particle number\cite{Mar96}.
Let us also mention mesoscopic rings for which a proper
description of the conductivity properties requires
that there be no fluctuations 
(i.e., a canonical distribution)
in the number of electrons\cite{Bou94}.
Likewise, the description of mesoscopic
superconducting islands\cite{DEU94} 
or metallic grains\cite{BRT96} demands a distribution with a
given parity of
the number of electrons. The recent
discovery of Bose-Einstein condensates 
in dilute alkali atomic gases is currently arousing great
interest\cite{GSS95} ; in this case also, the theoretical formulation 
requires us to use states with well-defined numbers of particles.

A standard formal method to assign a density operator
$\DDh$ to a system characterized by data of both types
proceeds as follows\cite{Bal91}.
The type~I data, referring to some conserved observables,
 determine a Hilbert space, or a
subspace $\HIu$ (for instance
the subspace with a given number of particles)
in which the density matrix $\DDh$ operates.
{On the other hand,} type II data
are {the} expectation values $\avOmk$  
of the remaining conserved observables
{which act} in $\HIu$.
{As a consequence, they} impose on $\DDh$ the constraints
\be\label{I010}
\avOmk=\Tru{\Omk\DDh}/\Tru{\DDh}\q,
\ee
where the trace $\Tru{{}}$ is meant to be taken on the
Hilbert subspace $\HIu$. 
{As we shall see, it is}
convenient to leave $\DDh$ unnormalized or, {equivalently}, not
to include the unit operator among the set $\Omk$.
Within the subspace $\HIu$ the state $\DDh$ is determined,
up to a normalization factor,
by maximizing the
von Neumann entropy
\be\label{I020}
S\equiv-\frac{\Tru{\DDh\,\log\DDh}}{\Tru{\DDh}}+\log\Tru{\DDh}\q,
\ee
under the constraints
(\ref{I010}).
Associating a Lagrangian multiplier $\lambda_k$ with each datum
$\avOmk$, one recovers the generalized
Boltzmann-Gibbs distribution
\be\label{I030}
\DDh\propto\hbox{e}^{-\sum_k\lambda_k\Omk}\q,
\ee
where $\lambda_k$
and $\avOmk$ are related to each other through
\be\label{I040}
\avOmk=-
\frac{\partial}{\partial\lambda_k}\log
\Tru{\hbox{e}^{-\sum_k\lambda_k\Omk}}\q.
\ee

In what follows, we take the form (\ref{I030})
of the density operator for
granted, irrespective of the mechanism that led to it.
Equilibrium distributions of the
form (\ref{I030}) can be the outcome of some
transport equation,
the detailed nature of 
which is beyond {the scope of this work}.
The form (\ref{I030}) can also be justified by various theoretical
arguments.
One of them is the above-mentioned maximum entropy criterion
which has its roots in information theory\cite{Jay57}.
Alternatively, the Gibbs argument provides a
 direct statistical derivation of the Boltzmann-Gibbs distribution.
An extension to non-commuting observables $\Omk$,
relying as in the classical case on Laplace's principle of
indifference, has been worked out in ref.~\cite{BBa87}.

Formally, the assignment
of the density operator
(\ref{I030}) to the system solves our problem
since the thermodynamic functions,
expectations values, fluctuations, correlations of the observables 
and all their cumulants are conveniently obtained
from the {\it characteristic function}
\be\label{I050}
\varphi(\xi)\equiv\log\Tru{\ADh(\xi)\DDh}\q,
\ee
where the operator $\ADh(\xi)$, depending on the
{sources} $\xi_\gamma$ associated with the observables
$\QDhg$,
is defined as
\be\label{I060}
\ADh(\xi)\equiv\exp(-\sum_\gamma\xi_\gamma\QDhg)\q.
\ee
Indeed, by expanding (\ref{I050}) in powers of the sources,
\be\label{I070}
\varphi(\xi)=\varphi(0)-\sum_\ggg\,\xi_\ggg\avQDhg+
{\textstyle\usd}\sum_{\ggg\delta}\,\xi_\ggg\xi_\delta\,C_{\ggg\delta}
-\cdots
\ee
one gets at first order
the expectation values $\avQDhg$ of the observables $\QDhg$.
At second order appear their correlations
\be\label{I080}
C_{\ggg\delta}\equiv{\textstyle\usd}
\langle\QDh_\ggg\QDh_\delta+\QDh_\delta\QDh_\ggg\rangle
-\avQDhg\langle\QDh_\delta\rangle
\ee
and fluctuations $\Delta Q^2_\ggg\equiv C_{\ggg\ggg}$, and in
higher orders appear the cumulants of higher rank.
Moreover, {since we chose not to enforce 
the normalization of} $\DDh$,
the generalized thermodynamic potential is given by the
zeroth-order term :
\be\label{I090}
\varphi\zz
=\log\Tru{\DDh}
=\log\Tru{{\rm e}^{-\sum_k\lambda_k\Omk}}
=S-\sum_k\lambda_k\avOmk
\q.
\ee

Unfortunately,
in most physical cases, the calculation above cannot be
 worked out explicitly.
Indeed, the statistical operator (\ref{I030}) is tractable
only if all the operators $\Omk$ are of the
single-particle type.
For systems of interacting
particles, this is no longer the case
{since, in general,} the data $\avOmk$ include the average
energy $\langle\HDh\rangle$.
Recourse to some approximation scheme then becomes unavoidable.
Many approaches start with the
replacement of the density operator
(\ref{I030}) by one with the form
\be\label{I100}
\DIh\propto \hbox{e}^{-\MDh}\q,
\ee
where $\MDh$ is a single-particle 
operator $\sum_{ij}M_{ij}\abt_i\abd_j$, or
more generally a single quasi-particle operator
which includes also pairs $\abt_i\abt_j$ and $\abd_i\abd_j$.
(In this article, we shall mainly deal with systems of
fermions ; for the boson case,
terms linear in the operators $\abt_i$ and $\abd_i$ 
{ should be added to} $\MDh$.)

However, the form (\ref{I100}) {\it entails
two difficulties}, both of which will be
adressed in this article.

(i) The first difficulty concerns {\it the optimum selection of the
independent-quasi-particle approximate state} (\ref{I100}).
One often uses a variational criterion, and this is
an approach that we shall follow also.
[Conventional many-body perturbation theory 
is not suited to the problems we have in mind
since no small parameter is available.]
The choice of this criterion, however,
raises a crucial question.
A standard option, which leads to
the Hartree-Fock (or
Hartree-Fock-Bogoliubov) solution,
 consists in choosing
the minimization of the
free energy or, at zero temperature, of the energy.
This procedure is clearly adapted to the evaluation of
the thermodynamic quantities that are given 
by $\varphi\zz$, the value of Eq.~(\ref{I050}) when the
sources $\xi_\gamma$ vanish (see Eq.~(\ref{I090})).
However, the resulting approximate state is not
necessarily suited to
an optimization of
the characteristic function $\varphi(\xi)$
for non-zero values of {the sources} $\xi_\gamma$.
For restricted trial spaces
such as (\ref{I100}) (or (\ref{I110})),
a variational method designed to evaluate
$\varphi(\xi)$, rather than $\varphi\zz$,
is expected to yield an optimal independent-particle density
operator which,
owing to its dependence on the intensity of the sources,
is better adapted to our purpose than
the ($\xi$-independent) Hartree-Fock solution.

(ii) The second difficulty is more familiar.
Let us for instance consider a system for which the
particle number is of type I and the energy of type II.
It should be described by 
a canonical equilibrium distribution
$\DDh\propto\hbox{e}^{-\beta\HDh}$ built up in the $\NO$-particle
Hilbert space $\HIu$. Since
our system is finite, the results are expected to differ
from those of a grand canonical equilibrium. 
However, perturbative and variational
approaches are often worked out more conveniently in the entire
Fock space $\HI$.
In particular, even an independent-particle state such as (\ref{I100})
where $\MDh=\sum_{ij}M_{ij}\abt_i\abd_j$ 
resides in this Fock space $\HI$.
Thus the simple Ansatz (\ref{I100}) requires 
leaving the space
$\HIu$ to which the exact canonical state belongs;
as a consequence, the
standard variational treatment introduces spurious components
into the trial state.
{\it Unphysical thermodynamic fluctuations} 
of the particle-number 
thus arise from the form of the trial state (\ref{I100}),
which would be better suited to a grand canonical than 
to the canonical equilibrium being approximated. 
These fluctuations should 
be eliminated, and it is natural to improve the approximation by 
introducing a {\it projection} 
onto states with the right number of particles.
This is {achieved} by replacing Eq.~(\ref{I100}) by the
more elaborate Ansatz
\be\label{I110}
\DIh\propto\PDh_\NO\,\,\hbox{e}^{-\MDh}\,\PDh_\NO\q,
\ee
where $\PDh_\NO$ is the projection onto the subspace
$\HIu$ with the exact particle number $\NO$.
The characteristic function (\ref{I050}) can now be
evaluated by taking the trace $\Tr{}$ in the full Fock space $\HI$.
Likewise, when the exact state has a non-fluctuating angular momentum
or spatial parity, any trial state of the form (\ref{I100})
introduces thermodynamic
fluctuations for these quantities, and
adequate projections $\PDh$ are required 
so as to suppress these fluctuations.
However, if the operator $\MDh$
commutes with $\PDh$, it is not necessary
in Eq.~(\ref{I110}) to insert the projection
$\PDh$ on both sides of the exponential.

Projections turn out to be very useful in another 
circumstance, namely when a symmetry is broken by the operator $\MDh$. 
Since in this case $\MDh$ does not commute with $\PDh$,
the projection should appear on both sides as in (\ref{I110}).
Although projection matters mostly for finite systems,
let us first recall, for the sake of comparison, the situation
in the infinite systems. 

Consider a set of interacting particles
which, in the thermodynamic limit, presents a 
{\it spontaneously broken symmetry}. One can think of a 
Bose liquid or a superconducting electron gas where the
$\NDh$-invariance is broken.
In such a system an essential property of the exact 
statistical operator $\DDh$ is
the long-range order displayed by the 
associated $n$-point functions.
For an interacting Bose fluid at low temperature, 
if we denote by $\psi^\dagger({\bf r})$
the creation operator at the point $\bf r$, 
expectation values over $\DDh$ such as 
$\langle\psi^\dagger({\bf r}_1)\psi({\bf r}_2)\rangle$,
$\langle\psi^\dagger({\bf r}_1)\psi^\dagger({\bf r}_2)
\psi({\bf r}_3)\psi({\bf r}_4)\rangle$
factorize {in terms of  a single function $\varphi({\bf r})$} as
$\varphi({\bf r}_1)^*\varphi({\bf r}_2)$,
$\varphi({\bf r}_1)^*\varphi({\bf r}_2)^*
\varphi({\bf r}_3)\varphi({\bf r}_4)$ when the various
points all move apart from one another.
{On the other hand, since} for any size of the system its
exact density operator $\DDh$ commutes with $\NDh$,
as does the Hamiltonian $\HDh$, {the quantity}
$\langle\psi({\bf r})\rangle$ vanishes.
In order to give a meaning
to the factors $\varphi({\bf r})$,
a standard procedure consists of adding to $\HDh$
small sources which do not commute with $\NDh$: for instance
terms in $\abd_i$ and $\abt_i$ for bosons, or in $\abt_i\abt_j$ 
and $\abd_i\abd_j$ for fermions.
The resulting equilibrium density operator $\DDh_{\rm B}$ 
displays therefore an explicit broken invariance.  
Despite the fact that such sources do not exist in nature,
the remarkable properties of $\DDh_{\rm B}$ 
make it an almost inevitable substitute for $\DDh$.
Indeed, in the limit when one first lets 
the size of the system grow to infinity and then lets 
the sources tend to zero,
$\DDh_{\rm B}$ is equivalent to $\DDh$ as regards all the observables
that commute with $\NDh$, such as 
$\psi^\dagger({\bf r}_1)\psi({\bf r}_2)$
or $\psi^\dagger({\bf r}_1)\psi^\dagger({\bf r}_2)
\psi({\bf r}_3)\psi({\bf r}_4)$; however,
the expectation value $\langle\psi({\bf r})\rangle_{\rm B}$ 
over $\DDh_{\rm B}$ does not any longer vanish and is interpreted 
as an {\it order parameter}. Furthermore, $\DDh_{\rm B}$ 
{satisfies} the {\it clustering property} 
for all operators, even those which do not commute with $\NDh$; 
in particular, we now have 
$\langle\psi^\dagger({\bf r}_1)\psi({\bf r}_2)\rangle_{\rm B}\rightarrow
\langle\psi^\dagger({\bf r}_1)\rangle_{\rm B}
\langle\psi({\bf r}_2)\rangle_{\rm B}$ when 
$\vert{\bf r}_1-{\bf r}_2\vert\rightarrow\infty$
so that $\langle\psi^\dagger({\bf r})\rangle_{\rm B}$ 
can be identified with $\varphi({\bf r})$.
The clustering property of $\DDh_{\rm B}$ as well as the
equivalence of $\DDh$ and $\DDh_{\rm B}$ for the physical observables
were recognized for the pairing of electrons as early as 
the inception of the BCS theory\cite{BCS57}. Thus,
for sufficiently large systems, the same physical
results are obtained not only for different canonical ensembles, but also 
with or without the inclusion of sources that  
break invariances.

The situation is different for  a finite system
governed by the same interactions as in the infinite case above,
under similar conditions of density and temperature.
Here again the expectation value 
$\langle\psi({\bf r})\rangle$ is zero in 
the exact equilibrium state, but now
no analogue to $\DDh_{\rm B}$ can be devised. Indeed, we 
can no longer
define an order parameter by means 
of factorization of $n$-point functions
since the relative coordinates cannot grow 
to infinity. Nor can we introduce sources
which would produce non-vanishing expectation values
$\langle\psi({\bf r})\rangle_{\rm B}$ 
without changing the physical $n$-point 
functions, as this requires the sources to
tend to zero only after the thermodynamic limit has been taken.
Even the translational invariance of a nucleus, an atom or a
cluster cannot be broken in the {\it exact} ground state
or in thermal equilibrium states
if the Hamiltonian includes no external forces and commutes
with the total momentum. Returning to the $\NDh$-invariance,
for a nucleus with pairing or for a  
superconducting mesoscopic island or metallic grain, 
it is no longer without physical consequences to break
the ${\rm e}^{i\phi\NDh}$ gauge invariance. This not only
introduces spurious off-diagonal elements in $\NDh$ for the
density matrix $\DDh$, but also affects observable 
matrix elements in the diagonal blocks in $\NDh$.

Thus, the breaking of invariances does not have the same status 
for finite and infinite systems. In the latter case, it
stands as a conceptual ingredient 
of the theoretical description which
correctly implements the clustering property of the $n$-point 
functions. For finite systems it is 
also essential, but now as a basic tool
for building {\it approximation schemes} which economically
account for some main features of the correlations.
Both in perturbative and variational 
treatments, the starting point is most often an 
approximate density operator of the form (\ref{I100}) which does
not commute with conserved observables such as $\NDh$.
This yields tractable and sometimes remarkably accurate 
approximations, for which the clustering of the correlation functions
arises naturally from Wick's theorem.
Remember the success of the
BCS theory in describing pairing 
correlations betweens nucleons in finite nuclei.
(One may wonder why 
nuclear physicists, who had acknowledged 
the existence of pairing effects long before 1957, did not 
anticipate the BCS theory ; it may be that, too obsessed by
the conservation of the particle number, 
they were inhibited in developping models which 
would break the $\NDh$-invariance.)

Nonetheless, more realistic descriptions of finite systems
require the restoration of the broken invariances. A
state of the form (\ref{I100}) involves
fluctuations in the particle number that
must be eliminated, whether they
arise from a grand canonical description,
or from unphysical off-diagonal elements {of} $\NDh$
{connecting} Hilbert spaces with different 
particle numbers.
A natural procedure to account for pairing correlations and at 
the same time enforce the constraint on the particle number
consists in using a projection.
The use of Eq.~(\ref{I110}) arises naturally in this context.
Likewise, breaking translational or rotational
invariance introduces spurious fluctuations 
of the linear or angular momentum, as well as 
spurious off-diagonal elements in the trial density operator 
(\ref{I100}). To remedy these defects we need again a trial
density operator of the form (\ref{I110})
involving the appropriate projection $\PDh$ on {\it both} sides.

The discussion above pertains to the breaking 
of a conserved quantity of type~I,
that is, one which is 
known with certainty , such as $\NDh$ in a 
canonical ensemble. For a quantity of type~II, known only
on average, such as $\NDh$ in a grand canonical ensemble, 
the invariance breaking again introduces off-diagonal elements 
{of} $\NDh$, which are spurious for a finite system.
One can suppress them, while letting $\NDh$ free to fluctuate, 
by means of a different type of projection, exhibited below 
in Eq.~(\ref{ha100}).

The problem of the restoration of broken symmetries has already had
a long history. Within the nuclear context, in the zero-temperature
limit, a general review of the litterature (up to 1980)
may be found in \cite{RSc80}. For a 
recent work about projection at finite
temperature and its connection with the static path approximation
see Ref.~\cite{RRi94}, 
where more references can be found. 
An earlier reference on parity projection is \cite{TSM81}. 
For a Monte-Carlo approach to the temperature dependence of
pair correlations in nuclei, see Ref.\cite{LDR96}.
In condensed matter physics, recent
experimental studies about isolated mesoscopic 
metallic islands\cite{DEU94} or small metallic grains\cite{BRT96}
have motivated several theoretical works about modified BCS 
theories with well defined number parity\cite{ANa92}-\cite{MLa97}. 

The two problems (i) and (ii) that we have
stated above will be 
worked out within a general variational setting described
in Sect.~2.
We shall write an action-like
functional, Eq.~(\ref{e040}), which yields as its stationary value the
characteristic function (\ref{I050}). This corresponds
to the static limit of a variational principle
proposed elsewhere to evaluate multi-time
correlation functions \cite{BVe93}.
The approach uses a method \cite{GRS83,BVe88}
which allows the construction of variational
principles (and variational approximations) tailored to
the optimal evaluation of the quantity of interest
(here the characteristic function).
This systematic and very general method encompasses several 
known variational principles, including
the Lippmann-Schwinger formalism \cite{LSc50}
(the quantity of interest being the $S$-matrix).
An important ingredient is the incorporation 
in the trial functional of 
the Lagrangian multipliers
associated with the primary equations, in our case  
the (symmetrized) Bloch equation 
determining the canonical density operator (\ref{I030}).
As a result, the number of variational 
degrees of freedom is doubled.
Here, the variational quantities entering our functional 
(\ref{e040}) consists of two matrices : 
{one akin to a density operator
and the other} to the observable $\ADh(\xi)$ defined
in (\ref{I060}). Within this
framework, one can state some general conditions on the trial spaces 
that ensure the preservation of the thermodynamic identities by the 
variational approximations.

Starting in Sect.~3, we apply the variational principle of Sect.~2
to systems of fermions for which pairing effects are significant.
Sect.~3 (as well as Sect.~4) deals with grand canonical
equilibrium.
The two variational operators introduced in
Sect.~2 are taken as
exponentials (\ref{f040}) of single-quasi-particle 
operators and
the resulting expression of the approximate action-like functional
is given in Sect.~3.2.
The equations
which express the stationarity
and determine the optimal characteristic function, are
derived in Sect~3.3. 
They couple, with
mixed boundary conditions, the variational
parameters characterizing the two trial matrices
and they generalize 
the Hartree-Fock-Bogoliubov (HFB) approximation\cite{Bog59}.

{In Sect.~4,} the coupled equations obtained in 
Sect.~3 are expanded in powers of the sources $\xi_\gamma$.
The HFB approximation is recovered at the zeroth order, while
the first order gives back the usual HFB
formula for the average values of observables (Sect.~4.1).
Fluctuations and correlations of observables are
provided by the second order.
In the ensuing formulas for conserved and 
non-conserved observables,
the RPA (random-phase approximation) kernel plays a central r\^ole.
In Sect.~4.3.3, through a diagrammatic analysis, we show 
the correspondence between our results and
the summation of the RPA (bubble) diagrams.

Sect.~5.1 addresses the projection problem
stated in (ii) above. The functional and the
coupled equations that result from variational Ans\"atze 
of the type (\ref{I110}) are derived
in Sects.~5.2 and 5.3, respectively.
Special attention is devoted to the case where the
variational approximation does not break the 
$\PDh$-invariance (Sect.~5.4).

Sect.~6 specializes the formalism of Sect.~5 to
the projection
on even or odd particle number.
The more detailed study of this relatively simple case 
is motivated by its relevance to the description of heavy nuclei and 
to the interpretation of recent
experiments on isolated superconducting islands\cite{DEU94}
or small metallic grains\cite{BRT96}.
Equations are obtained in Sect.~6.3 which 
should replace the BCS ones. 
Ensuing results are compared to those of Ref.~\cite{JSA94}.
In Section 6.4 we analyze some limiting 
situations with a special emphasis
on the low temperature limit. 
Numerical applications are presented in Sect.~6.5; they illustrate
physical situations encountered in nuclei, metallic islands or grains.

In the last Section we summarize our main results 
and present some suggestions for other applications
and for extensions.

Appendix~A introduces our notation
for the Liouville representation of the reduced 
{fermion quasi-particle} space.
In particular, it gives a proof of the 
factorization of the RPA 
kernel (used in Sect.~4) into
the stability matrix and the matrix
whose elements are the constants associated
with the Lie-Poisson structure {of the HFB equations}.

In Appendix~B, using the 
Liouville-space notation, we reexpress
the equations derived in Sect.~5
in a fashion which {exhibits} their formal similarity
with those obtained in Sect.~3.

The length of this paper is partly a consequence of our wish
to allow several reading options.
The contents of Sects.~5 and 6 are largely independent of that of
Sect.~4. The reader primarily interested by the variational
projection method can therefore bypass Sect.4, except for the 
brief Sect.~4.1. If he is mostly concerned by the projection on 
even or odd particle-number, and ready to admit a few formulas 
demonstrated in Sect.~5, he can directly jump to Sect.~6.
If his interest is focused upon the 
improvements over BCS introduced by parity-number
projection, he may even begin with Sect.~6.3. 
Alternatively, the reader mainly interested by 
the variational evaluation of correlations and fluctuations in the grand 
canonical formalism can skip Sects.~5 and 6.
\newpage

\section[S2]{The Variational Principle}
\setcounter{equation}{0}

Our purpose in this Section is to write a variational expression adapted
to the evaluation of the characteristic function (\ref{I050}), namely
\be\label{e010}
\exp\varphi(\xi)\equiv\Tru{\ADh(\xi)\DDh}\q.
\ee
The operator $\ADh$, defined in Eq.~(\ref{I060}), involves the observables
of interest $\QDhg$ and depends on the associated
sources $\xi_\gamma$.
The density operator $\DDh$ describes thermodynamic equilibrium. It is
an exponential of the conserved quantities
$\Omk$ of type II that are only
given on average (Eq.~(\ref{I030})).
From now on we assume that the Hamiltonian $\HDh$
is one of these conserved quantities, with $\beta$
as its conjugate variable.
We thus write $\DDh$ in the form
\be\label{e020}
\DDh=e^{\displaystyle -\bb\KDh}
\q,\q\q\q\bb\KDh\equiv\sum_k\lambda_k\Omk\q.
\ee
We have not introduced the unit operator among the set $\{\Omk\}$.
Then $\varphi\zz$ does not vanish but is identified
(Eq.~(\ref{I090})) with the
logarithm of the partition function
$\Tru{{\rm e}^{-\bb\KDh}}$,
for instance of the
canonical partition function for $\KDh\equiv\HDh$ or of
the grand canonical partition
function for $\KDh\equiv\HDh-\mu\NDh$. Thus, our variational principle
will yield not only the expectation values and cumulants of the set
$\{\QDhg\}$, but also the thermodynamic quantities.
As regards the conserved quantities of type~I, they define the Hilbert
space $\HIu$ over which the
trace $\Tru{}$ is taken in Eq.~(\ref{e010}).
According to the circumstances, $\HIu$ can be the Fock space, as in
Sects.~3 and 4, or a more restricted space.
Sects.~5 and 6 resort in principle to the
latter case; nonetheless, the introduction, as in Eq.~(\ref{I110}),
of a projection onto $\HIu$
will allow us to use again the trace $\Tr{}$ in the full Fock space.

\subsection[S21]
{The Action-like Functional and the Stationarity Conditions}

We wish to replace the density operator (\ref{e020})
by a variational approximation $\DIh\bbb$, since $\DDh$ itself
is untractable. With this in mind, we
note that $\DDh={\rm e}^{-\bb\KDh}$
can be constructed as the solution $\DIh\bbb$
of the Bloch equation, or rather of its symmetrized form
\be\label{e030}
\duh{\DIh} + \usd[\KDh\DIh+\DIh\KDh] = 0\q,
\ee
where $\DIh\uu$ is a $u$-dependent
operator ($0\leq u\leq\beta$) subject to the initial
condition $\DIh\zz=\uni$.
In accordance with the general method \cite{GRS83,BVe88}
devised to evaluate variationally a quantity of interest,
here (\ref{e010}),
we regard the equation (\ref{e030})
as a
set of constraints which impose 
the canonical form
$\DIh\bbb=e^{-\bb\KDh}$, and we associate
with it a Lagrangian multiplier $\AIh\uu$
which is also a $u$-dependent operator.
{\it The characteristic function (\ref{e010}) is then the stationary
value of the action-like functional}
\be\label{e040}\ba{l}
\II\{\DIh(u),\,\AIh(u)\}\equiv\Tru{\ADh\DIh\bbb}\\
\q\q\q\displaystyle
-\intzb
\Tru{\AIh\uu\pag \duh{\DIh\uu} +\usd[\KDh\DIh\uu+\DIh\uu\KDh]\pad}
\q,
\ea\ee
under arbitrary variations of the trial operators
$\AIh(u)$ and $\DIh(u)$,
the latter being subject to the boundary condition
\be\label{e050}
\DIh(0)=\uni\q.
\ee

The specifics of our problem
enter the functional (\ref{e040}) through the temperature $1/\beta$\,,
the operator
$\KDh$ and through the
operator $\ADh$ (a functional
of the observables).

A variational approximation for $\exp\varphi(\xi)$
will be generated
by restricting the trial class for  $\AIh(u)$ and $\DIh(u)$
and by looking for the corresponding stationary value of (\ref{e040}).
The error will be of second order in
the difference between the optimum trial operators and the
exact ones. An essential feature of
the method is the doubling of the number of variational
degrees of freedom, which include not only the trial state $\DIh(u)$
but also the Lagrangian multiplier operator $\AIh(u)$.
This doubling is beneficial only when approximations are
made; indeed, for unrestricted variations of $\AIh(u)$ the
stationarity conditions
on $\AIh(u)$ alone reduce to the Bloch equation
(\ref{e030}) and are sufficient to ensure that
$\II\{\DIh,\,\AIh\}$ yields the exact value
for $\exp\varphi(\xi)$.

When $\DIh(u)$ and $\AIh(u)$ are restricted to vary within some subset,
the stationarity conditions of the functional $\II\{\DIh,\AIh\}$
with respect to $\AIh\uu$
are read directly from Eq.~(\ref{e040}):
\be\label{e060}
\Tru{\delta\AIh\,\pag \duh{\DIh} + \usd[\KDh\DIh+\DIh\KDh]\pad}=0\q.
\ee
As regards the variation
with respect to $\DIh\uu$,
an integration by parts
allows us to rewrite the functional (\ref{e040}) in the form
\be\label{e070}\ba{l}
\II\{\DIh(u),\,\AIh(u)\}=\Tru{\AIh\zz}+\Tru{\DIh\bbb(\ADh-\AIh\bbb)}\\
\q\q\q\displaystyle +\intzb\Tru{\DIh(u)\,\pag
\frac{\hbox{d}\AIh(u)}{\hbox{d}u}
-\usd[\AIh(u)\KDh+\KDh\AIh(u)]\pad}\q.
\ea\ee
From Eq.~(\ref{e070}), for $0\le u<\bb$,
one obtains the stationarity
conditions:
\be\label{e080}
\Tru{\delta\DIh\,\pag \duh{\AIh} - \usd[\AIh\KDh+\KDh\AIh]\pad}=0\q,
\ee
and for $u=\bb$:
\be\label{e090}
\Tru{\delta\DIh\bbb(\ADh-\AIh\bbb)}=0\q.
\ee
In Eqs. (\ref{e060}) and (\ref{e080}) we have omitted
the $u$-dependence of $\DIh$, of $\AIh$, and of the allowed
variations $\delta\DIh$ and $\delta\AIh$.

By construction of the functional (\ref{e040}), Eq.~(\ref{e060}) 
gives back the symmetrized Bloch equation (\ref{e030})
for unrestricted variations of $\AIh$.
For unrestricted variations of $\DIh$,
Eqs.~(\ref{e080}) and (\ref{e090})
yield (in the Hilbert space $\HIu$) 
the Bloch-like equation for $\AIh\uu$
\be\label{e100}
\duh{\AIh} - \usd[\AIh\KDh+\KDh\AIh] = 0\q
\ee
in the interval $0\leq u <\beta$,
and the boundary condition
\be\label{e110}
\AIh\bbb=\ADh
\ee
at the final value $u=\bb$. The variable
$u$ is thus evolving {\it backward} from $\beta$ towards 0
in Eq. (\ref{e100}), the solution of which is
\be\label{e120}
\AIh\uu=\exp[{\scriptstyle \usd}
(u-\bb)\KDh]\ADh\exp[{\scriptstyle \usd}(u-\bb)\KDh]\q.
\ee

The exact equation (\ref{e100}),
which is analagous
to a backward Heisenberg equation in the ``time'' $u$,
merely duplicates the Schr\"odinger-like
equation (\ref{e030}).
However, in any physical application,
the trial spaces for $\DIh(u)$ and $\AIh(u)$
have to be restricted
to make the calculations tractable.
Then, in general, Eqs.~(\ref{e060}) and (\ref{e080}) are coupled
and must be solved simultaneously.
Moreover, the boundary conditions (\ref{e050}) on $\DIh\zz$ and
(\ref{e090}) on $\AIh\bbb$ imply that Eqs.~(\ref{e060})
and (\ref{e080}) should be solved in the forward and 
backward directions
in $u$, respectively.

\subsection[S22]{General Properties}

The stationary value $\IIs$ of the functional (\ref{e040})
depends on an ensemble of parameters $\eta$. Among them
are obviously the source parameters $\xi_\gamma$
[the dependence is through
the boundary condition (\ref{e090})] and the temperature $1/\beta$.
Other parameters $\oo$ may occur in the operator $\KDh$,
as when it includes
an external field $-\oo\JDh$.
In particular, for a grand canonical ensemble, where
$\KDh=\HDh-\mu\NDh$, the chemical potential $\mu$ is regarded as
one of the parameters $\oo$.
The symbol $\eta$ will therefore stand for
the set $\{\xi,\,\beta,\,\omega\}$.

We shall take advantage of a general
property of variational principles that we now state
in our notation.
The stationary value $\IIs(\eta)$ which is sought,
\be\label{e130}
\IIs(\eta)\equiv
\IIs\{\ADh(\xi),\,\beta,\,\KDh\}=
\II\{\DIh_\eta,\,\AIh_\eta,\,\eta\}\q,
\ee
depends on the parameters $\eta$ both directly and through
the optimal solution $\{\DIh_\eta$, $\AIh_\eta\}$
determined within our variational class by the stationarity conditions
$\delta\II\{\DIh,\,\AIh,\,\eta\}/\delta\DIh=0$ and
$\delta\II\{\DIh,\,\AIh,\,\eta\}/\delta\AIh=0$.
Owing to these conditions, the
full derivative of $\IIs$ with respect to $\eta$
reduces to its partial
derivative evaluated at the stationary point :
\be\label{e140}
\frac{\hbox{d}\ }{\hbox{d}\eta}\IIs(\eta)=
\left.\frac{\partial\ }{\partial\eta}\II\{\DIh,\,\AIh,\,\eta\}
\right\vert_{{\DIh=\DIh_\eta}\atop{\AIh=\AIh_\eta}}\q.
\ee
This relation holds not only for
the exact solution but also within any trial
class for $\AIh$ and $\DIh$.

For any variational space, the derivatives
$\hbox{d}\DIh/\hbox{d}u$ and
$\hbox{d}\AIh/\hbox{d}u$ belong to the set
of allowed variations
$\delta\DIh$ and $\delta\AIh\,$, respectively.
Writing Eqs.~(\ref{e060}) and (\ref{e080}) for these variations
and subtracting one equation from the other, we obtain
along the stationary path
the conservation law
\be\label{e150}
\du\Tru{\KDh\,[\AIh_\eta\uu\DIh_\eta\uu+\DIh_\eta\uu\AIh_\eta\uu]}=0\q.
\ee

In the following, we shall always use for $\AIh$
a trial space that contains the unit operator
and in which variations $\delta\AIh$ proportional to $\AIh$
($\delta\AIh\propto\AIh$) are allowed.
We then find from Eq.~(\ref{e060}) that the integrand of the
functional (\ref{e040})
vanishes, so that the stationary value
is given by
\be\label{e160}
\IIs(\eta)=\Tru{\ADh\DIh_\eta\bbb}\q,
\ee
as in the exact expression (\ref{e010}).

Moreover when, as will be the case in this work, the trial
space of $\DIh$ is chosen
such that variations
$\delta\DIh$ proportional to $\DIh$ ($\delta\DIh\propto\DIh$)
are allowed, another helpful property emerges
(which of course holds also for the
exact solution).
Indeed, substituting $\delta\AIh\propto\AIh$  in Eq.~(\ref{e060})
and $\delta\DIh\propto\DIh$ in Eq.~(\ref{e080}) and
combining these equations, we obtain
\be\label{e170}
\du\Tru{\AIh_\eta(u)\DIh_\eta(u)}=0\q.
\ee
Using $\delta\DIh\propto\DIh$ in Eq.~(\ref{e090}),
we also find
\be\label{e180}
\Tru{\ADh\DIh_\eta\bbb}=\Tru{\AIh_\eta\bbb\DIh_\eta\bbb}\q,
\ee
even when $\ADh$ does not belong to the
variational space for $\AIh$.
The relations (\ref{e170}) and (\ref{e180})
imply that the approximation (\ref{e160})
for $\exp\varphi=\Tru{\ADh\DIh_\eta\bbb}$ is equal
to $\Tru{\AIh_\eta(u)\DIh_\eta(u)}$ for any value $u$,
and in particular that
\be\label{e190}
{\rm e}^{\displaystyle\varphi(\xi)}=\Tru{\AIh_\eta\zz}\q.
\ee

Let us now use Eq.~(\ref{e140}) when $\eta$
stands for the source parameters $\xi_\gamma$ ; we find :
\be\label{e200}
\frac{\hbox{d}\ }{\hbox{d}\xi_\gamma}\IIs\{\ADh(\xi),\,\beta,\,\KDh\}=
{\rm e}^{\displaystyle\varphi(\xi)}
\frac{\hbox{d}\varphi}{\hbox{d}\xi_\gamma}=
\Tru{\DIh_\eta\bbb\frac{\hbox{d}\ADh}{\hbox{d}\xi_\gamma}}\q.
\ee
This relation is valid for all values of
the $\xi$'s. According to Eq.~(\ref{I070}),
in the limit where these $\xi$'s vanish, 
Eq.~(\ref{e200}) yields
\be\label{e210}
-\left.
\frac{\hbox{d}\varphi}{\hbox{d}\xi_\gamma}
\right\vert_{\xi=0}
=\langle\QDhg\rangle=
\frac{\Tru{\QDhg\DIh_{\xi=0}\bbb}}{\IIs\{\uni,\,\beta,\,\KDh\}}\q,
\ee
where, as in the rest of this Section, the subscript
$\xi=0$ is shorthand for $\{\xi=0,\,\beta,\,\omega\}$.
Since it follows from Eq.~(\ref{e160}) that
\be\label{e220}
\IIs\{\uni,\,\beta,\,\KDh\}=\Tru{\DIh_{\xi=0}\bbb}\q,
\ee
one has
\be\label{e230}
\langle\QDhg\rangle=
\frac{\Tru{\QDhg\DIh_{\xi=0}\bbb}}{\Tru{\DIh_{\xi=0}\bbb}}\q.
\ee
This result tells us that the calculation
of the expectation values of any operator
requires only the solution of Eqs.~(\ref{e060}) and
(\ref{e080}) for $\ADh=\uni$.
Equation (\ref{e230}) shows therefore that
{\it the density operator} $\DIh_{\xi=0}\bbb$
{\it suited to the calculation of the thermodynamic potentials
is also suited to the variational evaluation of
the expectation values of the
observables} $\QDhg$.
The knowledge of the $\xi$-dependence of
$\DIh_\xi\bbb$ is only needed for the calculation of the
higher-order cumulants.

To find the derivative of $\IIs$ with respect to
the inverse temperature $\beta$,
we rewrite the functional (\ref{e040}) as
\be\label{e240}\ba{rl}
\II=&\displaystyle
\int_0^\infty\,\hbox{d}u\,\hbox{Tr}_1
\acg\ADh\,\DIh\uu\delta(u-\beta)\,
\vphantom{\duh{\DIh\uu}}\right.\\
&\displaystyle \q\q-\left.\AIh\uu\pag
\duh{\DIh\uu} +\usd[\KDh\DIh\uu+\DIh\uu\KDh]
\pad\theta(\beta-u)
\acd\q.
\ea\ee
We then apply Eq.~(\ref{e140}) with $\eta$ replaced by $\beta$.
Noting that $\hbox{d}\DIh\uu/\hbox{d}u\vert_{u=\beta}$
belongs to the variational class
allowed for $\delta\DIh\bbb$ and using (\ref{e090}), we obtain
\be\label{e250}
\db\IIs=\hbox{e}^{\displaystyle\varphi}\,
\frac{\hbox{d}\varphi}{\hbox{d}\beta}=
-\usd\Tru{\KDh\,[\AIh_\eta\bbb\DIh_\eta\bbb
+\DIh_\eta\bbb\AIh_\eta\bbb]}\q,
\ee
in which $\AIh_\eta\bbb$ can be replaced by $\ADh$
when the variational space for $\AIh$ contains $\ADh\ $.
When $\ADh=\uni$, Eq.~(\ref{e250}) becomes
\be\label{e260}
-\frac{\hbox{d}\varphi}{\hbox{d}\beta}=
-\db\log\Tru{\DIh_{\xi=0}\bbb}=
\frac{ \Tru{\KDh\DIh_{\xi=0}\bbb}}
{ \Tru{\DIh_{\xi=0}\bbb} }=
\langle\KDh\rangle\equiv\EI \q,
\ee
where
we have used Eq.~(\ref{e230}) and introduced
the notation $\EI$ for the equilibrium energy
in the canonical case ($\KDh\equiv\HDh$), or the expectation value
$\langle\HDh-\mu\NDh\rangle$ in the grand-canonical case.
Thus, in the framework of our variational method
({ for any trial space allowing $\delta\AIh\propto\AIh$}),
the derivative with respect to $\beta$
of the approximate thermodynamic potential ($\ADh(\xi)=\uni$)
provides the same result as the expectation value of $\KDh$
obtained from Eq.~(\ref{e230}) with $\QDh=\KDh$
($\ADh(\xi)=\exp-\xi\KDh\,$). 
We can also define an approximate thermodynamics potential
\be\label{e261}
F_{\rm G}\equiv-\bb^{-1}\varphi(0)\q,
\ee
and an approximate entropy
\be\label{e262}
S\equiv\bb(E-F_{\rm G})=\varphi(0)
-\bb\frac{\hbox{d}\varphi(0)}{\hbox{d}\beta}\q,
\ee
which, owing to Eq.~(\ref{e260}), satisfy the relation
\be\label{e263}
S=-\frac{\partial F_{\rm G}}{\partial T}\q.
\ee
Our approximation is then
{\it consistent with the fundamental thermodynamic relations} between the
partition function, the temperature, the energy and the entropy.
Note that $S$ as defined by Eq.(\ref{e262}) is not necessarily
related to $\DIh_{\xi=0}\bbb$ by Eq.(\ref{I020}).

If we now use the relation (\ref{e140}) where
$\eta$ is identified with the parameter $\oo$
of an external field $-\oo\JDh$
entering  $\KDh$, we have
\be\label{e270}
\doo\IIs=\hbox{e}^{\displaystyle\varphi}
\frac{\hbox{d}\varphi}{\hbox{d}\oo}
=\usd\intzb\Tru{\JDh\,
[\DIh_\eta\uu\AIh_\eta\uu+\AIh_\eta\uu\DIh_\eta\uu]}\q.
\ee
When $\ADh=\uni$, i.e. when $\xi_\gamma=0$, we shall
see in Sect.~2.4 that under some rather general conditions
(satisfied for instance in Sects.~4.1, 5.4.2, 6.2 and 6.3), the products
$\DIh_\eta\uu\AIh_\eta\uu=\AIh_\eta\uu\DIh_\eta\uu$ do not depend on
the variable $u$. Then Eq.~(\ref{e270}) reduces to
\be\label{e280}
\frac{1}{\bb}\doo\log\Tru{\DIh_{\xi=0}\bbb}=
\frac{ \Tru{\JDh\,\DIh_{\xi=0}\bbb} }{ \Tru{\DIh_{\xi=0}\bbb} }
\q.
\ee
Here again, as for Eq.~(\ref{e260}), our
variational approximation is consistent
with the thermodynamic identities
expressing the expectation values of
observables $\JDh$ as derivatives
of the thermodynamic potential.

Let us finally consider the case when the variational classes for
$\DIh$ and $\AIh$ are such that
$\DIh^{-1}\delta\DIh$ and $\delta\AIh\,\AIh^{-1}$
belong to a common set of operators.
Then letting $\delta\DIh=\DIh\XDh$ and $\delta\AIh=\XDh\AIh$
in Eqs.~(\ref{e060}) and (\ref{e080})
and subtracting one equation from the other, we
find
\be\label{e290}
\Tru{\XDh\,( \duh{\AIh\DIh} + \usd[\AIh\DIh\,,\,\KDh])}=0\q.
\ee
Likewise, if we let
$\delta\AIh=\AIh\YDh$ and $\delta\DIh=\YDh\DIh$, we find
\be\label{e300}
\Tru{\YDh\,( \duh{\DIh\AIh} - \usd[\DIh\AIh\,,\,\KDh])}=0\q.
\ee
For unrestricted variations, $\XDh$ and $\YDh$ are arbitrary
and Eqs.~(\ref{e290}-\ref{e300}) imply that $\AIh\DIh$
and $\DIh\AIh$ both satisfy exact
equations of the Liouville-von Neumann type
(with a Hamiltonian $\pm i\KDh/2$) :
\be\label{e310}
\duh{\AIh\DIh} = \usd\,[\KDh\, ,\, \AIh\DIh]
\q,\q\q\q
\duh{\DIh\AIh} = -\usd\,[\KDh\, ,\, \DIh\AIh]
\q.
\ee
Note however
that the associated boundary conditions on $\AIh\DIh$
and $\DIh\AIh$ cannot be expressed in a simple form.
Indeed the conditions (\ref{e050}) and
(\ref{e110}) do not involve $\DIh$ and $\AIh$ 
evaluated at the same argument,
but $\DIh\zz$ and $\AIh\bbb$ evaluated
at the initial and final limits
of the integration interval.

In Sects.~3 and 4, $\AIh(u)$ and $\DIh(u)$ will be chosen within
the trial class
of exponentials of quadratic forms of creation and
annihilation operators.
All the properties of the variations $\delta\AIh$ and $\delta\DIh$
discussed above are then satisfied;
in particular, the quantities
$\delta\AIh\,\AIh^{-1}$, $\AIh^{-1}\,\delta\AIh$,
$\delta\DIh\,\DIh^{-1}$ or $\DIh^{-1}\,\delta\DIh$
span the set of quadratic forms.

\subsection[S23]{Cumulants of Conserved Observables}

The variational principle (\ref{e040}) provides
in general rather complicated expressions for the
correlations, and more generally for the higher order cumulants
of the observables $\QDhg$.
In this Subsection we consider a single observable
$\QDh$ which is conserved, that is, which commutes with the
operator $\KDh\,$.
{(The extension to several observables
commuting with $\KDh$, but not necessarily among themselves,
is readily performed by replacing $\xi\QDh$ by
$\sum_\ggg\xi_\ggg\QDhg$.)}
In the exact case, one has the obvious relation
\be\label{e320}
{\rm e}^{\displaystyle\varphi(\xi)}=
\Tru{ {\rm e}^{\displaystyle -\xi\QDh}\,
e^{\displaystyle -\bb\KDh}}=
\Tru{ {\rm e}^{\displaystyle -\bb(\KDh+\xi\QDh/\bb)} }\q,
\ee
which tells that the
exponential of the characteristic function
(\ref{I050}) associated with the
observable $\QDh$ is equal to the partition function
corresponding to the shifted operator
\be\label{e330}
\KDh^{\,\prime}(\xi)\equiv\KDh+\xi\QDh/\bb\q.
\ee

We shall see now that
the stationary solution of our variational
principle satisfies also the property (\ref{e320})
when the trial spaces
for the operators $\AIh$ and $\DIh$ are
invariant under left and
right multiplications by $\exp(-\lambda\QDh)$,
where $\lambda$ is any
c-number.
This invariance property is,
for instance, satisfied when
$\QDh$ is a single-quasi-particle
operator and when
$\AIh$ and $\DIh$ are, as in Sects.~3 and 4,
exponentials of a quadratic form. We shall also encounter
the property (\ref{e320}) in a different context in Sects.~5 and 6.

Let us indeed consider a trial set
$\{\AIh\uu,\,\DIh\uu\}$ satisfying
the boundary conditions (\ref{e050}) and (\ref{e110})
with $\ADh=\exp(-\xi\QDh)$.
We note that the two operators $\AIh'\uu,\,\DIh'\uu$
defined by
\be\label{e340}
\AIh'\uu=
{\rm e}^{\displaystyle u\xi\QDh/2\bb}
\,\AIh\uu\,
{\rm e}^{\displaystyle u\xi\QDh/2\bb}
\ ,\q
\DIh'\uu=
{\rm e}^{\displaystyle -u\xi\QDh/2\bb}
\,\DIh\uu\,
{\rm e}^{\displaystyle -u\xi\QDh/2\bb}
\ ,
\ee
obey the boundary conditions (\ref{e050}) and (\ref{e110})
with $\ADh=\uni$.
Moreover the value of the functional (\ref{e040}) calculated
with $\AIh'\uu,\,\DIh'\uu$,
$\ADh=\uni$ and $\KDh'$ is equal
to the value calculated with $\AIh\uu,\,\DIh\uu$,
$\ADh=\exp(-\xi\QDh)$ and $\KDh$.
By sweeping over
the trial set $\{\AIh\uu,\,\DIh\uu\}$ and using the 
above invariance of the variational spaces, we obtain
\be\label{e350}
\exp\varphi\{{\rm e}^{\displaystyle -\xi\QDh},\,\bb,\, \KDh\}
= \IIs\{{\rm e}^{\displaystyle -\xi\QDh},\,\bb,\, \KDh\}
=\IIs\{\uni,\,\bb,\,\KDh+\xi\QDh/\bb\}\q,
\ee
which, within our approach, is the variational
counterpart of the relation (\ref{e320}).
As a consequence, for any trial space satisfying this invariance,
{\it all the cumulants
at temperature $1/\bb$ of a conserved observable
can be obtained from a calculation of the partition function
for the shifted operator $\KDh^{\,\prime}$} defined in (\ref{e330}).
In particular, the mean value $\langle\QDh\rangle$ and the variance
$\Delta\QDh$ are given by
\be\label{e360}\ba{rl}
\langle\QDh\rangle
&\equiv\displaystyle\left.
-\frac{\partial}{\partial\xi}
\varphi\{{\rm e}^{\displaystyle -\xi\QDh},\,\bb,\,\KDh\}
\right\vert_{\xi=0}
=\displaystyle\left.
-\frac{1}{\bb}\frac{\partial}{\partial\lambda}
\varphi\{\uni,\,\bb,\,\KDh+\lambda\QDh\}
\right\vert_{\lambda=0}\q,\\
\\
\Delta\QDh^2
&\equiv\displaystyle\left.
\frac{\partial^2}{\partial\xi^2}
\varphi\{{\rm e}^{\displaystyle -\xi\QDh},\,\bb,\,\KDh\}
\right\vert_{\xi=0}
=\displaystyle\left.
\frac{1}{\bb^2}\frac{\partial^2}{\partial\lambda^2}
\varphi\{\uni,\,\bb,\,\KDh+\lambda\QDh\}
\right\vert_{\lambda=0}\q,
\ea\ee
which are {\it consistent with exact thermodynamical relations}

An example of special
interest concerns the particle-number operator
$\NDh$ in the
grand-canonical equilibrium ($\KDh\equiv\HDh-\mu\NDh$).
To obtain the cumulants of $\NDh$
(i.e. $\ADh=\exp(-\xi\NDh)$) it suffices
to consider the partition function
in which $\mu$ is replaced by $\mu-\lambda$
with $\lambda\equiv\xi/\bb$.
Then Eqs.~(\ref{e360}) become
\be\label{e370}
\langle\NDh\rangle=
\frac{1}{\bb}\frac{\partial}{\partial\mu}
\varphi\{\uni,\,\bb,\,\HDh-\mu\NDh\}\q,\q\q\q
\Delta\NDh^2=
\frac{1}{\bb}\frac{\partial}{\partial\mu}
\langle\NDh\rangle\q.
\ee
These equations, valid
in any variational space that
satisfies the aforementioned invariance,
coincide with the usual thermodynamic relations
for the expectation value and the fluctuations of
the particle-number operator.
Similar relations arise for any conserved observable
such as the momentum or the angular momentum.

\subsection[S24]{Connection with the Standard Variational Principle}

While for $\DIh\uu$ the exact solution $\exp({-u\KDh})$ has
a trivial, exponential dependence on $u$, the exact solution
(\ref{e120}) for $\AIh\uu$ depends on $u$ in a more
complicated fashion since  $\log\AIh\uu$ is a linear function
of $u$ only if the operators
$\ADh$ and $\KDh$ commute,  as was the case in Sect.~2.3.
For restricted variational spaces, due to
the coupling between the
differential equations (\ref{e060}) and (\ref{e080}),
we expect in general both the approximate solutions
$\DIh\uu$ and $\AIh\uu$ to display
a non-trivial $u$-dependence.
This is actually the feature
which provides its flexibility to the method.

Nevertheless, for $\xi_\gamma=0$
($\ADh=\uni$), that is, when we only want a
variational evaluation of the thermodynamic
potential $\varphi\zz$,
a simplification occurs whenever
the trial spaces satisfy the following properties
(which will hold in all the Sections~3 to 6 and Appendix A) :
(i) the trial sets for the operators
$\DIh\uu$ and $\AIh\uu$ are identical; (ii) if $\DIh$
belongs to this set, so does $c\,\DIh^{\,\alpha}$
where $c$ and $\alpha$ are any positive constants.
(The power $\alpha$ of the operator $\DIh$ is well-defined in a 
basis where it is diagonal, even when it has vanishing
eigenvalues. 
This remark will allow us to use the results
of the present Section not only for
independent-quasi-particle density operators (Sect.~4.1) 
but also in Sects.~5.4, 6.2 and 6.3 where we will
consider projections of these density
operators onto subspaces of the
Fock space.)
The properties above enable us to make the following Ansatz,
suggested by Eq.~(\ref{e120}) for $\ADh=\uni$ :
\be\label{e380}
\DIh_0\uu={\rm e}^{\displaystyle -u\HDh_0}\q,\q\q\q
\AIh_0\uu={\rm e}^{\displaystyle -(\bb-u)\HDh_0}\q,
\ee
where the trial operator $\HDh_0$ does not depend on $u$.
As a consequence, we have
\be\label{e390}
\AIh_0\uu\DIh_0\uu=
\DIh_0\uu\AIh_0\uu=
\DIh_0\bbb={\rm e}^{\displaystyle -\bb\HDh_0}\equiv\YI\Deltah\q,
\ee
where we defined $\YI$ as the normalization
$\Tru{{\rm e}^{-\bb\HDh_0}}$.
The variational expression (\ref{e040}) then reduces
to
\be\label{e400}
\II=\YI(1-\log\YI)-\YI\Tru{\Deltah(\log\Deltah+\bb\KDh)}\q.
\ee
By maximizing (\ref{e400})
with respect to $\YI$ , we obtain
\be\label{e410}
\log\II=\log\YI=-\Tru{\Deltah\log\Deltah}-\bb\,\Tru{\Deltah\KDh}\q,
\ee
which shows that, under the conditions of the present section,
the entropy $S$ defined by (\ref{e262}) is equal to
\be\label{e411}
S=-\Tru{\Deltah\log\Deltah}\q,
\ee
and is related to $\DIh\bbb$ by Eq.(\ref{I020}).
{\it The standard variational principle for
thermodynamic potentials is
thereby recovered} ; indeed we find
$\varphi\zz$ as the maximum with
respect to $\Deltah$ (under the constraint $\Tru{\Deltah}=1$)
of the right-hand
side of (\ref{e410}), in
agreement with the fact that $\log\YI$ is the Legendre transform 
of the entropy. This reduction, 
however, takes place only for $\ADh=\uni$ or $\xi_\gamma=0$.

The analysis of the previous Subsection
showed that for an operator $\QDh$
commuting with $\KDh$, the characteristic function
$\varphi(\xi)$ is equal to the thermodynamic potential
associated with the shifted operator 
$\KDh^{\,\prime}=\KDh+\xi\QDh/\bb$
when the trial spaces satisfy the invariance property
stated in Sect. 2.3.
In this case the cumulants of any order, and in
particular the correlations, can be deduced from
the standard variational principle (\ref{e410}) with $\KDh$
replaced by $\KDh^{\,\prime}$.

\subsection[S25]{Variational Principle for $\DIh\uu$ Alone}

The Ansatz (\ref{e380}), which restricts the parametrization
of the two
$u$-dependent variational quantities
$\DIh\uu$ and $\AIh\uu$ to a single
operator $\HDh_0$, is not suited to the evaluation
of the characteristic function $\varphi(\xi)$ for
$\xi_\gamma\neq 0$ (and therefore of cumulants of non-conserved
quantities).
In the following Sections
we shall therefore exploit the freedom allowed by
the full set of the variables $\DIh\uu$
and $\AIh\uu$. Nevertheless
the form of the exact solution (\ref{e120})
for $\AIh\uu$
suggests another possibility :
instead of letting $\AIh\uu$ and $\DIh\uu$ vary
independently, one can constrain them according to
\be\label{e420}
\AIh\uu=\DIh^{1/2}(\bb-u)\ADh\,\DIh^{1/2}(\bb-u)\q.
\ee
Inserting this trial form into the functional (\ref{e040}) leads to
the variational expression
\be\label{e430}\ba{l}\displaystyle
\Tru{\ADh\,\DIh\bbb}-\intzb
\Tru{\DIh^{1/2}(\bb-u)\ADh\,\DIh^{1/2}(\bb-u)
\pag \duh{\DIh\uu}\right.}\\
\q\q\q\left. \vphantom{\duh{\DIh\uu}}
\displaystyle +\usd[\KDh\DIh\uu+\DIh\uu\KDh]\pad
\q,
\ea\ee
in terms of a single variational quantity $\DIh\uu$
obeying the boundary condition (\ref{e050}).
Had we not symmetrized the Bloch equation, we
would have found the alternative functional
\be\label{e440}
\Tru{\ADh\,\DIh\bbb}-
\intzb
\Tru{\ADh\,\DIh(\beta-u)
\pag \duh{\DIh\uu} +\KDh\DIh\uu\pad}\q,
\ee
whose stationary value with respect to $\DIh\uu$ yields
the wanted quantity $\Tr{\ADh\DDh}$.
A special case of this variational principle
has already been proposed and applied
in connection with the Monte Carlo method \cite{WKC83};
it is obtained from (\ref{e440}) in the case of canonical
equilibrium $\KDh=\HDh$
by taking for $\ADh$ the dyadic $\vert R\rangle\langle R'\vert$,
where $R$ is a point in the $3N$-dimensional space.
The stationary value of (\ref{e440}) is then
the density matrix itself in the $R$-representation.

If we further specialize $\DIh\uu$ to depend exponentially on $u$,
as does the exact solution, in the form
\be\label{e450}
\DIh\uu={\rm e}^{\displaystyle -\MIh\, u/\bb}\q,
\ee
both functionals (\ref{e430}) and (\ref{e440})
become
\be\label{e460}
\Tru{\ADh\,{\rm e}^{\displaystyle -\MIh}\left(
\uni+\MIh-\int_0^\bb{\rm d}u\,
{\rm e}^{\displaystyle \MIh\, u/\bb}\,\KDh\,
{\rm e}^{\displaystyle -\MIh\, u/\bb}
\right)}\ .
\ee
We thus recover the variational expression (3.20) of Ref.\cite{BVe88}.
In (\ref{e460}) the trial quantity is the operator $\MIh$.
For unrestricted variations of $\MIh$, the stationary condition is
$\MIh=\bb\KDh$, and the stationary value is the characteristic function
$\Tru{\ADh{\rm e}^{-\bb\KDh}}$. It has been shown
in \cite{BVe88} that, unlike the previous variational functionals, 
(\ref{e460}) is maximum for
$\MIh=\bb\KDh$ under rather general conditions.
\newpage

\section[S3]{Extended Finite-Temperature HFB
Approximation for Characteristic Functions}
\setcounter{equation}{0}

From now on we will consider systems of interacting fermions
for which pairing effects are important.
In Sect.~3.2 we therefore apply the variational principle
of Sect.~2 to derive consistent coupled equations
which are adapted to the
evaluation of the characteristic function (\ref{e010})
and which account for pairing.
We deal, in this Section and Sect.~4, with grand canonical
equilibrium, letting
$\KDh\equiv\HDh-\mu\NDh$ and taking traces in the Fock space.
The formalism will also apply if $\KDh$ includes additional
constraints such as $-\oo\JDh$, where $\JDh$ is the
$z$-component of the angular momentum.
More generally,
for a finite system, one may want to control its position
in space by including in the set $\Omk$
the center-of-mass operator, or,
if it is deformed, its
orientation by the quadrupole-moment tensor.
We first introduce some notation
and definitions which will be used throughout the present article.

\subsection[S31]{Generalities and Notations}

For the sake of conciseness, it is convenient to introduce
$2n$-dimensional column and line vectors
defined as

\be\label{f010}
\gbd=\pag \ba{c}
\abd_1\\
\vdots\\
\abd_n\\
\abt_1\\
\vdots\\
\abt_n
\ea\pad
\q,\q
\gbt=\pag \abt_1\cdots\abt_n\,\abd_1\cdots\abd_n \pad
\q,
\ee
where $\abt_i$ and $\abd_i$
are the creation and annihilation operators associated
with a fixed single-particle basis ($i=1,n$).
If we denote by
$\gb_\la$ the $\la$-th component $\{\la=1,2n\}$ of the
column vector $\gbd$, the fermionic
anticommutation rules are given by either one of the relations
\be\label{f020}
[\gb_\la,\,\gb^\dagger_{\la'}]_+=\delta_{\la\la'}\ ,\q\q
[\gb_\la,\,\gb_{\la'}]_+=\si_{\la\la'}\q,
\ee
where $\si$ is the $2n\times 2n$ matrix
\be\label{f030}
\si=\pag\ba{cc}0&1\\1&0\ea\pad\q.
\ee

In what follows, as in the HFB theory, operators
having the form
\be\label{f040}
\TDh=\exp( -\li -{\scriptstyle \usd}\gbt\LI\gbd )
\q,\ee
where $\li$ is a c-number and $\LI$ is a $2n\times2n$ matrix,
play a crucial r\^ole.
As a consequence of the anticommutation relation (\ref{f020}),
we can restrict ourselves to matrices $\LI$
which satisfy the relation
\be\label{f050}
\si\LI\si = -\LI^{\rm T}\q,
\ee
where $\LI^{\rm T}$ denotes the transposed matrix of $\LI$.
Instead of $\li$ and $\LI$
one can alternatively use the c-number $\ttt$ and the $2n\times 2n$
matrix $\TI$ defined by
\be\label{f060}
\ttt={\rm e}^{\displaystyle -\li}\ ,\q\q
\TI={\rm e}^{\displaystyle -\LI}\q,
\ee
to specify the operator (\ref{f040}).
Another equivalent parametrization involves
the normalization $\ZI$,
\be\label{f070}
\ZI\equiv\Tr{\TDh}=
{\rm e}^{\displaystyle -\li}\pag \dtt{1+
{\rm e}^{\displaystyle -\LI}} \pad^\ud=
\ttt\exp\acg{\textstyle \usd}\trd{\log(1+\TI)}\acd\q,
\ee
and the generalized reduced density matrix $\RI$
(or matrix of the contractions),
\be\label{f080}
\RI_{\la\la'}\equiv
\frac{\Tr{\gb_\la\TDh \gb_{\la'}^\dagger}}{\Tr{\TDh}}=
\left(\frac{1}{{\rm e}^{\displaystyle \LI}+ 1}\right)_{\la\la'}=
\left(\frac{\TI}{\TI+ 1}\right)_{\la\la'}\q.
\ee
In Eq.~(\ref{f070}), the symbol $\hbox{tr}_{2n}$
denotes a trace in the space of $2n\times 2n$ matrices
while Tr is the trace in the Fock space.
The property (\ref{f050}) implies the relations :
\be\label{f090}
\si\TIm\si=\TI^{\rm T}
\ ,\q\q
\si\RI\si=1-\RI^{\rm T}
\q.
\ee
As a consequence, the $2n\times 2n$
matrix $\RI$ can be written in the form
\be\label{f100}
\RI=\pag \ba{cc}
\den&\kapm\\ \kappa_+& 1-\den^{\rm T}
\ea \pad\q,
\ee
where the $n\times n$ matrices $\den$, $\kapm$ and $\kapp$
correspond to the normal and abnormal
contractions :
\be\label{f110}\acg\ba{rl}
\den_{ij} =&\Tr{\abd_i \TDh \abt_j}/\Tr{\TDh}\q,\\
\kapm_{ij}=&\Tr{\abd_i \TDh \abd_j}/\Tr{\TDh}\q,\\
\kapp_{ij}=&\Tr{\abt_i \TDh \abt_j}/\Tr{\TDh}\q.
\ea\pod\ee
The matrices $\kapm$ and $\kapp$ are antisymmetric.
When the operator $\TDh$ is hermitian, one has $\den=\den^\dagger$,
$\kapp=\kapm^\dagger$ and the matrix $\RI=\RI^\dagger$
reduces to the usual HFB form.

Operators of the type
(\ref{f040}) allow the use of a
generalized Wick theorem,
{\it even when they are not hermitian} \cite{BBr69}.
They also form a non-abelian multiplicative {\it group},
the algebra of which is characterized by the relations
(\ref{f200}--\ref{f240}).
These simple algebraic properties (more details can be found in
\cite{BBr69} and
the Appendix of \cite{BVe85}) will be used extensively below.

\subsection[S32]{The Variational Ans\"atze and the Associated
Functional}

Let us come back to our problem. We assume that the
observables $\QDh_\ggg$ are operators of
the single-quasi-particle type
(this restriction is not essential and can be removed
at the cost of some formal complications (see \cite{BVe93})):
\be\label{f120}
\QDh_\ggg=\qi_\ggg+
{\scriptstyle \usd}\gbt\QI_\ggg\gbd\q,
\ee
so that the operator
$\ADh(\xi)$ (see Eq.~(\ref{I060})) belongs to the class (\ref{f040}).
We thus write it as
\be\label{f130}
\ADh(\xi)\equiv\TDh^\AP\equiv\exp( -\li^\AP -
{\scriptstyle \usd}\gbt\LI^\AP\gbd ) \q,
\ee
with
\be\label{f140}
\li^\AP\equiv\sum_\ggg\xi_\ggg\,\qi_\ggg\ ,\q\q
\LI^\AP\equiv\sum_\ggg\xi_\ggg\,\QI_\ggg\q.
\ee

Now we choose the trial spaces for the
operators $\DIh\uu$ and $\AIh\uu$ entering the functional
(\ref{e040}). This choice determines our
variational approximation.
We take as trial sets :
\be\label{f150}
\DIh\uu=\TDhd\uu\ ,\q\q\AIh\uu=\TDha\uu\q,
\ee
with
\be\label{f160}
\TDhd\uu\equiv\exp( -\lid\uu -
{\scriptstyle \usd}\gbt\LId\uu\gbd )\q,
\ee
\be\label{f170}
\TDha\uu\equiv\exp( -\lia\uu -
{\scriptstyle \usd}\gbt\LIa\uu\gbd )\q.
\ee
The $2n\times2n$ matrices
$\LId\uu$ and $\LIa\uu$ (constrained to satisfy
the relation (\ref{f050}))
and the c-numbers $\lid\uu$ and $\lia\uu$ constitute
a set of independent variational parameters.
Using Eqs.~(\ref{f060}-\ref{f080}),
we define the coefficients $\tta$, $\ttd$,
the normalization factors $\ZIa$, $\ZId$,
the matrices $\TIa$, $\TId$ and
the reduced density matrices $\RIa$, $\RId$ associated
with the operators $\TDha$ and $\TDhd$.

The boundary condition (\ref{e050}) on $\DIh\zz$ implies the relations:
\be\label{f180}
\lid\zz=0\ ,\q\q\LId\zz=0\q,
\ee
or equivalently
\be\label{f190}
\ttd\zz=1\ ,\q\q\TId\zz=1\q;\q\q\ZId\zz=2^n\ ,\q\q\RId\zz=\usd\q.
\ee

The products  $\DIh\uu\AIh\uu$ and $\AIh\uu\DIh\uu$
occur in the functionals (\ref{e040}) and (\ref{e070}).
Taking advantage of the group properties \cite{BBr69} of the
operators (\ref{f040}), we write
\be\label{f200}
\AIh\DIh=\TDhb\ ,\q\q\DIh\AIh=\TDhc\q,
\ee
with
\be\label{f210}
\TDhb=\exp( -\lib -
{\scriptstyle \usd}\gbt\LIb\gbd ) \q,
\ee
\be\label{f220}
\TDhc=\exp( -\lic -
{\scriptstyle \usd}\gbt\LIc\gbd ) \q,
\ee
where the $u$-dependence of
the c-numbers $\lib$ and $\lic$, and of the matrices
$\LIb$ and $\LIc$
has not been indicated;
using the notation (\ref{f060}),
one has the relations :
\be\label{f230}
{\rm e}^{\displaystyle -\lib}=\ttb={\rm e}^{\displaystyle -\lic}
=\ttc={\rm e}^{\displaystyle -\lia-\lid}=\ttd\tta\q,
\ee
\be\label{f240}
{\rm e}^{\displaystyle -\LIb}=\TIb=\TIa\TId\ ,\q\q
{\rm e}^{\displaystyle -\LIc}=\TIc=\TId\TIa\q.
\ee
According to Eqs.~(\ref{f070}) and (\ref{f230}-\ref{f240}),
the normalization of
$\DIh\AIh$ and $\AIh\DIh$, 
which we denote in shorthand by $\YI$, is
given at each ``time'' $u$ by
\be\label{f250}
\ba{rl}
\YI
&\equiv\Tr{\AIh\DIh}=\ZIb=\Tr{\DIh\AIh}=\ZIc\\
&={\rm e}^{\displaystyle -(\lia+\lid)}
\,\exp\acg\usd\trd{\log(1+
{\rm e}^{\displaystyle -\LIa}
{\rm e}^{\displaystyle -\LId}
)}\acd\\
&
=\tta\ttd
\,\exp\acg\usd\trd{\log(1+
\TIa\TId
)}\acd\q.
\ea
\ee
As we already know from Sect.~2.2,  the normalization
$\YI$ does not depend on $u$ at the stationary point
and it is a variational approximation for the quantity $\exp\varphi$
that we are looking for (see below Eqs.(\ref{f480}-\ref{f500})).
It thus plays the r\^ole of a generalized
partition function which reduces to the grand canonical partition
function when the sources $\xi_\ggg$ are set to zero.

We can now write the functional (\ref{e040})
(which provides the characteristic function (\ref{e010}))
in terms of the variational parameters $\ttd$, $\TId$, $\tta$, $\TIa$.
The term ${\rm Tr}\AIh\hbox{d}\DIh/\hbox{d}u$
is obtained by differentiating $\ttd$ and $\TId$ in
Eq.~(\ref{f250}) with respect to $u$, 
while holding $\tta$ and $\TIa$ fixed,
which yields
\be\label{f260}
\Tr{\AIh\duh{\DIh}}=
\YI\,( \duh{\log\ttd}+\usd\trd{\TIdm \RIc \duh{\TId}} )\q.
\ee

The last two terms in the integrand of the functional (\ref{e040})
are supplied by a straightforward application of the generalized
Wick theorem. They involve the operator $\KDh\equiv\HDh-\mu\NDh$
where the Hamiltonian $\HDh$ of the system is assumed to
have the form
\be\label{f270}
\HDh=\sum_{ij=1}^n\,\BO_{ij}\,\abt_i\abd_j+
{\textstyle\usq}\sum_{ijkl=1}^n\VO_{ijkl}\,\abt_i\abt_j\abd_l\abd_k\q,
\ee
and where
\be\label{f280}
\NDh=\sum_{i=1}^n\,\abt_i\abd_i\q.
\ee
The matrix elements $\BO_{ij}$ of the one-body part of $\HDh$ and
the antisymmetrized matrix elements $\VO_{ijkl}$
of the two-body interaction satisfy the relations
\be\label{f290}
\BO_{ij}=\BO_{ji}^*\ ,\q\q
\VO_{ijkl}=-\VO_{jikl}=-\VO_{ijlk}=\VO_{klij}^*\q.
\ee
As mentioned at the beginning of this Section,
the one-body part associated
with the matrix $\BO{}$ may include,
in addition to the kinetic energy,
some other single-particle operator whose average value is
kept fixed. One finds
\be\label{f300}
\Tr{\KDh\AIh\DIh}=\YI\,\EI\{\RIb\}\ ,\q\q
\Tr{\KDh\DIh\AIh}=\YI\,\EI\{\RIc\}\q,
\ee
with
\be\label{f310} \EI\{\RI\}=
\sum_{ij}\,(\BO_{ij}-\mu\delta_{ij})\,\den_{ji}
+{\textstyle\usq}\sum_{ijkl}\,\VO_{ijkl}\,(
2\den_{ki}\den_{lj}-\kapp_{ij}\kapm_{kl})
\q.\ee
When $\den{}=\den{}^\dagger$ and $\kapp{}=\kapm{}^\dagger$,
the expression (\ref{f310})
is formally identical to the HFB energy.
However, the quantities
$\EI\{\RIb\}$ and $\EI\{\RIc\}$ are not necessarily real
since $\RIb$ and $\RIc$ do not have to be hermitian.

Altogether, for the trial choices
(\ref{f160}) and (\ref{f170}), the functional (\ref{e040})
reads
\be\label{f320}\ba{l}
\II\{\AIh,\DIh\}=\displaystyle
\YI^\AP-\,\intzb\YI\left(
\frac{\hbox{d}{\log\ttd}}{\hbox{d}u}
\right.\\
\q\q\q\q\left.\displaystyle
\vphantom{\frac{\hbox{d}{\log\ZId}}
{\hbox{d}u}}
+\usd\trd{\TIdm \RIc \,\duh{\TId}}
+\usd[\EI\{\RIc\}+\EI\{\RIb\}] \right)
\q,\ea\ee
where use has been made of the relation
\be\label{f330}
\TIdm \RIc = \RIb \TIdm\q.
\ee
According to Eq.~(\ref{f250}), the boundary term $\YI^\AP$ is given
by 
\be\label{f340}\ba{rl}
\YI^\AP&\equiv\Tr{\ADh(\xi)\DIh\bbb}\equiv\Tr{\TDh^\AP\DIh\bbb}\\
&=\ttt^\AP\ttd\,\exp\acg
\usd\trd{ \log(1+\TI^\AP\TId\bbb) } \acd\q,
\ea\ee
where the c-number
$\ttt^\AP$ and the matrix $\TI^\AP$,
which specify the characteristic function we are looking for,
are expressed in terms of the sources $\xi_\ggg$
by Eqs.~(\ref{f120}-\ref{f140}) and (\ref{f060}).
Similarly the form (\ref{e070}) of $\II$ reads
\be\label{f350}\ba{l}
\II\{\AIh,\DIh\}=\displaystyle
\YI^\AP-\YI\bbb+\YI\zz+\intzb\YI\left( \duh{\log\tta}
\right.\\
\q\q\q\q\left.\displaystyle
+\usd\trd{\TIam  \RIb \,\duh{\TIa}}
-\usd[\EI\{\RIb\}+\EI\{ \RIc \}] \right)
\q .\ea\ee
In the functionals (\ref{f320}) and (\ref{f350}), the $u$-dependence
has been omitted in the variational parameters
$\ttd$, $\TId$, and $\tta$, $\TIa$,
as well as in the normalization $\YI$
given by Eq.~(\ref{f250}) and in the contractions $\RIc$, $\RIb$
associated with $\TIc$, $\TIb$, respectively;
the latter quantities are related to $\TIa$ and $\TId$
through (\ref{f240}).

\subsection[S33]{The Coupled Equations}

The next step is to write the equations expressing the
stationarity of the functional
(\ref{f320}) or (\ref{f350}).
We derive them directly from these expressions,
but could also have used the general
equations (\ref{e060}) and (\ref{e080}-\ref{e090}).

Requiring the stationarity of
(\ref{f320}) with respect to
$\tta\uu$ (in the functional
(\ref{f320}), $\tta\uu$ only appears
as a proportionality factor in $\YI\uu$)
gives us the equation of
evolution for $\lid\uu=-\log\ttd\uu$ :
\be\label{f360}
\duh{\lid} =
\usd\trd{\TIdm \RIc \,\duh{\TId}}
+\usd[\EI\{ \RIc \}+\EI\{ \RIb \}] \q,
\ee
which we shall solve explicitly below (Eq.(\ref{f460})).

We now require the stationarity of
the functional (\ref{f320}) with respect to $\TIa\uu$,
which appears through $\RIb\uu$, $\RIc\uu$ and $\YI\uu$.
Eq.(\ref{f360}) implies that we can disregard in
this calculation the variation of
$\YI\uu$. If we note
that the variations of $\RIb\uu$ and $\RIc\uu$ resulting from those of
$\TIa\uu$ satisfy $\delta\RIc=\TId\delta\RIb\TIdm$ as a
consequence of (\ref{f330}), we readily find the differential equation
\be\label{f370}
\duh{\TId}=-\usd[\HI\{\RIc\}\TId+\TId\HI\{\RIb\}]\q,
\ee
where $\HI\{\RI\}$ has the same formal definition,
\be\label{f380}
\delta\EI\{\RI\}={\textstyle\usd}\trd{\HI\{\RI\}\,\delta\RI}\q,
\ee
as the HFB Hamiltonian, although $\HI\{\RI\}$
is not necessarily hermitian.
For the Hamiltonian
(\ref{f270}), the $2n\times 2n$ matrix $\HI$
takes the form
\be\label{f390}
\HI\{\RI\}=\pag \ba{cc}
\hhh&\delm\\ \Delta_+& -\hhh^{\rm T}
\ea \pad\q,
\ee
with
\be\label{f400}\acg\ba{rl}
\hhh_{ij} =&(\BO_{ij}-\mu\delta_{ij})
+\sum_{kl}\VO_{ikjl}\den_{lk}\q,\\
\delm_{ij}=&\usd\sum_{kl}\VO_{ijkl}\kapm_{kl}\q,\\
\delp_{ij}=&\usd\sum_{kl}\kapp_{kl}\VO_{klij}\q.
\ea\pod\ee
As a consequence of Eq.~(\ref{f290}), the matrices $\delm$
et $\delp$ are antisymmetric.
This entails for $\HI$ the same relation as (\ref{f050}) :
\be\label{f410}
\si\HI\si=-\HI^{\rm T}\q,
\ee
which is consistent with $\si\delta\RI\si=-\delta\RI^{\rm T}$
and Eq.~(\ref{f380}).

Likewise, the stationarity conditions of
the functional (\ref{f350}) with respect
to $\ttd\uu$ and $\TId\uu$ provide, for $0\le u<\bb$,
the differential equations
\be\label{f420}
\duh{\lia} =
\usd\trd{\TIam \RIb \,\duh{\TIa}}
-\usd[\EI\{ \RIb \}+\EI\{ \RIc \}] \q,
\ee
for $\lia\uu=-\log\tta\uu$ and  
\be\label{f430}
\duh{\TIa}=\usd[\HI\{\RIb\}\TIa+\TIa\HI\{\RIc\}]\q,
\ee
for $\TIa\uu$.

Finally, by rendering the functional (\ref{f350}) stationary with
respect to $\ttd\bbb$ and $\TId\bbb$
(see Eqs.~(\ref{f130}) and (\ref{f060})), one obtains
the boundary conditions
for $\AIh\uu$ :
\be\label{f440}
\tta\bbb=\ttt^\AP\ ,\q\q
\TIa\bbb=\TI^\AP\q.
\ee

The main equations to be solved are the
differential equations (\ref{f370}) and
(\ref{f430}) for $\TId\uu$ and $\TIa\uu$
which are coupled through the matrices $\HI\{\RIb\}$ and
$\HI\{\RIc\}$. Moreover one has to deal
with the mixed boundary conditions
\be\label{f450}
\TId\zz=1\ ,\q\q\TIa\bbb=\TI^\AP\q.
\ee
All our variational
quantities depend upon the sources $\xi_\ggg$
(see Eqs. (\ref{f120}-\ref{f140}) and (\ref{f060}))
through the
boundary conditions (\ref{f440}).

One notes that the evolution equation (\ref{f430}) of $\TIa$ can be
deduced from Eq.~(\ref{f370}) for $\TId$
by an overall change in sign of the
right hand side and by the exchange $a\leftrightarrow d$,
as could have been inferred from the comparison between the forms
(\ref{f320}) and (\ref{f350}) of the functional $\II$.
(The same is true for the equations
(\ref{f360}) and (\ref{f420}) governing $\lid$
and $\lia$ if one takes
account of Eqs.~(\ref{f370}) and (\ref{f430}).)

Once Eqs.~(\ref{f370}) and (\ref{f430})
are solved, the c-numbers
$\lid\uu$ and $\lia\uu$ are obtained by integration of Eqs.
(\ref{f360}) and (\ref{f420}) whose r.h.s. depend only on
$\TId$ and $\TIa$.
Indeed, using Eqs. (\ref{f370}) and (\ref{f430})
with the boundary conditions $\lid\zz=0$ and
$\lia\bbb=\li^\AP$, we find
\be\label{f460}\ba{rl}
\lid\uu &={\displaystyle \int_0^u}
\,\hbox{d}u\,\acg
-\usq\trd{ [\RIc\,\HI\{\RIc\}+\RIb\,\HI\{\RIb\}] }
\pod \\
&
\q\q\q\pog
+\usd[\EI\{ \RIc \}+\EI\{ \RIb \}]
\acd\q,\\
\ea\ee
\be\label{f470}
\lia\uu =\li^\AP+\lid\bbb-\lid\uu\q.
\ee

From the equations for
$\TId$, $\TIa$,
$\lid$, $\lia$,
together with the definition (\ref{f250})
and the relation (\ref{f380}),
one can readily verify
the conservation laws (\ref{e170}) and (\ref{e150})
which now read, respectively :
\be\label{f480}
\duh{\YI}=0\ ,\q\q\q\du(\EI\{\RIb\}+\EI\{\RIc\})=0\q.
\ee
One can also verify that the integrands of the functionals
(\ref{f320}) or (\ref{f350}) vanish.
As was shown in Sect. 2, these properties directly result
from the feature that variations
$\delta\AIh\propto\AIh$ and $\delta\DIh\propto\DIh$
are allowed in the trial classes
(\ref{f160}-\ref{f170}).

We are looking eventually for
the characteristic function $\varphi\{\ADh(\xi),\bb,\KDh\}$
which is given by the logarithm of the stationary
value of the functional (\ref{f320}).
This value is provided by the boundary term $\YI^\AP$
defined in (\ref{f340}). In this formula,
the matrix $\TId\bbb$ and the c-number
$\ttd\bbb=\exp[-\lid\bbb]$ are furnished by the
solution of the coupled Eqs.~(\ref{f370}), (\ref{f430})
and by (\ref{f460}), respectively. 
Since, according to Eq.~(\ref{f480}),
the normalization $\YI$ (defined in (\ref{f250})) 
does not depend on $u$ we have the relation
\be\label{f490}
\varphi(\xi)\equiv\varphi\{\ADh(\xi),\bb,\KDh\}
=\log\YI^\AP=\log\YI\uu\ ,\ 0\leq u\leq\bb\q.
\ee
In particular, $\YI$ can be evaluated at $u=0$ (see Eq.~(\ref{e190})); 
from Eqs.~(\ref{f250}) and (\ref{f190})
we find
\be\label{f500}
\log\YI=-\lia\zz+{\textstyle \usd}\trd{\log(1+\TIa\zz)}\q,
\ee
where $\lia\zz=\li^\AP+\lid\bbb$ 
is explicitly given by Eqs.~(\ref{f460}--\ref{f470})
once the coupled 
equations (\ref{f370}) and (\ref{f430}) which yield $\TIa\zz$
have been solved.

Let us add a few remarks about the
properties of these last equations.
In the derivations of this Section (as well as in Sect.~2),
the hermiticity of the operators $\DIh\uu$ and $\AIh\uu$,
as defined in Eqs.~(\ref{f150}-\ref{f170}), was nowhere
assumed. If, for some value of $u$, the operators
$\DIh$ and $\AIh$ are hermitian, i.e. if
$\TId=\TId^\dagger$, $\TIa=\TIa^\dagger$,
$\ttd=\ttd^*$ and $\tta=\tta^*$, it follows that
$\RIb=\RIc^\dagger$ and hence that
$\HI\{\RIb\}=\HI\{\RIc\}^\dagger$.
When the operator $\ADh$ is hermitian, the
inspection of Eqs.(\ref{f370}), (\ref{f430}) and
( \ref{f450}-\ref{f470})
shows that these equations preserve
the hermiticity of $\TId$ and $\TIa$ and the
reality of $\ttd$ and $\tta$.
In this case,
the stationary value
$\YI$ given by (\ref{f500})
is also real.
If $\ADh$ is not hermitian,
as would happen for the characteristic functions
with imaginary $\xi$'s,
the boundary condition
(\ref{f440}) prohibits $\AIh\uu$ and $\DIh\uu$
from being hermitian.

Combining  Eqs.~(\ref{f370}) and (\ref{f430}), one obtains
the evolution equations for the
products $\TIb=\TIa\TId$ and $\TIc=\TId\TIa$\,,
and hence for the HFB-like contraction matrices
$\RIb$ and $\RIc$ associated (through
Eq.~(\ref{f080})) with the operators $\TDhb$ and $\TDhc$.
The calculation relies on the property
\be\label{f510}
\delta\RI=(1-\RI)\delta\TI(1-\RI)\q,
\ee
and it leads to
\be\label{f520}
\duh{\RIb}=\usd[\HI\{\RIb\},\,\RIb]\ ,\q\q
\duh{\RIc}=-\usd[\HI\{\RIc\},\,\RIc]\q.
\ee
These equations,
the single-quasi-particle reductions of Eqs.~(\ref{e310}),
involve commutators. They 
are reminiscent of the time-dependent-Hartree--Fock--Bogoliubov 
equations (see Sect.~5 of the Appendix A),  
$\pm i\,u$ replacing the time $t$. However, here
$\HI\{\RIb\}$ and $\HI\{\RIc\}$ are in general not hermitian
and $u$ is real.
A consequence of the simple form
(\ref{f520}) is that the eigenvalues of $\RIb$ and $\RIc$
remain constant with $u$, in contrast to those of $\RIa$ and $\RId$.
It also follows from Eqs.~(\ref{f520}) and (\ref{f380})
that each of the
quantities $\EI\{ \RIb \}$ and $\EI\{ \RIc \}$ is conserved,
a finer result than Eq.~(\ref{f480}).
In spite of their form, one should not infer from Eqs.~(\ref{f520})
that the matrices $\RIb$ and $\RIc$ are decoupled.
Actually the constraints (\ref{f190}) and (\ref{f440})
do not entail any explicit boundary conditions for
$\RIb$ and $\RIc$.
Moreover, Eqs. (\ref{f520}) are not equivalent to Eqs.
(\ref{f370}) and (\ref{f430}) since, except in very
particular cases, $\RIa$ and $\RId$
are not determined uniquely by
knowledge of $\RIb$ and $\RIc$.
\newpage

\section[S4]{Expansion of the Extended HFB
Approximation  : Fluctuations and Correlations}
\setcounter{equation}{0}

In this Section we derive the
thermodynamic quantities, the expectation value of
the observables, their fluctuations and
the correlations between these observables
from the characteristic function
as given by either one of the
formulae (\ref{f340}) or (\ref{f500}).
We shall see that this evaluation does not
in fact require
the full solution of the coupled equations
(\ref{f370}), (\ref{f430}), (\ref{f460}) and (\ref{f470})
 for finite values of the sources $\xi$.
Indeed these quantities can be
obtained through a prior expansion of
the coupled equations in powers of the sources
$\xi_\ggg$,  followed by a solution of
the resulting expanded equations order by order. This
procedure turns out to overcome the difficulties
associated with the mixed boundary conditions.
As expected, the usual HFB equations are recovered in
the case $\ADh=\uni$, when the sources $\xi_\ggg$ vanish
(Sect.~4.1.1). These equations are also relevant 
in the calculation of the
average values of single-quasi-particle operators (Sect. 4.1.2).
In Sects~4.2 and 4.3 we work out the second order in the sources ;
this {yields} an explicit variational expression for
the fluctuations and the correlations which differs
from the naive application of the HFB approximation.
In the present Section, the increasing orders
of the expansion with respect to the $\xi_\ggg$'s will be indicated
by lower indices (0, 1, ...) assigned
to the considered
quantities.

\subsection[S41]{The HFB Approximation Recovered }

\subsubsection[S411]{Thermodynamic Quantities}

To zeroth order in the sources (i.e., $\ADh(\xi)=\uni$)
our problem amounts to evaluating
the partition function
$\YI=\Tr{\DIhz\bbb}$.
We expect that the coupled equations of Sect.~3.3 will yield
the standard HFB approximation for $\YI$, since the trial
spaces for $\DIh\uu$ and $\AIh\uu$ satisfy the conditions
required in the discussion of Sect.~2.4. Let us verify this point.
If we associate with the contraction matrix
$\RIze$ the single-particle Hamiltonian $\HI_0$ defined by
\be\label{ga010}
\bb\HI_0\equiv\log\frac{1-\RIze}{\RIze}\ ,\q\q
\RIze\equiv\frac{1}{{\rm e}^{\bb\HI_0}+1 }\q,
\ee
the usual optimum HFB matrix $\RIze$
is given,
at temperature $1/\bb$, by
the self-consistent equation
\be\label{ga020}
\HI_0=\HI\{\RIze\}\q,
\ee
where $\HI\{\RI\}$ was defined in Eqs.~(\ref{f390}-\ref{f400}).
Because the relation (\ref{ga020}) entails the commutation
of $\HI\{\RIze\}$ with $\RIze$,
Eqs.~(\ref{f520})
are satisfied by the $u$-independent solution
\be\label{ga030}
\RIbze\uu=\RIcze\uu=\RIze\q.
\ee
In the equations (\ref{f370}) and (\ref{f430})
for $\TIdze\uu$ and $\TIaze\uu$,
the effective Hamiltonians do
not depend on $u$ since
$\HI\{\RIbze\}=\HI\{\RIcze\}=\HI_0$.
Moreover, the boundary conditions (\ref{f450})
reduce to $\TIdze\zz=1$, $\TIaze\bbb=\TI^\AP_0=1$. We thus find :
\be\label{ga040}
\TIdze\uu={\rm e}^{\displaystyle -u\HI_0}\ ,\q\q
\TIaze\uu={\rm e}^{\displaystyle -(\bb-u)\HI_0}\q.
\ee
Eqs.(\ref{ga030}-\ref{ga040}) are the single-quasi-particle reductions
of the general relations (\ref{e390}) and (\ref{e380}).
One also notes that $\RIze$
is equal to $\RIdze\bbb$, the contraction matrix
associated with the state $\DIhz\bbb$.

Using the solution (\ref{ga040}) of the coupled equations,
we can now
calculate $\lia\zz$ from Eqs.~(\ref{f460}-\ref{f470}),
and hence $\log\YI$ from Eq.~(\ref{f500}).
The integrations over $u$, which can be performed
explicitly, {yield} $\log\YI$ from which
according to Eqs.~(\ref{e261}) and (\ref{e262})
we define the thermodynamic grand potential $\FI_{\rm G}$
and the entropy $\SI\{\RIze\}$
\be\label{ga050}
\log\YI=\SI\{\RIze\} -\bb\EI\{\RIze\}
\equiv -\bb\FI_{\rm G}\{\RIze\}\q.
\ee
The entropy is given by
\be\label{ga060}
\SI\{\RI\}=-{{\textstyle\usd}}
\trd{[\RI\log\RI+(1-\RI)\log(1-\RI)]}\q.
\ee
It is thus the entropy of a
quasi-particle gas characterized by the contraction matrix $\RI$.
Owing to the symmetry relation (\ref{f090}),
the two terms under the trace in (\ref{ga060}) are equal.
The HFB energy $\EI\{\RIze\}=\langle\HDh-\mu\NDh\rangle$
is given by the expression
(\ref{f310}).

As expected from the discussion of
Sect.~2.4, Eq.~(\ref{ga050}) coincides with the outcome
of the variational principle (\ref{e410}), which reads here
\be\label{ga070}
\log\YI=\log\Tr{\DIhz\bbb}={\rm Max}_\RI\,[\SI\{\RI\}-\bb\EI\{\RI\}]
=-\bb\,{\rm Min}\,\FI_{\rm G}\{\RI\}\q.
\ee
The stationarity condition of (\ref{ga070})
is obtained from (\ref{f380}) and from
the first-order variation
\be\label{ga080}
\delta\SI\{\RI\}=\usd
\trd{\log\frac{1-\RI}{\RI}\,\delta\RI}
\ee
of the reduced entropy (\ref{ga060}) ; it thus
gives back the usual self-consistency
condition (\ref{ga020}).
Therefore, at zeroth order
in the sources, we have recovered
the standard self-consistent
HFB approximation for the thermodynamic potential.

\subsubsection[S412]{Expectation Values of Observables}

The variational approximation for
the expectation values $\langle\QDhg\rangle$
in grand canonical equilibrium
is in principle given by the
first-order terms in the sources $\xi_\gamma$.
{However}, the knowledge of the zeroth-order
solution (\ref{ga010}), (\ref{ga020}) is here sufficient
since, according to Eq.~(\ref{e230}) and to
the definitions (\ref{f120}) and (\ref{f080}), we have
\be\label{gb010}
\langle\QDhg\rangle=
\frac{\Tr{\QDhg\DIhz\bbb}}{\Tr{\DIhz\bbb}}=
\qi_\gamma+\usd\trd{\QI_\gamma\RIze}\q.
\ee

For the expectation value of the
operator $\KDh$, we can apply
Eq.~(\ref{e260})
which was obtained by derivation with respect
to the parameter $\beta$.
Using Eqs.~(\ref{f300}) at zeroth order and (\ref{ga030}), we find
\be\label{gb020}
\langle\KDh\rangle=\EI\{\RIze\}=
-\frac{\partial\ }{\partial\bb}\log\YI\q,
\ee
where, as in Eq.~(\ref{gb010}),
the expectation value is taken over $\DIhz\bbb$.

For the calculation of $\langle\NDh\rangle$, we can use
either (\ref{gb010}) or
the discussion in Sects.~2.2 and 2.3.
From Eq.~(\ref{e280}) (or (\ref{e370})) we get
\be\label{gb030}
\langle\NDh\rangle=
\tru{\rho}=
\frac{n}{2}+\usd\trd{\NI\RIze}=
\frac{1}{\bb}\frac{\partial}{\partial\mu}\log\YI\q,
\ee
where, as in Eq.~(\ref{f120}), the operator $\NDh$ can be represented
by the c-number $n/2$ and the $2n\times 2n$ matrix
\be\label{gb040}
\NI=\left(
\ba{cc}
1&0\\0&-1
\ea
\right)\q.
\ee

The formulas (\ref{gb010}-\ref{gb030}) are those given by the standard
approximation, which consists in evaluating expectation values
in the HFB state $\DIhz\bbb$. Our variational procedure thus does
not yield anything new when the characteristic function is expanded
to first order in the sources. This is not the case
at the next order.

\subsection[S42]{Fluctuations and Correlations of 
Conserved Observables}

\subsubsection[S421]{The Particle Number Operator}

As a first example of a conserved observable (i.e., which
commutes with $\KDh$),
we consider the particle
number $\NDh$ and we evaluate its variance $\DNh^{\,2}$.
A naive method would consist in taking
the difference between the expectation value
$\langle\NDh^2\rangle$ and the square $\langle\NDh\rangle^2$,
both evaluated by means of the Wick's theorem
applied to the state $\DIhz\bbb$.
This would yield
\be\label{gc010}
\langle\NDh^2\rangle-\langle\NDh\rangle^2=
\usd\trd\NI(1-\RIze)\NI\RIze,
\ee
where $\RIze$ is the contraction matrix (\ref{ga010}-\ref{ga020})
corresponding to $\DIhz\bbb$.

However, the state
$\DIhz\bbb$ is variationally fitted to the evaluation
of the thermodynamic potential $\FI_{\rm G}$, not to the
evaluation of $\DNh^{\,2}$.
In agreement with our general philosophy,
we should rather rely on a variational principle
suited to the evaluation of the characteristic function $\varphi$.
The variational approximation for the
fluctuation $\DNh^{\,2}$ is then supplied by the second derivative
of $\varphi$ with respect to the source associated with the observable
$\NDh$. The result is
expected to differ from (\ref{gc010}), since
the optimal values of the trial operators $\DIh\uu$
and $\AIh\uu$ are thus dependent on the sources $\xi_\gamma$.

Actually, under conditions on the trial classes that are
obviously satisfied here, we have already determined in Sect.~2.3
the result of this evaluation :
\be\label{gc020}
\DNh^{\,2}=\frac{1}{\bb}\frac{\partial\langle\NDh\rangle}{\partial\mu}=
-\frac{1}{\bb}
\frac{\partial^{\,2}{\displaystyle \FI_{\rm G}}}{\partial\mu^2}\q,
\ee
where $\FI_{\rm G}$ and $\langle\NDh\rangle$ are the outcomes of the
variational calculation at zeroth and first orders in the sources,
given by (\ref{ga050}) and (\ref{gb030}), respectively.
The identities (\ref{gc020}), satisfied by our approximation for
$\DNh^2$,
{agree with exact thermodynamic relations} which {are}
{\it violated by the naive result} (\ref{gc010}). Inserting (\ref{gb030})
into (\ref{gc020}), we have
\be\label{gc030}
\DNh^{\,2}=\frac{1}{2\bb}\trd{\NI\frac{\partial\RIze}{\partial\mu}}\q.
\ee
The derivative $\partial\RIze/\partial\mu$ is supplied by
Eq.~(\ref{ga020}) which defines $\RIze$ self-consistently;
the parameter $\mu$ enters this equation through the contribution
$-\mu\NI$ to $\HI\{\RIze\}$. The expression (\ref{gc030}) reduces
to (\ref{gc010}) only for a system of non-interacting particles, with
$[\NI\,,\,\RIze]=0$, and is otherwise non trivial due
to self-consistency. We shall write it explicitly below
(Eq.~(\ref{gd120})) in a more general context.

The result (\ref{gc030}) arising from
of our variational treatment is more
satisfactory than the naive expression (\ref{gc010}).
We noted that it meets
the thermodynamic {identities} (\ref{gc020}).
Moreover, for a system whose pairing
correlations do not disappear at zero temperature,
Eq.~(\ref{gc010}) has a spurious finite limit as $\bb$ goes to $\infty$,
whereas the variational estimate (\ref{gc030})
for the particle-number dispersion
vanishes in this limit as $1/\bb$, with a coefficient
equal to the density of single-particle states at the Fermi level.
{It is therefore consistent with the expectation
that, for finite systems
with or without pairing, results from the}
grand-canonical and canonical {ensembles
must} coincide at zero temperature.

\subsubsection[S422]{Characteristic Function
for Conserved Single-Quasi-Particle Observables}

The ideas above are readily extended to correlations between
{any pair of single-quasi-particle observables $\QDhg$ and
$\QDh_\delta$ commuting} with the operator $\KDh$.
Indeed, it follows from the discussion of Sect.~2.3 that the evaluation
of the characteristic function, and therefore of the
cumulants, can be reduced in this case to the calculation
of a thermodynamic potential
if the trial spaces for $\AIh$ and $\DIh$ are invariant
with respect to the transformation (\ref{e340}).
Within the spaces (\ref{f160}) and (\ref{f170})
considered in this Section,
this invariance is satisfied for
operators $\QDhg$ of the type (\ref{f120}).

The identity (\ref{e350}) then shows us that the variational
evaluation of the characteristic function $\varphi(\xi)$
amounts to the variational evaluation of the
partition function associated with the shifted Hamiltonian
$\KDh^{\,\prime}=\KDh+\sum_\gamma\xi_\gamma\QDhg/\bb$. The calculation
is thus the same as in Sect.~4.1.1, except for the
change of $\KDh$ into $\KDh^{\,\prime}$,
and Eq.~(\ref{ga070}) replaced by
\be\label{gd010}
\varphi(\xi)={\rm Max}_{\RI}
[\SI\{\RI\}-\bb\EI\{\RI\}-
\sum_\ggg\xi_\ggg(\qi_\ggg+{\textstyle\usd}\trd{\QI_\ggg\RI)}]\q.
\ee
The optimum matrix $\RIp$ is characterized by the
extremum condition (\ref{gd010}),
which yields the self-consistent HFB
equations
\be\label{gd020}
\RIp=\frac{1}{
{\rm e}^{\displaystyle \,\bb\HIp\{\RIp\}}+1 }\ ,\q\q
\HIp\{\RIp\}\equiv\HI\{\RIp\}+\frac{1}{\bb}\sum_\ggg\xi_\ggg\,\QI_\ggg\q.
\ee

The characteristic function (\ref{gd010}) depends on the $\xi$'s both
directly and through the matrix $\RIp$. From its
first derivatives we recover the expectation values (\ref{gb010}).
The second derivatives, taken at $\xi=0$, yield the correlations
between the $\QDhg$'s.
In order to write their explicit expression,
setting $\RIp=\RIze+\delta\RI$,
we shall expand $\SI\{\RIp\}$,
$\EI\{\RIp\}$ and $\FI_{\rm G}\{\RIp\}$ around $\RIze$ up
to second order. To this aim, we introduce a condensed notation.

Up to now, we have regarded $\RI_{\lambda\lambda'}$,
$\HI_{\lambda\lambda'}$ as well as other quantities denoted by a script
capital ($\QI$, $\LI$, $\TI$, $\delta\RI$, $\HI_0$, etc.)
as $2n\times 2n$ matrices.
We  find it convenient to consider them alternatively
as {\it vectors in the
Liouville space}, with the pair $(\lambda\lambda')$
playing the r\^ole of a single
index (see Appendix, Sect.~A.1). Thus, the first-order
variations (\ref{ga080}) of $\SI\{\RI\}$ and (\ref{f380})
of $\EI\{\RI\}$ appear now as scalar products
that we shall denote by the colon sign :
which stands for $\usd\trd{{}}$. Likewise
the second-order variations will generate {\it matrices}
in the Liouville space, with two pairs of indices $(\lambda\,\lambda')$.

With this notation we can write,
for $\RIp=\RI_0+\delta\RI$,  the expansions of the HFB entropy
$\SI\{\RI\}$, energy $\EI\{\RI\}$ and grand
potential $\FI_{\rm G}\{\RI\}$, defined by Eqs.~(\ref{ga060}),
(\ref{f310}) and (\ref{ga050}), respectively, as
(see Appendix, Sect.~A.2) :
\be\label{gd030}
\SI\{\RIp\}\simeq\SI\{\RIze\}+\bb\HI_0:\delta\RI+
{\textstyle\usd}\delta\RI:\SL_0:\delta\RI\q,
\ee
where $\bb\HI_0$ stands for $\log(1-\RIze)/\RIze$ as in (\ref{ga010}) ;
\be\label{gd040}
\EI\{\RIp\}\simeq\EI\{\RIze\}+\HI\{\RIze\}:\delta\RI+
{\textstyle\usd}\delta\RI:\ELl_0:\delta\RI\q,
\ee
where $\HI\{\RIze\}$ is defined by (\ref{f380}) ;
\be\label{gd050}
\FI_{\rm G}\{\RIp\}\simeq\FI_{\rm G}\{\RIze\}+
{\textstyle\usd}\delta\RI:\FLl_0:\delta\RI\q,
\ee
where we made use of the stationary condition (\ref{ga020})
and where
\be\label{gd060}
\FLl_0\equiv\ELl_0-\frac{1}{\bb}\SL_0\q.
\ee
In Eqs.~(\ref{gd030}-\ref{gd050}),
the second derivatives
$\SL_0$, $\ELl_0$ and $\FLl_0$, as well as
the first derivatives $\bb\HI_0$
and $\HI\{\RIze\}$, are functions of
the HFB density $\RI_0$.
In the $\lambda$-basis of Sect.~3.1,
$\SL_0$, $\ELl_0$ and $\FLl_0$
are 4-index tensors~; we regard them here
as (symmetric) matrices in the Liouville
space ($\lambda\lambda'$). The explicit expression
of $\SL_{0\,(\lambda\lambda')(\mu\mu')}$ is given in Eq.~(\ref{ab070}).
Contraction of such a matrix with a vector proceeds as
\be\label{gd070}
(\SL_0 :\delta\RI)_{\lambda\lambda'}\equiv
{\textstyle\usd}\sum_{\mu\mu'}
\SL_{0\,(\lambda\lambda')(\mu\mu')}\,\delta\RI_{\mu'\mu}\q.
\ee
In the Appendix A (Sect.~A.4), we endow $-\SL_0$
with a geometric interpretation.
It appears naturally as a Riemannian metric
tensor which defines a distance between two neighbouring states
characterized by the contraction matrices
$\RIze$ and $\RIze+\delta\RI$.

The correlation $\Cgd$ between the two
observables $\QDhg$ and $\QDh_\delta$ (defined as in 
Eqs.~(\ref{I070}-\ref{I080}))
is obtained from the expansion
of the expression (\ref{gd010}) around $\RI_0$ :
\be\label{gd080}
\varphi(\xi)\simeq\log\YI-\sum_\ggg\xi_\ggg\langle\QDhg\rangle
-{\rm Min}_{\delta\RI}
[{\textstyle\usd}\bb\,\delta\RI:\FLl_0:\delta\RI+
\sum_\ggg\xi_\ggg\QI_\ggg:\delta\RI]\ .
\ee
The extremum condition,
\be\label{gd090}
\bb\,\FLl_0:\delta\RI=-\sum_\ggg\xi_\ggg\QI_\ggg\q,
\ee
determines $\delta\RI$ as function of the $\xi$'s.
From Eq.~(\ref{gd080}), one has
\be\label{gd100}
\Cgd=
\pog
\frac{\partial^2\varphi }{\partial\xi_\gamma\,\partial\xi_\delta}
\right\vert_{\xi=0}=
-\QI_\gamma : \pog
\frac{\partial\  }{\partial\xi_\delta}\delta\RI
\right\vert_{\xi_=0}\q.
\ee
Inserting in Eq.~(\ref{gd100})
the formal solution
of Eq.~(\ref{gd090}),
\be\label{gd110}
\pog\frac{\partial\ }{\partial\xi_\delta}\delta\RI\right\vert_{\xi=0}=
-\frac{1}{\bb}\FLl_0^{-1}:\QI_\delta\q,
\ee
one obtains
\be\label{gd120}
\Cgd=\frac{1}{\bb}\,\QI_\ggg:\FLl_0^{-1}:\QI_\delta\q.
\ee
We shall discuss the problem of the vanishing eigenvalues of $\FLl_0$ in
Sect.~4.2.3. The formula (\ref{gd120}) encompasses the
fluctuations $\Delta\QDh_\ggg^2$, obtained for $\ggg=\delta$.

\noindent{\it Remark }:
The above results have been obtained indirectly by using the
identity (\ref{e350}).
We could also have
found them directly by solving the coupled equations of Sect.~3.3.
This is feasible, but not straightforward. Indeed, having
defined $\RIp$ and $\HI\{\RIp\}$ by Eq.~(\ref{gd020}), we
can first check
that the solution of Eqs.~(\ref{f520}) is given by
\be\label{gd130}
\RIb\uu={\rm e}^{\displaystyle -u\LI^\AP/2\bb}\,\RIp\,
{\rm e}^{\displaystyle u\LI^\AP/2\bb}\,,\q
\RIc\uu={\rm e}^{\displaystyle u\LI^\AP/2\bb}\,\RIp\,
{\rm e}^{\displaystyle -u\LI^\AP/2\bb}\ .
\ee
The proof uses the identity (\ref{ge030}) below
where $\UI$ is replaced by $\exp u\LI^\AP$. We can
then solve the coupled equations (\ref{f370}) and (\ref{f430}),
together with the required boundary conditions $\TId\zz=1$ and
$\TIa\bbb=\exp(-\LI^\AP)$, which results in
\be\label{gd140}\ba{rl}
\TId\uu&={\rm e}^{\displaystyle u\LI^\AP/2\bb}\,
{\rm e}^{\displaystyle -u\HIp\{\RIp\}}
\,{\rm e}^{\displaystyle u\LI^\AP/2\bb}\q,\\
\TIa\uu&={\rm e}^{\displaystyle -u\LI^\AP/2\bb}\,
{\rm e}^{\displaystyle -(\bb-u)\HIp\{\RIp\}}
\,{\rm e}^{\displaystyle -u\LI^\AP/2\bb}\q.
\ea\ee
We finally recover the characteristic function (\ref{gd010}) by
means of (\ref{f460}--\ref{f470}) and (\ref{f490}--\ref{f500}). If
$\LI^\AP$ commutes with $\RIp$, and therefore with
$\HIp\{\RIp\}$, the expressions (\ref{gd130}) and (\ref{gd140})
simplify. Otherwise, when some $\QDhg$--invariances are broken,
the $u$-dependence in the solution (\ref{gd140})
of the evolution equations
(\ref{f370}) and (\ref{f430}) is no longer as simple as
it is for $\xi=0$ or
for the exact solution [given by $\DIh\uu={\rm e}^{-u\KDh}$
and by Eq.~(\ref{e120}) for $\AIh(u)$].
In particular $\log\TId\uu$ is not proportional here to $u$,
but contains contributions in $u^3$, $u^5$, ... .
Although, in the case
of commuting observables,
we were able  to indirectly solve
the coupled equations (\ref{f370}) and (\ref{f430}),
it would not have been easy to guess a priori
the complicated $u$-dependence of the solutions
(\ref{gd130}-\ref{gd140}).

\subsubsection[S423]{Broken Invariances}

Since equilibrium is described by the minimum of 
the grand potential $\FI_{\rm G}\{\RI\}$,
given by (\ref{gd050}),
the eigenvalues of the matrix $\FLl_0$ defined in Eq.~(\ref{gd060})
are non-negative. We have,
however, to consider the case of vanishing eigenvalues
for which the inversion of Eq.~(\ref{gd090}) raises a problem.
A characteristic situation where this
{happens} is that of
broken invariances.
For instance, in the mean-field description
of pairing correlations, the particle-number $\NDh$
symmetry is broken. For convenience,
in this Subsection, we shall use this operator
as the representative of all broken symmetries,
although the whole discussion applies to any {\it conserved}
single-quasi-particle observable.

By definition
of a conserved observable, $\NDh$ satisfies the
commutation relation $[\KDh\,,\,\NDh]=0$. When the
$\NDh$-invariance is not broken, $\NDh$ commutes with the
approximate density operator $\DIhz\bbb$ and we have
both $[\RIze\,,\,\NI]=0$ and $[\HI_0\,,\,\NI]=0$,
where the matrix $\NI$ has been defined in (\ref{gb040}).
However, when the $\NDh$-invariance is broken,
$\NI$ does not commute with
$\RIze$ and $\HI_0$ despite the commutation of $\NDh$ with
$\KDh$ in the Fock space.
{In the HFB theory, this non-commutation manifests itself } by the
occurence of non-zero off-diagonal elements
in (\ref{f100}) and (\ref{f390}).

Let us show how the $\NDh$-invariance, whether or not it is broken,
is reflected on our variational objects. We can associate with
$\NDh$ a continuous set of unitary transformations
$\UDh\equiv\exp\,i\,\epsilon\NDh$ which leave $\KDh$ invariant.
They are represented in the $2n$-dimensional $\lambda$-space of
Sect.~3.1  by unitary matrices
\be\label{ge010}
\UI\equiv{\rm e}^{\displaystyle i\,\epsilon\NI}\q.
\ee
From the definition $\EI\{\RI\}=\Tr{\DIh\KDh}/\Tr{\DIh}$ and
the commutation $[\KDh\,,\,\NDh]=0$, we find
\be\label{ge020}
\EI\{\RI\}=
\frac{ \Tr{ \UDh\DIh\UDh^{-1}\KDh } }{ \Tr{ \UDh\DIh\UDh^{-1} } }
=\EI\{\UI\RI\,\UI^{-1}\}\q,
\ee
for any $\RI$
satisfying the symmetry condition (\ref{f090}).
We also have $\SI\{\RI\}=\SI\{\UI\RI\,\UI^{-1}\}$ since
$\SI\{\RI\}$, defined by (\ref{ga060}), depends only on
the eigenvalues of $\RI$. Thus, if $\FI_{\rm G}\{\RI\}$ reaches
its minimum for some $\RIze$ satisfying (\ref{ga020}) and if
the $\NDh$--invariance is broken, 
the transformation $\UI\RIze\,\UI^{-1}$
generates a set of distinct solutions which give the same value
$\FI_{\rm G}\{\UI\RI_0\,\UI^{-1}\}=\FI_{\rm G}\{\RI_0\}$.
{This implies that} the second derivative
$\FLl_0$ of $\FI_{\rm G}\{\RI\}$ has
a vanishing eigenvalue. To find the {associated
eigenvector}, we note that the first-order variation of
(\ref{ge020}) with respect to $\RI$, together with
the definition (\ref{f380}) of
$\HI\{\RI\}$, yields
\be\label{ge030}
\UI\HI\{\RI\}\UI^{-1}=\HI\{\UI\RI\UI^{-1}\}\q.
\ee
Hence, letting $\RI=\RIze$, taking an
infinitesimal transformation $\UI$
and using $\delta\HI=\ELl_0:\delta\RI$ (a consequence of the
definition (\ref{gd040}) of $\ELl_0$), we find
\be\label{ge040}
[\NI\,,\,\HI_0]=\ELl_0:[\NI\,,\,\RIze]\q.
\ee
The same argument holds when $\EI\{\RI\}$ is
replaced by $\FI_{\rm G}\{\RI\}$, but then the first derivatives
of $\FI_{\rm G}$ vanish at $\RI=\RIze$
and Eq.~(\ref{ge040}) is replaced by
\be\label{ge050}
\FLl_0:[\NI\,,\,\RIze]=0\ ,\q\q[\NI\,,\,\RIze]:\FLl_0=0\q.
\ee
These equations express
that the commutator $[\NI\,,\,\RIze]$, which does not vanish
when the $\NDh$-invariance is broken by $\RIze$, is a right or
left eigenvector of the matrix $\FLl_0$ for the eigenvalue 0.

We are now in position to discuss the expression (\ref{gd120})
for the correlation $\Cgd$ of two {conserved}
observables $\QDhg$ and $\QDh_\delta$ when some invariance (here
the $\NDh$--invariance) is broken.
Depending on the nature of
the observables, we can distinguish three cases. 
\begin{itemize}
\item (i) {\it If both $\QDhg$ and $\QDh_\delta$ are such that}
\be\label{ge060}
\QI:[\NI\,,\,\RIze]=\langle[\QDh\,,\,\NDh]\rangle=0\q,
\ee
in particular if $\QDhg$ and $\QDh_\delta$ commute with $\NDh$
(as well as with $\KDh$), their expectation values (\ref{gb010})
are well defined, although $\RIze$ is not. Indeed, 
all the density matrices $\,\UI\RIze\,\UI^{-1}$
for which $\FI_{\rm G}\{\RI\}$
is minimum, will provide the same value for $\langle\QDhg\rangle$.
The condition (\ref{ge060}) ensures that the equation
(\ref{gd090}) for
$\partial\,\delta\RI/\partial\xi_\delta\vert_{\xi=0}$ has a finite
solution, that we can denote $-\bb^{-1}\FLl_0^{-1}:\QI_\delta$.
Due to (\ref{ge050}) this
solution is not unique ; it is only defined
within the addition of $i\epsilon\,[\NI\,,\,\RIze]$, where
$\epsilon$ is arbitrary.
The condition (\ref{ge060}), however, implies that this
addition is irrelevant. Hence
{\it the correlation (\ref{gd120}) is finite and well-defined},
in spite of the vanishing eigenvalue (\ref{ge050}) of $\FLl_0$.
This conclusion holds in particular for
the fluctuation $\DNh^{\,2}$ obtained when $\QDhg=\QDh_\delta=\NDh$.

\item (ii) {\it If $\QDh_\delta$, but not $\QDhg$,
satisfies (\ref{ge060})},
the expectation value $\langle\QDhg\rangle$ depends on the specific
solution $\RIze$ of Eq.~(\ref{ga020}).
{This feature is characteristic of a situation with}
broken $\NDh$-invariance. Moreover, the additive arbitrary contribution
$i\epsilon\,[\NI\,,\,\RIze]$ to
$\partial\,\delta\RI/\partial\xi_\delta\vert_{\xi=0}$ also
contributes to (\ref{gd120}), so that the correlation between
$\QDhg$ and $\QDh_\delta$ is {\it finite, but ill-defined.}

\item (iii) Finally,
{\it if both $\QDhg$ and $\QDh_\delta$ violate (\ref{ge060})}, the
equation (\ref{gd090}) has no finite solution. The fluctuation
of $\QDhg$, in particular, is {\it infinite},
consistently with the fact that $\langle\QDhg\rangle$ is
ill-defined.
\end{itemize}
These conclusions hold without any changes for the breaking
of invariances other than $\NDh$.

\subsection[S43]{
Fluctuations and Correlations of Non-Conserved Observables
}

In this Section we consider the general case of single-quasi-particle
observables $\QDhg$ which do not commute with the operator $\KDh$.

\subsubsection[S431]{Kubo Correlations}

When the observables $\QDhg$ do not commute
with $\KDh$,
the expression (\ref{gd010}) is still the variational approximation for
$\log\Tr{\exp[-\bb\KDh-\sum_\gamma\xi_\gamma\QDhg]}$. By expansion
in powers of the $\xi_\gamma$'s, this
expression generates cumulants of the Kubo type.
In particular, while its zeroth and first-order contributions
coincide with (\ref{ga070})
and (\ref{gb010}), respectively,
the second-order contribution,
\be\label{gf010}
\displaystyle
C_{\ggg\delta}^{\rm K}=C_{\delta\ggg}^{\rm K}\equiv
\frac{\displaystyle \Tr{
{\rm e}^{\displaystyle -\bb\KDh}\,
\int_0^\bb\,{\rm d}u\,
{\rm e}^{\displaystyle u\KDh}\,\QDh_\gamma\,
{\rm e}^{\displaystyle -u\KDh}\QDh_\delta}}
{\bb\,\Tr{{\rm e}^{\displaystyle -\bb\KDh}}}
-\langle\QDh_\gamma\rangle\langle\QDh_\delta\rangle\q,
\ee
yields the {\it Kubo correlations} between the $\QDhg$'s,
which are thus {\it given variationally} by (\ref{gd120}), that is
\be\label{gf020}
C_{\ggg\delta}^{\rm K}
=\frac{1}{\bb}\,\QI_\ggg:\FLl_0^{-1}:\QI_\delta\q.
\ee

\subsubsection[S432]{Standard Correlations}

In order to obtain {our variational approximation
for the} true correlations
$\Cgd=\usd\langle\QDhg\QDh_\delta+\QDh_\delta\QDhg\rangle-
\langle\QDhg\rangle\langle\QDh_\delta\rangle$
between $\QDhg$ and $\QDh_\delta$,
we have to return to the coupled equations of Sect.~3.3,
facing their $u$--dependence and expanding
them in powers of the sources $\xi_\gamma$.
Sect.~4.1 gave the zeroth-order solution, which happened to
furnish both the thermodynamical potential
and the expectation values.
Fluctuations and correlations are provided by the next order.

We start from the fundamental relation (\ref{e200}) that
involves the stationary operator $\DIh\bbb$
and the operator $\ADh(\xi)$ defined in (\ref{I060}) ;
more explicitly it reads~:
\be\label{gg010}
\frac{\partial\varphi}{\partial\xi_\gamma}=
-\frac{\displaystyle \Tr{ \DIh\bbb\,
\int_0^1{\rm d}v\,
{\rm e}^{\displaystyle -v\sum_\delta\xi_\delta\QDhd}
\,\QDhg\,
\,{\rm e}^{\displaystyle -(1-v)\sum_\delta\xi_\delta\QDhd}}}
{\Tr{\DIh\bbb\,e^{\displaystyle -\sum_\delta\xi_\delta\QDhd}}}\q.
\ee
As discussed in Sect.~4.1.2, Eq.~(\ref{gg010})
taken at $\xi=0$ yields $\langle\QDhg\rangle$
as the expectation value
(\ref{gb010}) of $\QDhg$ in the state $\DIhz\bbb$
already determined in Sect.~4.1.1.
The correlations $\Cgd$ and the fluctuations
$\Delta\QDh_\ggg^2=C_{\ggg\ggg}$ are obtained by expanding (\ref{gg010})
up to first order in the $\xi$'s
which appear both explicitly and through the expansion
\be\label{gg020}
\DIh\bbb\equiv\DIhz\bbb+\DIhu\bbb+\ldots\equiv
\DIhz\bbb+\sum_\delta\xi_\delta\,\DIh_\delta\bbb+\ldots\q
\ee
of the stationary state $\DIh\bbb$.
For shorthand we denote with an index $\delta$ the coefficient
\be\label{gg030}
\DIh_\delta\bbb\equiv
\pog\frac{\partial\DIh\bbb}{\partial\xi_\delta}\right\vert_{\xi=0}
\ee
of $\xi_\ggg$ in the first-order contribution to $\DIh\bbb$.
Note that, thanks to the variational
nature of the method, only this first-order contribution
is needed to evaluate the
wanted second derivatives of the characteristic function.

By using the formalism of Sects.~3.1 and 3.2, we can rewrite
(\ref{gg010}) as
\be\label{gg040}\ba{rl}
\frac{\displaystyle \partial\varphi}{\displaystyle \partial\xi_\gamma}
&=\displaystyle
-\qi_\gamma-{\textstyle\usd}\trd{
\RIb\bbb\,
\int_0^1{\rm d}v\,
{\rm e}^{\displaystyle -v\LI^\AP}\,\QI_\gamma\,
{\rm e}^{\displaystyle v\LI^\AP}}\\
&=\displaystyle
-\qi_\gamma-{\textstyle\usd}\trd{
\RIc\bbb\,
\int_0^1{\rm d}v\,
{\rm e}^{\displaystyle v\LI^\AP}\,\QI_\gamma\,
{\rm e}^{\displaystyle -v\LI^\AP}}
\q.
\ea\ee
The correlation $\Cgd$ is the derivative of this
expression with respect
to $\xi_\delta$, evaluated at $\xi=0$.
Contributions arise here
from the dependences in $\xi_\delta$ 
of both $\RIb\bbb$ (or $\RIc\bbb$)
and $\LI^\AP\equiv\sum_\ggg\xi_\ggg\QI_\ggg$.
We expand as in (\ref{gg020})
the matrices $\RI^\ip(\xi)$ and $\TI^\ip(\xi)$
around $\xi=0$, introducing the same notations
$\RIig\equiv\partial\RI^\ip/\partial\xi_\ggg\vert_{\xi=0}$
and $\TIig\equiv\partial\TI^\ip/\partial\xi_\ggg\vert_{\xi=0}$
as in (\ref{gg030})
(the symbol $\ip$ stands for $\ap$, $\ep$, $\bp$ or $\cpp$).
We obtain
\be\label{gg050}\ba{rl}
\Cgd&=-\usd\trd{\QI_\gamma\,\RIbd\bbb}+
\usq\trd[\QI_\delta\,,\,\QI_\gamma]\RIze\\
&=-\usd\trd{\QI_\gamma\,\RIcd\bbb}-
\usq\trd[\QI_\delta\,,\,\QI_\gamma]\RIze\\
&=-\usq\trd{\QI_\gamma\,\{\RIbd\bbb+\RIcd\bbb\}}\q.
\ea\ee

We now work out the coupled equations of Sect.~3.3 so as to find
an explicit expression for (\ref{gg050}). First we note,
from the boundary conditions (\ref{f440}), that we have
\be\label{gg060}
\TIaze\bbb=1\ ,\q\q\TI^\ap_\delta\bbb=-\QI_\delta\q,
\ee
and hence, using (\ref{f080}), (\ref{f240}) and (\ref{ga030}) :
\be\label{gg070}\ba{rl}
\RIbd\bbb=&\RIdd\bbb-(1-\RIze)\,\QI_\delta\,\RIze\q,\\
\RIcd\bbb=&\RIdd\bbb-\RIze\,\QI_\delta\,(1-\RIze)\q.
\ea\ee
The insertion of
(\ref{gg070}) into the expression (\ref{gg050}) of $\Cgd$ yields
\be\label{gg080}
\Cgd=
{\textstyle\usq}\trd{[
\QI_\gamma\RIze\QI_\delta(1-\RIze)+
\QI_\gamma(1-\RIze)\QI_\delta\RIze]}
-{\textstyle\usd}\trd{ \QI_\gamma\,\RIdd\bbb}
\q.\ee
The first term in the right hand side is
what would emerge from
a naive calculation using Wick's theorem with respect
to the state $\DIh_0\bbb$.
The departure from this result, given by the last term,
arises from the variational nature of the calculation
which modifies $\DIh\bbb$ and hence the associated
contraction matrix $\RId\bbb$ when { sources are present}.

Rather than evaluating $\RIdd\bbb$
we shall transform Eq.~(\ref{gg050})
so as to bring the ``time'' $u$ down to 0 and
thus find a more explicit
expression for the correlation
$\Cgd$. To this aim we expand the equations of motion
(\ref{f520}) for $\RIb$ and $\RIc$ around the lowest-order contribution
(\ref{ga030}). This leads to
\be\label{gg090}
\duh{\RIbd}=\usd\KLl_0\,:\,\RIbd\ ,\q\q
\duh{\RIcd}=-\usd\KLl_0\,:\,\RIcd\q,
\ee
where the kernel $\KLl_0$ is defined,  for
$\RI=\RIze+\delta\RI$,  by
\be\label{gg100}\ba{rl}
\KLl_0:\delta\RI&\equiv\delta[\HI\{\RI\}\,,\,\RI]\\
&=\displaystyle [\HI\{\RIze\}\,,\,\delta\RI]+
[\,\sum_{\mu\mu'}
\left.\frac{\partial\HI\{\RI\}}{\partial\RI_{\mu\mu'}}
\right\vert_{\RI=\RIze}
{\delta\RI}_{\mu\mu'}
\,,\,\RIze]\q.
\ea\ee
The $(2n\times2n)\times(2n\times2n)$ tensor $\KLl_0$ is nothing but
the
{\it usual RPA tensor associated with the HFB Hamiltonian}
$\HI\{\RI\}$
and
customarily introduced to describe small
deviations around the equilibrium state
$\RIze$. As in Eq.~(\ref{gd070}), the colon symbol in 
$\KLl_0:\delta\RI$ stands for
half a trace (with a twist in the indices)
involving the last pair of indices of $\KLl_0$ and the 
pair of indices
of $\delta\RI$; equivalently, $\KLl_0:\delta\RI$ 
can be considered
as a scalar product in the $2n\times2n$ vector space of $\delta\RI$.
It is shown in the
Appendix (Sect.~A.3) that $\KLl_0$ is related simply to the matrix
$\FLl_0$ which, according to (\ref{gd050}), characterizes the
second-order variations of the thermodynamic
potential $\FI_{\rm G}\{\RI\}$ around
$\RIze$. More precisely if we define, for any matrix $\MI$, the
tensor $\CLl_0$ as the commutator with the HFB density matrix $\RIze$,
\be\label{gg110}
\CLl_0:\MI\equiv[\RIze\,,\,\MI]\q,
\ee
the matrix $\KLl_0$ is (in the Liouville space) {the product}
\be\label{gg120}
\KLl_0=-\CLl_0\FLl_0\equiv-\CLl_0 :\FLl_0\q,
\ee
again defined as in (\ref{gd070}).
The occurence of the simple relation (\ref{gg120})
and its meaning are explained in Sects.~3 and 5 of Appendix A.

The boundary condition $\TId\zz=1$ implies
\be\label{gg130}
\RIbd\zz=\RIcd\zz=\RIad\zz\equiv\XI\q.
\ee
A formal solution of Eqs.~(\ref{gg090}), together
with the boundary condition (\ref{gg130}),
gives
\be\label{gg140}
\RIbd\uu={\rm e}^{{\textstyle\usd}\displaystyle u\KLl_0}:\XI\ ,\q\q
\RIcd\uu={\rm e}^{-{\textstyle\usd}\displaystyle u\KLl_0}:\XI\q,
\ee
where $\XI$ is still unknown. The correlation (\ref{gg050})
can now be written as 
\be\label{gg150}
\Cgd=-\QI_\ggg:\cosh{\textstyle\usd}\bb\,\KLl_0:\XI\q.
\ee
With the aim to find $\XI$, using
(\ref{gg070}) and again (\ref{gg140}), we can obtain
\be\label{gg160}
\RIbd\bbb-\RIcd\bbb=[\RIze\,,\,\QI_\delta]=
2\,\sinh{\textstyle\usd}\bb\,\KLl_0\,:\,\XI\q.
\ee
If $\KLl_0$ had no vanishing eigenvalue, we could eliminate
$\XI$ between (\ref{gg150})
and (\ref{gg160}) and write:
\be\label{gg170}
\Cgd=-{\textstyle\usd}\QI_\gamma\,:\,\coth
{\textstyle\usd}\bb\,\KLl_0\,:\,
[\RIze\,,\,\QI_\delta]\q.
\ee
Unfortunately,
due both to the vanishing
eigenvalues of $\CLl_0$ associated with all one-quasi-particle
operators that commute with $\usd\gbt\HI_0\gbd$ and to the
possible vanishing eigenvalues of $\FLl_0$ associated with broken
invariances {(see Sect.4.2.3)},
the operator $\KLl_0=-\CLl_0\FLl_0$ has no inverse.

In order to determine the unknown quantity $\XI\equiv\RIad\zz$
entering (\ref{gg150}), we return to the equation (\ref{f430})
which governs {the evolution of} $\TIad\uu$.
{Eq.(\ref{gg060}) gives the boundary condition on  $\TIad\bbb$
while, in agreement with Eq.~(\ref{f510}),
$\TIad\zz$ is related to $\XI$ through
\be\label{gg180}
\TIad\zz=
(1-\RIze)^{-1}
\XI
(1-\RIze)^{-1}\q.
\ee

We now relate $\TIad\bbb$ to $\TIad\zz$ by expanding
Eq.~(\ref{f430}) to first order in the sources around the zeroth-order
solution (\ref{ga010}-\ref{ga040}). This yields
the differential equation
\be\label{gg190}
2\frac{{\rm d}\TIad}{{\rm d}u}-
\HI_0\TIad+\TIad\HI_0=
(\ELl_0  :\RIbd){\rm e}^{-(\bb-u)\HI_0}+
{\rm e}^{-(\bb-u)\HI_0}(\ELl_0  :\RIcd)\q,
\ee
where the $u$-dependence has been omitted in $\TIad$,
$\RIbd$ and $\RIcd$.
We used the expansion
$\HI\{\RIze+\delta\RI\}\simeq\HI_0+\ELl_0:\delta\RI$,
entailed by the definition (\ref{gd040}) of the matrix
$\ELl_0$. The quantities $\RIbd$ and $\RIcd$ are given
by Eq.~(\ref{gg140}) in terms of $\XI$
and as functions of $u$ .
{In order to transform the left hand side
into a derivative, we multiply  Eq.~(\ref{gg190})
 by $\exp\usd(\bb-u)\HI_0$
on the left and on the right. As a result, terms of the
form ${\rm e}^{v\HI_0}\MI{\rm e}^{-v\HI_0}$} appear on the 
r.h.s. of this equation.
Using the fact (see Appendix, Sect.~A.3) that the matrix
$\SL_0\CLl_0$ in the Liouville space
describes the commutation with $\bb\HI_0$,
we rewrite these terms as}
\be\label{gg200}
{\rm e}^{\displaystyle v\,\HI_0}\MI
{\rm e}^{\displaystyle -v\,\HI_0}=
{\rm e}^{\displaystyle \frac{v}{\bb}\,\SL_0\CLl_0}:\MI\q.
\ee
Relying now on (\ref{gg140}) we note that,
{after this transformation, the
two terms on the r.h.s. 
of Eq.(\ref{gg190})  depend on $u$
only through} exponentials of $\SL_0\CLl_0$ and $\KLl_0$.
{In addition,
these terms can be explicitly} integrated over $u$ by means of the
identity
\be\label{gg210}\ba{l}
{\rm e}^{\pm\usd\frac{\bb-u}{\bb}\SL_0\CLl_0}
 : \ELl_0 :
{\rm e}^{\pm\usd u\KLl_0} =\\
\q\q
\mp 2{\displaystyle\frac{{\rm d}\ }{{\rm d}u}}[
({\rm e}^{\pm\usd\frac{\bb-u}{\bb}\SL_0\CLl_0}-1)
\CLl_0^{-1}
{\rm e}^{\pm\usd u\KLl_0}
+
\CLl_0^{-1}
({\rm e}^{\pm\usd u\KLl_0}-{\rm e}^{\pm\usd\bb\KLl_0})
]\q,
\ea\ee
which is checked by {
using Eqs.~(\ref{gd060}) and (\ref{gg120})}.
Although $\CLl_0$ has no inverse,
the r.h.s.  is {nevertheless}
well defined because $\CLl_0^{-1}$ is always multiplied either
on the left side by $\SL_0\CLl_0$ or on the right side by
$\CLl_0\FLl_0$ when
the exponentials are expanded. This amounts to
extending to operators the natural convention 
$({\rm e}^{ax}-{\rm e}^{bx})x^{-1}=a-b$
for $x=0$. Integrating the transformed differential
equation for $\TIad\uu$,
with the boundary condition $\TIad\bbb=-\QI_\delta$ given in
(\ref{gg060}), we find
\be\label{gg220}\ba{l}
{\rm e}^{\usd(\bb-u)\HI_0}\,\TIad\uu\,{\rm e}^{\usd(\bb-u)\HI_0}
+\QI_\delta=\\
\q\q\crg
  (1-{\rm e}^{\usd\frac{\bb-u}{\bb}\SL_0\CLl_0})
\CLl_0^{-1}{\rm e}^{\usd u\KLl_0}
- (1-{\rm e}^{-\usd\frac{\bb-u}{\bb}\SL_0\CLl_0})
\CLl_0^{-1}{\rm e}^{-\usd u\KLl_0}
\pod\\
\q\q\q\q\pog
+ 2 \CLl_0^{-1}
(\sinh\usd\bb\KLl_0-\sinh\usd u\KLl_0)
\crd : \XI\q.
\ea\ee
For $u=0$, Eqs.~(\ref{gg180}) and (\ref{gg220}) provide
the required equation for $\XI$.
We can simplify the result by rewriting (\ref{gg180})
as
\be\label{gg230}\ba{rl}
{\rm e}^{\usd\bb\HI_0}\,\TIad\zz\,{\rm e}^{\usd\bb\HI_0}
&=4\cosh[{\textstyle\usd}\bb\HI_0]\,\XI\,
\cosh[{\textstyle\usd}\bb\HI_0]\\
&=-2\,(\sinh{\textstyle\usd}\SL_0\CLl_0)\CLl_0^{-1} :\XI\q,
\ea\ee
an equality that is derived by using the
expressions (\ref{ab080}) and (\ref{ac050}) of $\SL_0$
and $\CLl_0$ in the basis where $\HI_0$ is diagonal.
Again the r.h.s. of (\ref{gg230})
is well defined, despite the
vanishing eigenvalues of $\CLl_0$.
We eventually find
\be\label{gg240}
\QI_\delta=
2\CLl_0^{-1}\,\sinh({\textstyle\usd}\bb\KLl_0):\XI
\q.
\ee
Assuming that $\FLl_0^{-1}$ exists, Eq.~(\ref{gg240}) can be inverted
in the Liouville space; using the relation (\ref{gg120}), this gives
\be\label{gg250}
\XI=-\usd\,\frac{\KLl_0}{\sinh\usd\bb\,\KLl_0}\FLl_0^{-1} : \QI_\delta\q.
\ee
Hence, from (\ref{gg150}) we find
\be\label{gg260}
\Cgd={\textstyle\usd}
\QI_\ggg:(\KLl_0\coth{\textstyle\usd}\bb\,
\KLl_0)\FLl_0^{-1}:\QI_\delta\q.
\ee

This final result is our {\it
variational approximation for the correlations $\Cgd$ or
for the fluctuations $\Delta\QDh^2_\ggg$}.
The symmetry $\Cgd=C_{\delta\gamma}$ is a simple
consequence of the relation $\KLl_0=-\CLl_0\FLl_0$,
whereas this symmetry was not obvious from our starting
equations (\ref{gg050}), (\ref{gg080}) or (\ref{gg170}).
The formula
(\ref{gg260}) gives a meaning to the formal
expression (\ref{gg170})
which, due to the vanishing eigenvalues of the
commutator $\CLl_0$ in $\KLl_0=-\CLl_0\FLl_0$ (Eq.~(\ref{gg120})),
was not defined.
As for the vanishing eigenvalues of $\FLl_0$,
our discussion of Sect.~4.2.3
still holds
for (\ref{gg260}) as well as for
(\ref{gd120}), and all its conclusions remain valid when
$\QDhg$ and $\QDh_\delta$ are
arbitrary single-quasi-particle operators.

Finally, when one of the observables,
say $\QDhg$, is a conserved
quantity (i.e., $[\QDhg\,,\,\KDh]=0$),
we find from (\ref{ge050})
that
\be\label{gg270}
\QI_\ggg:\KLl_0\equiv-\QI_\ggg:\CLl_0\FLl_0=0\q,
\ee
even if the $\QDhg$--invariance is broken (if it is not,
we have $\QI_\ggg:\CLl_0=0$). Then the {left-action of the operator}
$\ \KLl_0\coth\usd\bb\,\KLl_0\ $ on $\QI_\ggg$
{amounts to a multiplication by
the c-number} $2/\bb$, so that from
Eq.~(\ref{gg260}) we recover 
the result (\ref{gd120}) of Sect.~4.2.2 for the correlations
when one of the two observables
$\QDhg$ or $\QDh_\delta$ is conserved.

\subsubsection[S433]{Diagrammatic Interpretation}

The expression (\ref{gg260}) can alternatively be obtained
by means of a {\it perturbative expansion}, where we take
$\DIhz\bbb$ as the unperturbed state and
$\usd\gbt\HI_0\gbd$
as the unperturbed Hamiltonian which
generates the quasi-particle or quasi-hole propagator.
Indeed, this expression coincides with the sum of the
contributions of all the
{\it bubble (or open ring, or chain) diagrams}.
These diagrams, which start from $\QDh_\delta$ and end up at $\QDhg$,
iterate the interaction 
$\VDh$ (see Eqs.~(\ref{f270} and (\ref{f290}))
with a pair of quasi-particle or quasi-hole lines.
A partial result \cite{BVe92} was already obtained
in the special case when (i) there is no broken invariance
and (ii) the observables $\QDh_\ggg$ and $\QDh_\delta$ are 
conserved. Moreover,
the derivation disregarded the difficulty caused by the vanishing
eigenvalues of the kernel $\KLl_0$.

Let us sketch here a proof which takes {into account this}
difficulty and applies to systems
with pairing. A
single line in a diagram
carries two indices $\lambda$ and $\lambda'$,
each of which is associated with either a creation or an
annihilation operator.
Like the contraction matrix $\RIze$, the associated propagator
is thus a matrix in the $2n$--dimensional space 
introduced in Sect.~3.1 ;
it depends moreover on the difference in ``times'' $u$,
with $0<u<\bb$. In bubble diagrams, the propagator
which connects two successive interactions involves
a pair of lines.
It appears as a $(2n\times 2n)\times(2n\times 2n)$ tensor,
depending again on a single time-difference and having the period
$\beta$.
As usual, the Fourier representation replaces this time difference
by an index $\omega$ which takes the values
$\omega=2\pi\,{\rm i}\,\bb^{-1}\,m$, where $m$ is an integer.
Again, it is convenient to regard the resulting propagator
$\Gamma(\omega)$ as a matrix in the Liouville space.
In a basis where $\RIze$ and $\HI_0$ are diagonal, 
a pair of lines is represented by
\be\label{gh010}
\Gamma_{(\lambda\lambda')(\mu\mu')}(\omega)=
2\delta_{\lambda\mu'}\delta_{\mu\lambda'}
\frac{\RI_{0\lambda}-\RI_{0\mu}}{\HI_{0\mu}-\HI_{0\lambda}-\omega}\q.
\ee
The numerator comes from the pair of statistical factors
$\RIze$ or $1-\RIze$ associated with the lines
$\lambda$ and $\mu$~; the denominator
accounts for the energies of these lines.
For $\HI_{0\mu}=\HI_{0\lambda}$, which implies
$\RI_{0\mu}=\RI_{0\lambda}$, the propagator
$\Gamma(\omega)$ vanishes, except for $\omega=0$
in which case it is equal to
\be\label{gh020}
\Gamma_{(\lambda\lambda')(\mu\mu')}(0)=
2\delta_{\lambda\mu'}\delta_{\mu\lambda'}
\bb\,\RI_{0\lambda}(1-\RI_{0\lambda})\q.
\ee
We can formally rewrite (\ref{gh010})
in an arbitrary basis of the Liouville
space by using the
{expressions (\ref{ad030}-\ref{ad040}) }
for $\SLl_0^{-1}$
and the definition (\ref{gg110}) of the commutator $\CLl_0$.
This provides
\begin{eqnarray}\label{gh030}
\Gamma(\omega)=&\displaystyle
-\frac{1}{\bb^{-1}\CLl_0\SLl_0+\omega}\CLl_0\ ,&\q(\omega\neq0)\\
\label{gh040}
\Gamma(0)=&\displaystyle -\bb\SLl_0^{-1}\q.
\end{eqnarray}
The expressions (\ref{gh020}) and (\ref{gh040}) for the
limiting value $\Gamma(0)$ are consistent with the value (\ref{ad030})
of {the} metric tensor $-\SLl_0^{-1}$
for $\RI_{0\mu}=\RI_{0\lambda}$.

Bubble diagrams iterate a vertex associated with
the two-body interaction $V$
by means of the propagator $\Gamma(\omega)$.
Regarded as a matrix in the $2n\times 2n$ Liouville space,
the interaction $V$ can be identified with
the matrix $\ELl_0$ (see Eq.~(\ref{gd040}))
of the second derivatives of $\EI\{\RI\}$,
as seen from the expression (\ref{f310}).
Finally, the initial and final vertices associated with the
operators $\QDh_\ggg$ and $\QDh_\delta$ (the correlation of which
we want to calculate) give rise to factors $\QI_\ggg$ and $\QI_\delta$,
regarded here as vectors of the Liouville space.
Altogether, the contribution of the $n$-th order bubble
diagram to $\Cgd$, evaluated with $\DIh_0\bbb$ as the unperturbed
density matrix, equals
\be\label{gh050}
C_{\gamma\delta}^{(n)}=\QI_\ggg :\bb^{-1}
\sum_\omega\left[
-\Gamma(\omega)\ELl_0
\right]^n\Gamma(\omega) : \QI_\delta\q.
\ee
The summation over $\omega=2\pi\,{\rm i}\,\bb^{-1}\,m$
accounts for the fact that
$\QI_\ggg$ and $\QI_\delta$ are both taken at the ``time''
$u=0$.
The ordering of $\QDh_\ggg$ and $\QDh_\delta$
is relevant in the evaluation of the zeroth-order term,
where the summand behaves as $-\CLl_0/\omega$
for $\omega\rightarrow\pm\infty$ ;
the average over both orderings as in (\ref{I080}) is achieved
by taking a principal part in the summation over $\omega$.

The contribution of all bubble diagrams to
the correlation $\Cgd$
is thus
\be\label{gh060}
C_{\ggg\delta}^{({\rm b})}=\QI_\ggg :
\frac{1}{\bb}\sum_\omega
\frac{1}{1+\Gamma(\omega)\ELl_0}\Gamma(\omega) :\QI_\delta\q.
\ee
The sum is meant as a principal part for
$\omega\rightarrow\pm\infty$.
By using the expression (\ref{gh030}) of
$\Gamma(\omega)$ we find for $\omega\neq0$ :
\be\label{gh070}
\frac{1}{1+\Gamma(\omega)\ELl_0}\Gamma(\omega)=
-\frac{1}{
\bb^{-1}\CLl_0\SLl_0+\omega-\CLl_0\ELl_0
}\CLl_0
=-\frac{1}{\omega+\KLl_0}\CLl_0\ ,
\ee
where we utilized the relations (\ref{gd060}) and (\ref{gg120})
which introduces $\KLl_0$.
The summation over $\omega=2\pi\,{\rm i}\,\bb^{-1}\,m$
is then performed by means of
\be\label{gh080}
\frac{1}{\bb}{\sum_\omega}'\frac{1}{\omega+z}=
\usd\coth\usd\,\bb\,z -\frac{1}{\bb\,z}\q,
\ee
where $\sum'_\omega$ is meant as a principal part
for $\omega\rightarrow\pm\infty$,
the value $\omega=0$ being excluded.
The expression (\ref{gh080}) approaches a finite limit for $z\rightarrow0$,
which we take as the value for $z=0$.
Using now the expression (\ref{gh040}) for $\Gamma(0)$, we obtain
\be\label{gh090}
C_{\ggg\delta}^{({\rm b})}=
\QI_\ggg :\left([
-{\textstyle\usd}\coth{\textstyle\usd}\bb\KLl_0 +(\bb\KLl_0)^{-1}
]\CLl_0+\left[\bb\FLl_0\right]^{-1}\right) :\QI_\delta\q.
\ee
The first term is defined even if $\KLl_0$ has
vanishing eigenvalues, since
these give no contribution to the bracket.
The second term may diverge for vanishing eigenvalues of $\FLl_0$,
in agreement with the discussion of Sect.~4.2.3. Finally,
using again Eq.~(\ref{gg120}), we find
\be\label{gh100}
C_{\ggg\delta}^{({\rm b})}=
\QI_\ggg :
({\textstyle\usd}
\coth{\textstyle\usd}\bb\KLl_0)\KLl_0\FLl_0^{-1} :\QI_\delta\q,
\ee
which is the same expression as our variational result (\ref{gg260}).

Likewise, if we wish to evaluate the contribution of the bubble diagrams
to the Kubo correlation (\ref{gf010})
, we have to associate with the
final operator $\QDh_\ggg$ of the chain of
bubbles a ``time'' $u$ running from
$0$ to $\bb$, instead of fixing this time at $0$.
This amounts to keeping only the term $\omega=0$
in the sum (\ref{gh050})
associated with the $n$-th-order diagram as well as in the sum
(\ref{gh060}) for the overall set of diagrams.
The result reduces therefore
to the last term of (\ref{gh090}), which is identical to
our variational approximation (\ref{gf020}).

Thus, both for the {standard} and
the Kubo correlations of single-quasi-particle
observables,
{\it our variational approach leads to the same result
as the summation of bubble diagrams}
in a perturbation theory where the HFB state is taken as the
unperturbed state.
The occurence of the RPA kernel is
consistent with the summation of
this set of bubble diagrams,
and {the} outcome (\ref{gg260})
shows how {\it the RPA appears naturally in a variational
frame} suited to the evaluation of correlations.
{The results
of Sect.~4.1 show also}
that the bubble diagrams, in our variational treatment, contribute
neither to the thermodynamic quantities nor to the
expectation values $\langle\QDh_\ggg\rangle$,
which are both given by the standard HFB mean-field answers.
(Diagrammatically this corresponds to the summation of
the instantaneous part of the self-energy insertions.)

Another variational principle, based
on the maximization of the entropy
rather than on the Bloch equation (\ref{e030}) for the
characterization of the state, has been
previously used \cite{BVe92} to evaluate
the correlations $\Cgd$. While the formula
obtained for the Kubo correlations was the same as (\ref{gf020}),
the result for ordinary correlations differed
from (\ref{gg260}). Its perturbative interpretation also involved
the set of bubble diagrams. However, starting from second order,
only the $\omega=0$ contribution to each diagram
was included (see Sect.4.1 of \cite{BVe92}).
In contrast, our result (\ref{gg260}) includes the summation
over all values $\omega=2\pi\, {\rm i}\,\bb^{-1}m$.
This appears as an advantage of the present
variational principle based on Bloch's equation.
\newpage

\section[S5]{Projected Extension of the Thermal
HFB Approximation}

\setcounter{equation}{0}

\subsection[S51]{Generalities and Notations}

In Sections~3 and 4 we have dealt with equilibriums in a
grand-canonical ensemble. However, as discussed
in the Introduction, many problems involve data
of type I that prohibit fluctuations
for the associated conserved operators.
For instance, in the case of the canonical ensemble, the state is required 
to have a well-defined particle number
$N_0$. Trial states of the type (\ref{f040}) that were
used in Sects.~3 and 4, namely
exponentials $\TDh$ of single-quasi-particle operators,
violate this requirement. We want, nonetheless, to keep the 
convenience associated with the operators $\TDh$ and the Fock space.
To resolve this dilemma, in conjunction with the variational 
functional (\ref{e040})
we shall use density operators $\DIh$
involving projections, such as in Eq.~(\ref{I110}).
For instance, a natural choice for a trial state with a {\it well-defined
particle number} is
\be\label{ha010}
\DIh=\PDh_\Nz\,\TDhd\,\PDh_\Nz\q,
\ee
where $\PDh_{N_0}$ is the projection
\be\label{ha020}
\PDh_\Nz=\intfi\,
{\rm e}^{-{\rm i}\theta\Nz}\,{\rm e}^{{\rm i}\theta\NDh}\q,
\ee
onto the subspace with the particle number $N_0$.
As in the preceding Sections, the 
variational parameters associated with the choice ({\ref{ha010})
are contained in the operator
$\TDhd$ which belongs to the class (\ref{f040}).

We shall consider successively the two types of situations
that were analyzed in the Introduction.
(i) The first type (Sects.~5.2 and 5.3) 
corresponds to the case where 
the choice (\ref{ha010}) involves an operator 
$\TDhd$ which {\it breaks a symmetry} of the Hamiltonian.
An example is that of a system in canonical equilibrium when 
pairing sources are added.
Then, the operator $\TDhd$ does not commute with
$\PDh_\Nz$, although the exact state $\DDh={\rm e}^{-\bb\KDh}$ does.
In this case, projections are necessary on both
sides of $\TDhd$. 
(ii) In the second type (Sect.~5.4),  
no symmetry is broken by the operator
$\TDhd$. This situation is
illustrated by the thermal Hartree-Fock approximation.
Then, the operator $\TDhd$ does not violate
the $N$-invariance and $\DIh$ can be defined 
with only one projection on either side of $\TDhd$;
this projection is needed 
because $\TDhd$ {\it acts in the
unrestricted Fock space}.

Whatever the case, projected trial states
are not easy to handle in the form (\ref{ha010}).
In particular, they
do not satisfy directly Wick's theorem which has been
an important tool in Sects.~3 and 4. In
order to retain the simple
properties of the exponential operators $\TDh$,
we wish to express each projection as
a sum of such exponentials.
This is feasible when the projection is associated
with some underlying group of operators $\TDhgu$
having single-quasi-particle operators as generators.
In this case the projection
can be expressed as a weighted sum of the
operators $\TDhgu$ over the elements $g$
of this group.
For instance, the projection $\PDh_\Nz$ in (\ref{ha020}) is built from
the group elements
$\TDhgu\equiv{\rm e}^{{\rm i}\theta\NDh}$
$(g\equiv\theta)$
{which are indexed} by the parameter
$0\leq\theta < 2\pi$.
As a consequence, {\it our projected trial states
will appear as sums of simple exponential operators}
of the $\TDh$ type, each of which is easy to handle. 

Likewise, if we {only} wish to eliminate
the {\it odd (or even) particle-number} components,
we should introduce a projection on even ($\eta=1$) or odd
($\eta=-1$) numbers, {which} can be expressed as the sum
\be\label{ha030}
\PDh_\eta={\textstyle \usd}\pag
1+\eta\,{\rm e}^{{\rm i}\pi\NDh}\pad\ ,\q\q\eta=\pm 1
\ee
over the two elements of the group ${\rm e}^{{\rm i}\theta\NDh}$
with $\theta=0$ and  $\theta=\pi$.

The projection on a given value $M$ of the
$z$-component of the {\it angular momentum},
\be\label{ha040}
\PDh_M=\intfi
{\rm e}^{-{\rm i}\theta M}\,{\rm e}^{{\rm i}\theta\JDh_z}\q,
\ee
is similar to (\ref{ha020}).
To project in addition on a given value of
the total angular momentum $\hat J$, we should introduce
the rotation operators $\TDhgu=\RDh(\Omega)$ in the Fock space ;
here the group index $g$ denotes the three Euler angles $\Omega$,
and $\RDh(\Omega)$ has again the form of an
exponential of single-particle operators. Integrating over
the Euler angles with the rotation matrices $D^J_{MK}(\Omega)$
yields the operator
\be\label{ha050}
\PDh^J_{MK}=\frac{2J+1}{8\pi^2}\int\,{\rm d}\Omega\,
D^{J\,*}_{MK}(\Omega)\,\RDh(\Omega)=
\sum_\gamma\vert \gamma\,J\,M\rangle\langle\gamma\,J\,K\vert\q,
\ee
where $\vert\gamma\,J\,M\rangle$ denotes a basis
in the Fock space labelled by $J$, $M$ and another set of quantum numbers
$\gamma$. When $K=M$, Eq.~(\ref{ha050})
defines the projection
on given values of $J$ and $M$.

The projections $\PDh_\pm$
over the spaces with a given {\it spatial parity} can be expressed as
\be\label{ha060}
\PDh_\pm={\textstyle\usd}\,\pag
1\pm{\rm e}^{{\rm i}\pi \NDh_-}
\pad\q,
\ee
where the single-particle operator
\be\label{ha070}
\NDh_-\equiv{\textstyle\usq}\,\sum_\sigma\int\,{\rm d}^3{\bf r}\,
\crg\psi^\dagger_\sigma({\bf r})-\psi^\dagger_\sigma(-{\bf r})\crd
\crg\vphantom{\psi^\dagger_\sigma}
\psi_\sigma({\bf r})-\psi_\sigma(-{\bf r})\crd\q,
\ee
counts the number of particles lying in the single-particle states
which are odd in the exchange 
${\bf r},\sigma\rightarrow -{\bf r},\sigma$
(the factor $\usq$ in Eq.~(\ref{ha070}) accounts {both for} the
normalization of the basic
antisymmetric states
$
[\psi^\dagger_\sigma({\bf r})-\psi^\dagger_\sigma(-{\bf r})]
\vert 0\rangle/\sqrt{2}$
and for the double-counting
in the integration over $\bf r$).

In order to encompass formally all these
situations, we shall denote by
\be\label{ha080}
\PDh=\ssf\,\TDhgu\q,
\ee
the decomposition of our relevant projection as a
linear combination of the group operators  $\TDhgu$. 
The symbol
$\ssf\ldots$ denotes a weighted sum or integral
over the elements of the group. In particular,
for (\ref{ha020}), it stands for
$(2\pi)^{-1}\int_0^{2\pi}{\rm d}\theta
{\rm e}^{-{\rm i}\theta\Nz}\ldots$ while
for (\ref{ha030}), it stands for $\usd\sum_{j=0,1}\eta^j$. The operators
$\TDhgu$ have the form (\ref{f040}) of an exponential of a
single-quasi-particle
operator, which we shall denote as
$-\li^\gup-\usd\gbt\LI^\gup\gbd$.
The coefficients $\li^\gup$ and $\LI^\gup$ depend on the
group index $g$ ; for instance, in (\ref{ha020}),
we have 
$\li^{[\theta]}=-{\rm i}\theta n/2$ and 
$\LI^{[\theta]}=-{\rm i}\theta\NI$ ($0\le\theta<2\pi$),
and in (\ref{ha030}),
$\li^{[j]}=-{\rm i}\,j\,\pi n/2$ and 
$\LI^{[j]}=-{\rm i}\,j\,\pi\NI$ ($j=0,1$),
where $\NI$ is the $2n\times 2n$ matrix defined by (\ref{gb040}).

As in Sect.~3.1, we denote the $2n\times 2n$ matrix $\exp(-\LIgu)$
as $\TIgu$ and the scalar $\exp(-\ligu)$ as $\ttgu$.
We shall have to deal with products of two to four  $\TDh$-operators,
such as $\TDhgu\TDhd\TDhgv$ or $\TDhgu\TDhd\TDhgv\TDha$ ; from
the algebra properties (\ref{f230}-\ref{f240}),
these products {\it have still the same }$\TDh$-{\it form}
and they are characterized by matrices such as
\be\label{ha090}
\TIf\equiv\TIgu\TId\TIgv\q,
\ee
and scalars $\ttf=\ttgu\ttd\ttgv$.
There is a contraction matrix $\RIf$ associated
with the matrix $\TIf$ through (\ref{f080});
$\RIf$, as $\TIf$,
depends on the ordering of the
original operators.

The equations that we shall derive in the next
Subsections can also handle
cases which are more general than pure
projections. For instance, in the description
of rotational bands of 
axial nuclei, it is often a good approximation
to assume that the quantum number $K$
associated with the component of the angular momentum
on an internal axis of inertia has a well-defined value $K_0$.
Then rotational states with given $J$ and $M$ values
can be sought with the trial form
\be\label{ha091}
\DIh=\PDh^J_{M\,K_0}\TDhd\PDh^J_{K_0\,M}\q,
\ee
where the operators $\PDh^J_{MK}$ defined by Eq.~(\ref{ha050}) 
are not projections when $M\neq K_0$. Thermal properties of triaxial 
nuclei which involve sums over the $K$ quantum number,
can also be handled by our method.
Another example is the case where a symmetry
(such as $\NDh$) is broken. 
Suppose that we still want to work 
in the grand canonical ensemble but also want to suppress the
off-diagonal contributions between subspaces with different 
values of $N$.
We are led to consider operators of the form 
\be\label{ha100}\ba{rl}\displaystyle
\sum_\NO\, \PDh_\NO\DDh\,\PDh_\NO&\displaystyle=
\frac{1}{4\pi^2}\int_0^{2\pi}\,{\rm d}\phi\,
\int_0^{2\pi}\,{\rm d}\phi'\,\sum_\NO\,
{\rm e}^{i(\phi-\phi')\NO}\,
{\rm e}^{-i\phi\NDh}\DDh\,{\rm e}^{i\phi'\NDh}\ ,\\
&\displaystyle
=\frac{1}{2\pi}\int_0^{2\pi}\,{\rm d}\phi\,{\rm e}^{-i\phi\NDh}
\DDh\,{\rm e}^{i\phi\NDh}\q,
\ea\ee
which are also amenable to our treatment.

\subsection[S52]{The Variational Ans\"atze and the Associated 
Functional}

Now we must choose the trial classes for the operators $\DIh\uu$
and $\AIh\uu$ to be
inserted into the functional (\ref{e040}) which, at its stationary point,
supplies the characteristic function of interest. For the operator
$\DIh\uu$ we already made the choice
\be\label{hb010}
\DIh\uu=\PDh\,\TDhd\uu\PDh=\sff\TDhgu\TDhd\uu\TDhgv
\equiv\sff\TDhf\uu\q,
\ee
where $\TDhd\uu$ is defined by Eq.~(\ref{f160}).
For the operator $\AIh\uu$ we keep the same choice,
\be\label{hb020}
\AIh\uu=\TDha\uu\q,
\ee
as in Eq.~(\ref{f170}). Again, as in Sect.~3.2, the quantities
$\lid\uu$, $\LId\uu$, $\lia\uu$, $\LIa\uu$ form a set
of independent variational parameters. The exact
density operator
$\DDh$ satisfies $\PDh\DDh=\DDh\PDh=\DDh$, and when
defined in the full Fock space the
operator $\KDh$ commutes with $\PDh$. 
Therefore, it makes no difference if we project $\DIh\uu$ or $\AIh\uu$
in the functional
(\ref{e040}); the choice $\DIh\uu=\TDhd\uu$, $\AIh\uu=\PDh\TDha\uu\PDh$
is equivalent to (\ref{hb010}-\ref{hb020}).

The results of Sect.~3.2 are readily adapted
through a mere replacement of $\TDhd$ by $\TDhf$ and a
summation over the group indices $g$ and $g'$.
Thus the normalization 
$\YB$, given by
\be\label{hb030}
\YB=\Tr{\AIh\uu\DIh\uu}=\sff\YIgg\uu\q
\ee
with
\be\label{hb040}
\log\YIgg\uu=-[\lia\uu+\lid\uu+\ligu+\ligv]
+{\textstyle\usd}\,\trd{\log[1+\TIa\uu\TIgu\TId\uu\TIgv]}\ ,
\ee
{replaces $\YI$ as defined by (\ref{f250}).
We recall that, according to Eq.~(\ref{e170}),  
$\YB$ does not depend on $u$
when $\AIh\uu$ and $\DIh\uu$ are stationary
solutions.} 

The action-like functional (\ref{f320}), is now replaced by
\be\label{hb050}\ba{l}\displaystyle 
\IB=\YB^\AP-\intzb\sff\YIgg\,
\left(-\duh{\lid\uu}
\right.\\
\q\q\q\left.\displaystyle
+\usd\trd{\TIdm\,\RIcdp\,\duh{\TId} }
\vphantom{\duh{\lid\uu}}
+\usd[\EI\{\RIcdp\}+\EI\{\RIbup\}]\,\right)\q.
\ea\ee
The boundary term $\YB^\AP=\Tr{\ADh\DIh\bbb}$ is obtained
from Eqs.~(\ref{hb030}-\ref{hb040})
by letting $u=\bb$ in $\lid\uu$
and $\TId\uu$ and replacing $\lia\uu$ and
$\LIa\uu$ by the given quantities $\li^\AP$ and $\LI^\AP$
(see Eq.~(\ref{f140})).
We have used the analogue of the property (\ref{f330}) to transform
${\TIgu}^{-1}\RIcup\TIgu$ into $\RIcdp$. Moreover, from the
commutation $[\KDh\,,\,\TDhgu]=0$ and in agreement with
(\ref{ge020}), we find that for any $\RI$ 
the energy (\ref{f310}) satisfies
\be\label{hb060}
\EI\{\TIgu\RI{\TIgu}^{-1}\}=\EI\{\RI\}\q,
\ee
a property that we have also used to write $\EI\{\RIcup\}$
instead of $\EI\{\RIcdp\}$ {(similarly,
one has $\EI\{\RIbup\}=\EI\{\RIbdp\}$)}.

The expression (\ref{hb050}) exhibits the fact that we might
as well have projected $\AIh\uu$ instead of $\DIh\uu$. Indeed,
taking into account the property (\ref{hb060}), this expression does not
change under the replacement of $gdg'$ by $d$ and of $a$ by $g'ag$.
Actually, we can readily check
that if one projects
$\AIh$ instead of $\DIh$, Eq.~(\ref{f320}) is directly
transposed into (\ref{hb050}) through the replacement of $a$
by $g'ag$.

Likewise, by an obvious extension of (\ref{f350}) where
$d$ is replaced by $gdg'$, we find for the functional $\IB$
the alternative form:
\be\label{hb070}\ba{l}
\IB=\displaystyle \YB^\AP-\YB\bbb+\YB\zz +\intzb\sff\YIgg\,
\left(-\duh{\lia\uu}\right.\\
\q\q\q\displaystyle\left.
+\usd\trd{\TIam\,\RIbup\,\duh{\TIa} }
-\usd[\EI\{\RIcdp\}+\EI\{\RIbup\}]\,\right)\ .
\ea\ee

The expressions (\ref{hb050}) and (\ref{hb070})
generalize (\ref{f320}) and (\ref{f350}) through
the averaging over $g$ and $g'$ with the weight $\YIgg/\YB$.
{They lead us to introduce} the matrices
\be\label{hb080}
\RBcu\equiv\YB^{-1}\,\sff\,\YIgg\,\RIcdp \ ,\q\q
\RBbu\equiv\YB^{-1}\,\sff\YIgg\,\RIbup \q,
\ee
and the pseudo-energies
\be\label{hb090}
\EBc\equiv\YB^{-1}\,\sff\YIgg\,\EI\{\RIcdp\}
\ ,\q\q
\EBb\equiv\YB^{-1}\,\sff\YIgg\,\EI\{\RIbup\}\q,
\ee
(remember that $\EI\{\RIcup\}=\EI\{\RIcdp\}$
and that $\EI\{\RIbup\}=\EI\{\RIbdp\}$).
With these notations
we can rewrite (\ref{hb050}) and
(\ref{hb070}) in a more condensed fashion,
formally closer to the functionals (\ref{f320}) and (\ref{f350}) :
\be\label{hb100}\ba{l}
\IB=\displaystyle \YB^\AP
-\intzb\,\YB\,\left(-\duh{\lid}\pod\\
\q\q\q\displaystyle \pog +\usd\trd{\TIdm\,\RBcu\,\duh{\TId}}
+\usd[\EBc+\EBb]\right)
\q,\ea\]
\[\ba{l}\displaystyle
\IB=\YB^\AP -\YB\bbb+\YB\zz+\intzb\,\YB\,\left(-\duh{\lia}\pod\\
\q\q\q\displaystyle \pog
+\usd\trd{\TIam\,\RBbu\,\duh{\TIa}}
-\usd[\EBc+\EBb] \right)
\q,\ea\ee
where $\RBcu$, $\RBbu$, $\EBc$ and $\EBb$ replace
$\RIc$, $\RIb$, $\EI\{\RIc\}$ and $\EI\{\RIb\}$
while $\YB$ replaces $\YI$.
An important change between the present
functional and the one of Sect.~3.2
{lies in that}
$\EBc$ differs {in general} from $\EI\{\RBcu\}$
{(as well as from $\EI\{\RBcd\}$)
when the operator $\KDh$ contains a two-body interaction}.
It implies that the variational outcome $\YB^\AP$
yielded by the functional (\ref{hb100})
will differ from that which would be obtained
by projecting the optimum state found in Sect.~3~;
this holds even for the case $(\ADh=\uni)$ of
thermodynamic potentials.

\subsection[S53]{The Coupled Equations}

As in Sect.~3.3 we find the equations governing the evolution
of $\lid\uu$ and $\TId\uu$ by requiring that the functional (\ref{hb050})
be stationary with respect to $\lia\uu$ and $\TIa\uu$.
The resulting differential equation for $\lid\uu$,
\be\label{hc010}
\duh{\lid}=\usd\trd{\TIdm\,\RBcu\,\duh{\TId}}
+\usd[\EBc+\EBb] \q,
\ee
is similar to Eq.~(\ref{f360}),
with $\RBcu$, $\EBc$ and $\EBb$ replacing 
$\RIc$, $\EI\{\RIc\}$ and $\EI\{\RIb\}$.
{The differential equation for $\TId\uu$
differs from} Eq.~(\ref{f370})
{by an explicit summation over the group indexes. It reads :}
\be\label{hc020}\ba{l}
\displaystyle\sff\YIgg\TIgu(1-\RIcdp)\acg
\duh{\TId}\pod\\
\displaystyle\pog\vphantom{\duh{\TId}}
\q\q\q\q+\usd[\HIscd\TId+\TId\HIsbd]
\acd(1-\RIbdp)\TIgv=0\q,
\ea\ee
{with}
\be\label{hc030}
\HI^{[i]}\equiv
\HI\{\RI^{[i]}\}+ 
\frac{\ksdgg}{1-\RI^{[i]}}\q,
\ee
{where $i$ stands for $dg'ag$ or $g'agd$ and where
the c-number $\ksdgg$ is defined by}
\be\label{hc040}\ba{l}
\displaystyle
\ksdgg\equiv\usd\acg\trd{[\TIdm(\RIcdp-\RBcu)\duh{\TId}}]\pod\\
\displaystyle\pog\vphantom{\duh{\TId}}\q\q\q\q\q\q\q\q
+\EI\{\RIcdp\}-\EBc+\EI\{\RIbdp\}-\EBb
\acd\q.
\ea\ee
{To derive the equation (\ref{hc020})
we} made use of (\ref{f510}), of (\ref{hc010}) and of the fact
that $\HI\{\RI\}$ satisfies for any $\RI$ the relation
\be\label{hc050}
\HI\{\exup\RI\exum\}=\exup\HI\{\RI\}\exum
\q,\ee
a consequence of the definition
(\ref{f380}) and of (\ref{hb060}).
Note that Eq.~(\ref{hc010}) does not imply the vanishing of the
integrand of (\ref{hb050}) before summation over $g$ and $g'$.

Likewise the differential equation for $\lia\uu$
is obtained by varying the functional (\ref{hb070})
with respect to $\lid\uu$ :
\be\label{hc060}
\duh{\lia}=\usd\trd{\TIam\,\RBbu\,\duh{\TIa}}
-\usd[\EBb+\EBc] \q,
\ee
while the equation for $\TIa\uu$ is provided by
varying (\ref{hb070}) with respect to $\TId\uu$
for $0\leq u<\bb$ :
\be\label{hc070}\ba{l}
\displaystyle\sff\YIgg\exdp(1-\RIbup)\acg
\duh{\TIa}\pod\\
\displaystyle\pog\vphantom{\duh{\TIa}}
\q\q\q\q-\usd[\HIsbu\TIa+\TIa\HIscu]\acd(1-\RIcup)\exup
=0\q,
\ea\ee
{where $\HIsbu$ and $\HIscu$ are defined 
according to Eq.~(\ref{hc030}) with $i$
standing for $agdg'$ or $gdg'a$
and with the c-number $\ksdgg$ replaced by $\ksagg$
given by}
\be\label{hc080}\ba{l}
\displaystyle
\ksagg\equiv\usd\acg -\trd{[\TIam(\RIbup-\RBbu)\duh{\TIa}}]\pod\\
\displaystyle\pog\vphantom{\duh{\TId}}\q\q\q\q\q\q\q\q
+\EI\{\RIbup\}-\EBb+\EI\{\RIcup\}-\EBc
\acd\q.
\ea\ee
Note that $\lia+\lid$ disappears from
Eqs.~(\ref{hc020}) and (\ref{hc070})
if we divide them by $\YB$.
These equations thus couple
only the matrices $\TId$ and $\TIa$. Once they are solved, $\lid$
and $\lia$ are obtained by integration of Eqs.~(\ref{hc010})
and (\ref{hc060}). 
In Appendix B, using the Liouville
formulation and the notation (\ref{hb080}-\ref{hb090}),
Eqs.~(\ref{hc020}) and (\ref{hc070})
are rewritten in the forms (\ref{ba100}-\ref{ba110})
which look similar to Eqs.~(\ref{f370}) and (\ref{f430}).

Finally, in agreement
with (\ref{e180}), the stationarity of (\ref{hb070}) with respect to
$\lid\bbb$ and $\TId\bbb$ provides the equations
\be\label{hc090}
\YB\bbb=\YB^\AP\ ,\q\q\RIcdp\bbb=\RI^{[dg'Ag]}\bbb\q,
\ee
which are satisfied by the same boundary conditions as in Sect.~3 :
\be\label{hc100}
\lia\bbb=\li^\AP\ ,\q\q\TIa\bbb=\TI^\AP\q.
\ee
The boundary conditions at $u=0$ are
$\lid\zz=0$, $\TId\zz=1$ as in Sect.~3.
Again, all our variational quantities $(\lia\uu$, $\lid\uu$, $\TIa\uu$,
$\TId\uu)$ depend upon the sources $\xi_\ggg$ through the
boundary conditions (\ref{hc100}).

Despite their somewhat complex form, the coupled equations
(\ref{hc020}) and (\ref{hc070}) entail some simple consequences.
{First, using  (\ref{hc010}) and (\ref{hc060}),
one notes} that $\YIgg$ varies with $u$ according to
\be\label{hc110}\ba{rl}\displaystyle
\duh{\ }\log\YIgg&=\ksdgg-\ksagg\\
&\displaystyle
=-\usd\duh{\ }
\trd{\log(1-\RIcdp)}-{\bar c}\uu\q,
\ea\ee
{where the function ${\bar c}\uu$ is given by the average}
\be\label{hc120}
{\bar c}\uu=
-\frac{1}{2\,\YB}\,\sff\,\YIgg\,\duh{\ }
\trd{\log(1-\RIcdp)}
\q.
\ee
Summation over
$g$ and $g'$ of Eq.(\ref{hc110}) leads to
\be\label{hc130}
\duh{\YB}=0\q,
\ee
which was expected from (\ref{e170}).
Thus {\it the ultimate quantity of interest},
{\it the characteristic function} $\varphi$
{\it provided by the stationary value of} $\,\log\II$, 
{\it is equal to}
$\log\YB\uu$ {\it for any} $u$. One can also check from (\ref{hc020}),
(\ref{hc070}) and (\ref{hc110}) that
\be\label{hc140}
\duh{\ }(\EBc+\EBb)=0\q,
\ee
which is a special case of the general relation
(\ref{e150}). The relations (\ref{hc130}) and (\ref{hc140})
generalize (\ref{f480}).

When the operator $\ADh$ is hermitian, the coupled
equations (\ref{hc020}) and (\ref{hc070})
admit solutions with hermitian matrices $\TId\uu$ and $\TIa\uu$,
and the resulting c-numbers $\lid\uu$ and $\lia\uu$
are real. This is seen by grouping
in all the equations
the terms corresponding to $\TDhgu$, $\TDhgv$ respectively with
the associated terms corresponding to $\TDh^{\gvp -1}$,
$\TDh^{\gup -1}$,
as is shown at the end of Appendix B.

\subsection[S54]{Unbroken $\PDh$-Invariance}

The above approach is general and covers situations
in which the trial operators $\TDhd\uu$ and $\TDha\uu$
do not commute with the projection $\PDh$. It applies for instance
to finite systems of fermions in {\it canonical}
equilibrium in which pairing correlations are operative.
In this case the operator
$\KDh$ denotes the Hamiltonian $\HDh$ alone without the term
$-\mu\NDh$, the projection
$\PDh$ is given by (\ref{ha020}) and
the independent quasi-particle trial state $\TDhd\uu$ as well as the
trial operator $\TDha\uu$ do not commute with $\NDh$, nor with the
operations $\TDh^{[\theta]}={\rm e}^{{\rm i}\theta\NDh}$ 
of the associated group. 

\subsubsection[S541]{Notation and General Formalism}

As discussed in Sect.~5.1 and in the Introduction, 
there are other situations
which require projection, although 
the $\PDh$-invariance is not broken by the
variational approximation. 
In these cases $\TDhd\uu$
and $\TDha\uu$
commute with all the group elements $\TDhgu$ and therefore with $\PDh$.
Hence {\it only one projection} is needed on either side of $\TDhd\uu$ and,
instead of (\ref{hb010}), the trial state can be written as
\be\label{hd010}
\DIh\uu=\PDh\,\TDhd\uu\,\PDh=\TDhd\uu\,\PDh
=\PDh\,\TDhd\uu=\ssf\TDhgu\,\TDhd\uu\q.
\ee
A single summation over the group
index $g$ is sufficient in Eq.~(\ref{hb030}) :
\be\label{hd020}
\YB=\Tr{\AIh\uu\DIh\uu}=\ssf\YIg\uu\q,
\ee
while Eq.~(\ref{hb040})) becomes
\be\label{hd030}
\log\YIg\uu=-[\lia\uu+\lid\uu+\ligu]+
{\textstyle\usd}\trd{\log[
1+\TIa\uu\exup\TId\uu
]}\q.
\ee
Likewise, the double averaging 
in (\ref{hb080}) or (\ref{hb090})
is replaced by a single one over $g$
with the weight $\YIg/\YB$ ; for instance, we have~:
\be\label{hd040}
\RBc\equiv\YB^{-1}\ssf\YIg\,\RIdag
\ ,\q\q
\RBb\equiv\YB^{-1}\ssf\YIg\,\RIadg
\q.
\ee

The expressions of the functional $\II$ and
of the coupled equations 
are readily derived from those obtained
in Sects.~5.2 and 5.3 by noting that
the matrix $\exup$ commutes with $\TId\uu$ and $\TIa\uu$
and that the factor $\exdp$ can everywhere be replaced
by $\TI^{[0]}=1$.
Since $g'$ disappears and
the ordering of $g$ with respect to $a$ and $d$
is irrelevant, there remain only two (instead of four)
different contraction matrices:
$\RIdag=\RIdga=\RIgda$ and $\RIadg=\RIagd=\RIgad$.
The functional (\ref{hb050}) then simplifies to
\be\label{hd050}\ba{l}
\IB=\YB^\AP-\displaystyle \intzb\ssf\YIg\,
\pag-\duh{\lid\uu}
\pod\\
\q\q\q\pog\displaystyle
+\usd\trd{\TIdm\,\RIdag\,\duh{\TId} }
+\usd[\EI\{\RIdag\}+\EI\{\RIadg\}]\,\pad\q.
\ea\ee
As before, the stationary value of $\log\IB$ (our
variational approximation for the characteristic function)
is equal to $\log\YB$ which does not depend on $u$.

Accordingly, the coupled equations of Sect.~5.3
take a simpler form. Note, however, 
that because in general $k_d^{0g}$ 
(Eq.~(\ref{hc040})) differs from $k_a^{g0}$ (Eq.~(\ref{hc080})),
the matrix $\HI^{[gd0a]}$ is different from 
$\HI^{[d0ag]}$ and $\HI^{[agd0]}$ different from $\HI^{[0agd]}$ 
(Eq.~(\ref{hc030})).

By combining the forms taken by Eqs. (\ref{hc020}), 
(\ref{hc070}) and
(\ref{hc110}) in this commutative case and by using (\ref{f510}), we find
that the averaged contraction
matrices (\ref{hd040}) satisfys
\be\label{hd060}\ba{rl}
\displaystyle \duh{\ }\RBb=&
{\displaystyle\,\ \ \frac{1}{2\YB}\,\ssf}\YIg\,
[\HI\{\RIadg\}\,,\,\RIadg]
\q,
\\
\displaystyle \duh{\ }\RBc=&
{\displaystyle\, -\frac{1}{2\YB}\,\ssf}\YIg\,
[\HI\{\RIdag\}\,,\,\RIdag]
\q.
\ea\ee
This was expected from the general equations (\ref{e290}) and (\ref{e300}),
since their validity conditions are satisfied when
the operators $\TDhd\uu$ and $\TDha\uu$ commute with $\PDh$.
While Eqs.~(\ref{f520}), which are appropriate to the unprojected
variational treatment, had the same form as the time-dependent
HFB equation (with an imaginary time), Eqs.~(\ref{hd060}) deal
with {the larger set of} matrices $\RIdag$ and $\RIadg$ and involve a
summation over the group index $g$.
The apparent decoupling of Eqs.~(\ref{hd060}) is again fictitious
because the boundary conditions cannot be
expressed explicitly in terms of $\RBb$ and $\RBc$.

\subsubsection[S542]{Partition Function, Entropy}

In the rest of this Subsection, still
assuming that the $\TDhgu$-invariance is not broken,
we focus on the case $\ADh=\uni$ for which our variational
principle provides an approximation for the
exact partition function
\be\label{hd070}
\YI_P=\Tr{\PDh{\rm e}^{\displaystyle -\bb\KDh}}\q.
\ee
We have shown in
Sect.~2.3 that by shifting appropriately
the operator $\KDh$, the variational principle (\ref{e040})
used with $\ADh=\uni$ also yields
{the characteristic function and therefore}
the cumulants of the
one-body operators which commute {\it both} with $\KDh$ and
with the projection $\PDh$.

As seen in Sect.~2.4,
the boundary condition $\AIh\zz=\uni$ allows us to restrict
the trial classes (\ref{hb010}) and (\ref{hb020}) to
\be\label{hd080}
\DIh_0\uu=\PDh{\rm e}^{\displaystyle -u\HDh_0}\ ,\q\q
\AIh_0\uu={\rm e}^{\displaystyle -(\bb-u)\HDh_0}\q;
\ee
the subscript 0 refers to the zero values of the sources $\xi_\ggg$.
The trial operator 
$\HDh_0$ is a ($u$-independent) single-quasi-particle operator
commuting with $\PDh$,
\be\label{hd090}
\HDh_0= h_0+{\textstyle\usd}\gbt\,\HI_0\,\gbd\q,
\ee
{which depends on the parameters}
$h_0$, a real number, and  $\HI_0$, a
$2n\times 2n$ hermitian matrix obeying (\ref{f410}) and commuting
with $\TIgu$.
The choice (\ref{hd080}) entails
\be\label{hd100}
\DIh_0\uu\AIh_0\uu=\PDh{\rm e}^{\displaystyle -\bb\HDh_0}
=\AIh_0\uu\DIh_0\uu=\DIh_0\bbb\q,
\ee
and the $u$-integration becomes trivial in the functional (\ref{e040}),
which reduces to
\be\label{hd110}\ba{rl}
\IB&=\Tr{\PDh{\rm e}^{-\bb\,\HDh_0}(1+\bb\,\HDh_0-\bb\KDh)}\\
&=\displaystyle\ssf\YI_0^g\crg
1+\bb h_0+{\textstyle\usd}\bb\trd{
\HI_0\RIzeg} -\bb\EI\{\RIzeg\}
\crd
\q,\ea\ee
with
\be\label{hd120}
\log\YI_0^g=-\bb h_0 -\ligu +{\textstyle\usd}\trd{\log[
1+{\rm e}^{-\bb\,\HI_0}\TIgu
]}\q,
\ee
where the trial quantities $h_0$ and $\HI_0$
must be determined variationally. 
In the generalized contraction matrices
\be\label{hd130}
\RIzeg\equiv\frac{\exup}{{\rm e}^{\bb\,\HI_0}+\exup}
=\frac{\RIze\exup}{1+\RIze(\exup-1)}
\q,
\ee
the matrices $\HI_0$, $\exup$ and $\RIze$ all commute;
for the unity element of the group $(\TDh^{[0]}=\uni\,;\,
\li^{[0]}=0\,,\,\TI^{[0]}=1$),
the matrix $\RI_0^0$ given by (\ref{hd130})
reduces to the contraction matrix $\RIze$ of
the usual finite-temperature HFB formalism (Eq.~(\ref{ga010})).
As in Eq.~(\ref{e400}), we recognize in the
simple expression (\ref{hd110}) the standard variational
principle associated with the maximization of the entropy 
compatible with the
constraint on $\langle\KDh\rangle$. 
If one introduces the quantities
\be\label{hd140}
\YB\equiv\ssf{\YI_0^g}\ ,\q\q
{\bar\RI}\equiv\frac{1}{\YB}\ssf\YI_0^g\,\RIzeg\ ,\q\q
\EIzb\equiv\frac{1}{\YB}\ssf\YI_0^g\,\EI\{\RIzeg\}\q,
\ee
the functional (\ref{hd110}--\ref{hd120}) takes the 
more compact form
\be\label{hd150}
\IB=\displaystyle\YB\crg
1+\bb h_0+{\textstyle\usd}\bb\trd{
\HI_0{\bar\RI}} -\bb\EIzb
\crd\q.
\ee

Within the selected variational space,
the best approximation for the exact partition function $\YI_P$
(Eq.~(\ref{hd070})) is supplied by the maximum of $\IB$ since the 
r.h.s. of (\ref{hd110}) is always smaller than $\YI_P$.
The maximization
with respect to the number $h_0$ yields
\be\label{hd160}
h_0=-{\textstyle\usd}\trd{\HI_0{\bar\RI}}+\EIzb\q.
\ee
{This equation ensures that, according to the discussion
of Sect.~2.4,
the maximum of} 
$\IB$ reduces to $\YB$.
Requiring {now} the functional 
$\IB$ to be maximum with respect to the matrix $\HI_0$
provides
\be\label{hd170}
\ssf\YI_0^g\,\RIzeg\,(\HI_0-\HI^g)(1-\RIzeg)=0\q,
\ee
with
\be\label{hd180}
\HI^g\equiv\HI\{\RIzeg\}+\frac{k^g}{1-\RIzeg}\q
\ee
where
\be\label{hd190}
k^g\equiv-\usd\,\trd{\HI_0(\RIzeg-{\bar\RI})}
+\EI\{\RIzeg\}-\EIzb\q.
\ee
{The equation (\ref{hd170}), together with the definitions
(\ref{hd180}) and (\ref{hd190}), }
determines the trial matrix $\HI_0$ to which
$\RIzeg$ is related through (\ref{hd130}).
The $\lambda$-basis where $\HI_0$ and $\RIzeg$ are diagonal
(with eigenvalues $\HI_{0\lambda}$
and $\RI_{0\,\lambda}^g$, respectively) is thus defined by the equations
\be\label{hd200}
\ssf\YI_0^g\,
\RI^g_{0\,\lambda}\HI\{\RIzeg\}_{\lambda\mu}(1-\RI^g_{0\,\mu})=0
\ ,\q\q\lambda\neq\mu
\ee
while the diagonal eigenvalues of the matrix $\HI_0$
are given by
\be\label{hd{210}}
\HI_{0\mu}=\sum_\lambda\,(M^{-1})_{\mu\lambda}\,
\crg\ssf\YI_0^g\,\acg
\RI^g_{0\,\lambda}\HI\{\RIzeg\}_{\lambda\lambda}(1-\RI^g_{0\,\lambda})
+\RI^g_{0\,\lambda}(\EI\{\RIzeg\}-\EIzb)
\acd\crd\q,
\ee
where $M$ is the symmetric matrix
\be\label{hd220}
M_{\lambda\mu}\equiv\ssf\YI_0^g\,\acg
\delta_{\lambda\mu}\RI^g_{0\,\lambda}(1-\RI^g_{0\,\lambda})
+{\textstyle\usd}\RI^g_{0\,\lambda}(\RI^g_{0\,\mu}-{\bar\RI}_\mu)
\acd\q.
\ee
Therefore the projection generates a
more complicated relation (Eqs. (\ref{hd170}) and (\ref{hd130}))
between $\HI_0$ and the $\RIzeg$'s
than the relation between $\HI_0$
and $\RIze$ (Eqs.~(\ref{ga010}) and (\ref{ga020})) given by the grand
canonical HFB formalism .

{
The equations (\ref{hd170}--\ref{hd190})
can also be obtained directly from the evolution equations
(\ref{hc020}) and (\ref{hc070}) when one takes into account
the simplifications induced by the choice (\ref{hd080}--\ref{hd090})
and by the commutation properties resulting from
Eqs.~(\ref{hd060}).
Indeed, $k^g$ is then the $u$-independent common value of 
the quantities $k_d^{0g}=k_a^{g0}=k^g$
introduced in (\ref{hc040}) and (\ref{hc080}),
while the matrices $\HI^{[i]}$ defined
in (\ref{hc030}) also become $u$-independent and all reduce to 
$\HI^{[agd0]}=\HI^{[0agd]}=\HI^{[d0ag]}=\HI^{[gd0a]}=\HI^g$.}

Several consistency properties can 
be derived from the formulation above.
When the functional (\ref{hd150}) is stationary with respect to the
number $h_0$, the approximate entropy $\SIzb$ 
associated  with the approximate
normalization $\YB$ and projected HFB energy
$\EIzb$ by the thermodynamic relation
\be\label{hd240}
\SIzb=\log\YB+\bb\EIzb\q,
\ee
(a specialisation of Eq.~(\ref{e262})
to the restricted variational spaces defined by 
Eqs.~(\ref{hd080}) and (\ref{hd090}))
is also related to the approximate density operator 
$\DIh_0\bbb=\PDh{\rm e}^{-\bb\HDh_0}$ by the von Neumann
relations~(\ref{I020}) and (\ref{e411}).
{Using Eqs.~(\ref{hd120}), (\ref{hd140}) and ({\ref{hd160}}),
one obtains the more explicit form}
\be\label{hd250}\ba{rl}
\SIzb&=-{\textstyle\usd}\trd{
[{\bar\RI}\,\log\RIze\,+\,(1-{\bar\RI})\log(1-\RIze)]
}\\
&\q\q\q\q\q\q\q
+\log\ssf{\rm e}^{-\ligu}[\det\{1+\RIze(\exup-1)\}]^{1/2}\q.
\ea\ee
{As expected, $\SIzb$ 
reduces to the HFB entropy $\SI\{\RIze\}$ (Eq.~(\ref{ga060}))
when the group of operators
$\exup$ is restricted to its unity element.
Furthermore, in accordance with Eq.~(\ref{e260}),
the stationarity of (\ref{hd110}) 
with respect to the matrix $\HI_0$
ensures that the equilibrium energy,
$-\partial\log\YB/\partial\bb$,
is equal to $\EIzb$.  
According to the discussion of Sect.~2.2, this guarantees that, 
in agreement with thermodynamics, the projected entropy $\SIzb$ 
also satisfies the relation} (\ref{e263})
where $\FIgb\equiv-T\,\log\YB$.

\noindent{\it Remark :}
Through the Ansatz (\ref{hd080}) we have
restricted the trial
class by choosing an exponential $u$-dependence
for the operators $\DIh\uu$ and $\AIh\uu$ and by relating their
exponents. Therefore, it
is not obvious that the maximum of $\IB$ as given by Eq.~(\ref{hd110})
is equal to the stationary value
of the functional (\ref{hd050})
which is defined within the more general class 
(\ref{hb010}-\ref{hb020}) of operators
whose $u$-dependences are not limited a priori.
In the light of Sect.~2.4, 
the commutation of $\PDh$ with $\TDha$ and $\TDhd$
is sufficient to ensure this equality. Indeed,}
if we associate a projection 
$\PDh$ to $\AIh\uu$ as well as to $\DIh\uu$,
the conditions i) and
ii) of Sect.~2.4 are both satisfied.

Nevertheless let us give a more explicit proof 
{along the line already followed in} Sect.~4.1 {for the
HFB case}.
We have to check that the Ansatz (\ref{hd080}) with $h_0$ and $\HI_0$
given respectively 
by (\ref{hd160}) and (\ref{hd170}) satisfies the set of
coupled equations (\ref{hc010}) and (\ref{hc020}), or
(\ref{hc060}) and (\ref{hc070})
which express the unrestricted stationarity.
We first recall that, for the choice (\ref{hd080}), 
the matrix $\RIdag\uu$ is $u$-independent
and given by (\ref{hd130}).
As $\lid\uu=uh_0$ and $\lia\uu=(\bb-u)h_0$, the functional
(\ref{hd050}) is identified with (\ref{hd110}) and
both equations (\ref{hc010}) and (\ref{hc060})
are identical with (\ref{hd160}).
The equation (\ref{hc020}) then becomes
\be\label{hd270}
\displaystyle\ssf\YI_0^g
\RIzeg\left(\HI_0-{\textstyle\usd}\HI^g-
{\textstyle\usd}{\rm e}^{\displaystyle u\HI_0}
\HI^g
{\rm e}^{\displaystyle -u\HI_0}\right)(1-\RIzeg)
=0\q.
\ee
Moreover, when $\HI_0$ is solution
of Eq.~(\ref{hd170})
(which implies Eq.~(\ref{hd200}) in the basis 
where $\HI_0$ is diagonal)
one has the commutation relation
\be\label{hd280}
\left[\HI_0\,,\,\ssf\,\YI_0^g\,\RIzeg\,\HI^g(1-\RIzeg)\right]=0\q,
\ee
so that Eq.~(\ref{hd270}) is satisfied for all values of $u$.
This last property can also be 
demonstrated by means of Eqs.~(\ref{hd060}).
Likewise, within the
replacement of $u$ by $\bb-u$, the stationarity condition
(\ref{hc070}) also has the form (\ref{hd270})
and this is a consequence of (\ref{hd170}). 
Therefore,
in the case $\ADh=\uni$ and unbroken $\PDh$-invariance
the result of the optimization of the functional (\ref{hb050})
is identical to the approximation $\YB$  of the
exact parition function $\YI_P$ (Eq.~(\ref{hd070}))
obtained  by the
maximization of (\ref{hd110}).
\newpage

\section[S6]{Projection on Even or Odd Particle Number}

\setcounter{equation}{0}

In this Section we consider finite systems of fermions, such as 
atomic nuclei or metallic grains, for which 
the Hamiltonian favors pairing correlations. In such cases 
there exists a {\it qualitative
difference between the properties of
systems with even and with odd
numbers of fermions}, whereas systems with $N$, $N\pm2$, 
$N\pm4$ \ldots particles {display strong similarities}.
At zero temperature the HFB approximation, which
is known to provide a good description
of pairing correlations in nuclei, involves 
a sum of components with the
same particle-number parity.
On the other hand, at non-zero temperature, 
any number, odd and even, of quasi-particle 
configurations are allowed by the standard HFB solution.
As a consequence the HFB thermal equilibrium state 
has both even and odd particle-number components.
A completely realistic description of finite systems 
with fixed particle number $N_0$ would require
a full projection as defined by Eqs.~(\ref{ha010}-\ref{ha020}).
(See Chapter 11 of Ref.\cite{RSc80} for the litterature,
starting with Refs.\cite{BKM60}, about the particle-number projection
at zero temperature.)
Nevertheless, the similar properties of systems with the same 
particle-number parity suggest that a first significant step 
towards this goal consists in projecting the independent
quasi-particle trial state onto the space with even or odd particle 
(or equivalently quasi-particle) number by means of the projection
$\PDh_\eta$ defined in Eq.~(\ref{ha030}). 

Using the formalism of Sect.~5
we want therefore 
to approximate variationally the characteristic function
$\varphi_\eta(\xi)$ of a system with a given parity
of the particle number (or $N$-parity).
In particular, for $\ADh=\uni$, we shall thus
obtain an approximation for
the {\it number-parity projected grand-partition function}
\be\label{he010}
\YI_\eta\equiv{\sum_{\vphantom{n}\atop N}}^{\,\,\prime}
{\rm e}^{\bb\mu N}\,{\rm Tr}_{\vphantom{N}\atop N}\,{\rm e}^{-\bb\HDh}\q,
\ee
{where the prime indicates
that the sum runs only} over even ($\eta=1$) or odd ($\eta=-1$)
values of $N$. We still enforce the particle-number average 
by means of the chemical potential $\mu$
entering the operator $\KDh=\HDh-\mu\NDh$.

In the specialization of the
equations of Sect.~5 to the present case, 
the sum $\ssf$  in the general 
formula~(\ref{ha080}) reduces to a sum of two terms:
\be\label{he020}
\PDh_\eta=\textstyle{\usd}(\TDh^\cze+\eta\,\TDh^\cpi)
\equiv\textstyle{\usd}(\uni+\eta\,{\rm e}^{{\rm i}\pi\NDh})\q.
\ee
In the $2n\times2n$ representation, we
have simply
\be\label{he030}
\li^\cze=0\ ,\q\TI^\cze=1\ ;\q
\li^\cpi=-{\textstyle\usd}{\rm i}\pi\,n\ ,\q
\TI^\cpi={\rm e}^{{\rm i}\pi\NI}=-1\q.
\ee
The commutation of $\TI^\cze$
and $\TI^\cpi$ with any matrix $\TI$
of the algebra introduced in Sect.~3.1 is obvious.
This reflects the fact that all operators of the single-quasi-particle
type (\ref{f120}), or of the trial forms $\TDhd$ or $\TDha$ given
by (\ref{f040}), commute with $\TDh^\cze$ and $\TDh^\cpi$ and with
the projection $\PDh_\eta$ because
they involve only an even number of
creation or annihilation operators.
Thus, although a BCS type of state (\ref{f160}) breaks the particle
number invariance ${\rm e}^{i\theta\NDh}$, it {\it does not break
the} $N$-{\it parity invariance} ${\rm e}^{i\pi\NDh}$, and
we can work with the variational spaces
\be\label{he040}
\DIh\uu=\PDh_\eta\TDhd\uu\ ,\q\q\AIh\uu=\TDha\uu\q,
\ee
where only one projection $\PDh_\eta$ is needed.

\subsection[S61]{Characteristic Function 
in the Particle-Number Parity Projected HFB 
Approximation}

In the present Section we shall therefore
rely on the formalism developed in Sects.~5.3.
The reader who has chosen to skip Sect.~5, is requested to 
accept Eqs.~(\ref{hd020}-\ref{hd050}) which, after the replacement of
$\ssf$ by a sum and the simplifications
associated with Eq.~(\ref{he030}), furnish the 
expression of the functional corresponding to
the $N$-parity projection.
The matrices $\RIadg\uu$ and $\RIdag\uu$ which occur
in the functional
(\ref{hd050}) 
can now be written explicitly:  
\be\label{he050}
\RI^{[i0]}=\RI^\ip\ ,\q\q
\RI^{[i1]}=\frac{\RI^\ip}{2\RI^\ip-1}\q,
\ee
with $i=ad$ or $da$ and 
with $\RIb$ or $\RIc$ defined as in Sect.~3.2.
The weight factors $\YI^\fiu$, given by Eq.~(\ref{hd030}),
take now the compact form
\be\label{he060}
\YI^0=\displaystyle 
{\rm e}^{-\lia-\lid}\acg{\rm det}_{2n}(1-\RIb)\acd^{-1/2}\q,
\ee
\be\label{he070}
\YI^1=\displaystyle r\,\YI^0\ ,\q\q
r=\acg{\rm det}_{2n}[\NI(1-2\RIb)]\acd^{1/2}
\q,
\ee
and hence the normalization $\YB_\eta$ (Eq.~(\ref{hd020})) reads
\be\label{he080}
\YB_\eta={\textstyle\usd}\YI^0(1+\eta r)\q.
\ee

According to the general formalism of Sect.~2, the evaluation
of the approximate
characteristic function 
$\varphi(\xi)=\log\YB_\eta=\ln{\rm Tr}\ADh(\xi)\DIh(\beta)$
requires the solution of the coupled equations which determine
$\DIh\uu$ and $\AIh\uu$.
To this aim we will use Eqs.~(\ref{hc020}-\ref{hc040}), 
(\ref{hc070}-\ref{hc080}) and (\ref{hc110}-\ref{hc120}) with
the following simplifications~: 
i) suppression of the integration on the group index
$g'$ which is everywhere set to 0 ($\TDh^{[g']}=\TDh^{[0]}=\uni)$, 
(ii) replacement of $\ssf$ by the sum $\sum_j$ with $j=0$ or 1, 
(iii) commutation of $\TI^{[j]}$ with
$\TIa$ and $\TId$ whenever it is required. 
Thanks to these simplifications
we can express the evolution equations in a more 
explicit form than in Sect.~5.3 since
the group elements $\TIgu$ are now trivial. 
For instance, Eq.~(\ref{hc020}) for $\TId\uu$ writes:
\be\label{he090}
\displaystyle
\frac{{\rm d}\,\TId}{{\rm d}u}
-\eta\,r\frac{1}{2\RIc-1}\,
\frac{{\rm d}\,\TId}{{\rm d}u}
\,\frac{1}{2\RIb-1}
=-\usd[{\bar \HI}_d^{(da)}\TId+\TId{\bar \HI}_d^{(ad)}]\q,
\ee
where we have introduced
the generalized HFB Hamiltonians {
${\bar \HI}_d^{(ad)}$  and ${\bar \HI}_d^{(da)}$ defined by
\be\label{he100}\ba{rl}
{\bar \HI}_d^{(ad)}\equiv&
\displaystyle
\HI^{[0a0d]}-\eta\,r
\frac{1}{2\RI^{[ad]}-1}\,
\HI^{[0a1d]}\,
\frac{1}{2\RI^{[ad]}-1}\q,\\
{\bar \HI}_d^{(da)}\equiv&
\displaystyle
\HI^{[d0a0]}-\eta\,r
\frac{1}{2\RI^{[da]}-1}\,
\HI^{[d0a1]}\,
\frac{1}{2\RI^{[da]}-1}\q,
\ea\ee
in terms of the matrices given in Eq.~(\ref{hc030}).
Taking into account that 
the quantities $k_d^{g'\,g}$ defined in (\ref{hc040}) now
satisfy the relation
\be\label{he110}
k_d^{00}=-\eta\,r\,\ksdzu
\ee
 with
\be\label{he120}\ba{l}
\displaystyle
\ksdzu=\frac{1}{2(1+\eta\,r)}\left\{
2\trd{[\TIdm\frac{\RIc(1-\RIc)}{2\RIc-1}
\frac{\displaystyle {\rm d}\TId}{\displaystyle {\rm d}u}]}\right.\\
\q\q\q+\displaystyle
\left.\EI\{\frac{\RIc}{2\RIc-1}\}-\EI\{\RIc\}+
\EI\{\frac{\RIb}{2\RIb-1}\}-\EI\{\RIb\}\right\}\ ,
\ea\ee
one can also express the matrices ${\bar \HI}_d^{(ad)}$  
and ${\bar \HI}_d^{(da)}$ as
\be\label{he130}
{\bar \HI}_d^{(i)}=\HI\{\RI^{[i]}\}-\eta\,r\,
\frac{1}{2\RI^{[i]}-1}\,
\HI\{\frac{\RI^{[i]}}{2\RI^{[i]}-1}\}
\,\frac{1}{2\RI^{[i]}-1}
+\frac{2\eta\,r\,\ksdzu}{2\RI^{[i]}-1}
\q,
\ee
where $i$ stands as above for $ad$ or $da$. Likewise, the
evolution equation (\ref{hc070}) 
for $\TIa\uu$ can be put in a form
similar to (\ref{he090}).
In Eqs.~(\ref{he090}-\ref{he100}) and (\ref{he130}), 
the subscript $d$ in the matrices ${\bar \HI}_d^{(i)}$
accounts for the fact that the 
c-numbers $k_a^{10}$ and $\ksdzu$ and therefore
the matrices ${\bar \HI}_d^{(i)}$
and ${\bar \HI}_a^{(i)}$ are in general different.

For $g=g'=0$,
the specialization of Eqs.~(\ref{hc110}) and 
(\ref{hc120}) to the
$N$-parity projection yields
a compact form
for the evolution equation of $\log\YI^0$,
\be\label{he140}
\frac{\displaystyle {\rm d}\log\YI^0}{\displaystyle {\rm d}u}
=-\frac{\eta\,r}{1+\eta\,r}
\trd{
\{\frac{1}{2\RIc-1}
\frac{{\rm d}\RIc}{{\rm d}u}\}}\q,
\ee
in terms of $\RIc$ (or of $\RIb$). This equation could
also have been obtained from the general
property ${\rm d}\YB_\eta/{\rm d}u=0$, using Eq.~(\ref{he070})
and the derivative of $r$ drawn from (\ref{he080}).
To complete the set of equations
which may be useful for the calculation of 
the characteristic function, we
can rewrite Eqs.~(\ref{hd060}) in terms of
the matrices $\RIc$ and $\RIb$. For the evolution of $\RIc$
we get:
\be\label{he150}\ba{l}
\displaystyle
\frac{\displaystyle {\rm d}\RIc}{\displaystyle {\rm d}u}
-\eta\,r\frac{1}{2\RIc-1}\,
\frac{\displaystyle {\rm d}\RIc}{\displaystyle {\rm d}u}
\,\frac{1}{2\RIc-1}
+2\frac{\RIc(\RIc-1)}{2\RIc-1}
\frac{\displaystyle {\rm d}\log\YI^0}{\displaystyle {\rm d}u}
\\
\q\q\q\q\q\q\q\displaystyle
=-\usd[{\bar \HI}_d^{(da)}\,,\,\RIc]\q,
\ea\ee
where ${\rm d}\log\YI^0/{\rm d}u$ is given by Eq.~(\ref{he140}).
A similar equation, with $da$ replaced by $ad$ and
a plus sign before the commutator in the right hand side,
holds for $\RIb$.
{These equations for $\RIc$ and $\RIb$} 
are coupled through the
boundary conditions, as were Eqs.(\ref{f520}) which they generalize.

{\it Remark~:} 
Due to the square roots occuring in Eqs.~(\ref{he060}-\ref{he070}),
we {must carefully fix} the determinations
which control the signs of $\YI^0$ and $\YI^1$.
In Eq.~(\ref{he060}) we can rely on the fact that, for
small values of the sources, the contraction matrix
$\RIb$ is nearly real, lying between 0 and
1, which makes the determination unambiguous.
In Eq.~(\ref{he070}) we have introduced the factor 
$\NI$ to account
for the term $-\li^\cpi$ of (\ref{hd030}), 
but both the matrices $\NI$ and $1-2\RIb$ have
positive and negative eigenvalues ; moreover,
they do not commute when pairing takes place.
In order to overcome this difficulty and to define unambiguously the
sign of $r$, we proceed by continuity, starting from the
unperturbed state without pairing. In this case, when
$\NI$ and $\RIb$ are simultaneously diagonalized,
the two diagonal blocks of $\NI(1-2\RIb)$ are identical,
due to (\ref{f090}), and provide twice the same factors in (\ref{he070}).
We then choose the determination of $r$ as
\be\label{he160}
r=\prod_{\vphantom{n}\atop\lambda}(1-2\RIb_\lambda)\q,
\ee
where $\RIb_\lambda$ are the diagonal elements of $\RIb$ lying in the
block $\NI=1$. Hence, $r$ is positive if the
number of eigenvalues of the single-particle Hamiltonian that are smaller
than the chemical potential
$\mu$ is even, while $r$ is negative if this number is odd. 
In the low temperature limit, for $\ADh=\uni$ and
arbitrary $\mu$, one obtains for the grand potential
$\FI_{\rm G}=-\bb^{-1}\,\log\YB_\eta$
the expected {result},
namely the minimum of $(\EI-\mu\langle\NDh\rangle)$ for either
even $N$ ($\eta=+1$) or odd $N$ ($\eta=-1$).
When the interactions are switched on and pairing 
takes place, we note
that within the determinant of Eq.~(\ref{he070}) the particle
number matrix $\NI$ can be replaced by the
matrix $\NI'$ (associated with the quasi-particle number)
which commutes with $\RIb$.
We can thus define $r$ by continuity, starting from 
(\ref{he160}) and keeping
the same form  (\ref{he160}) 
where the $\RIb_\lambda$ are now the eigenvalues of
$\RIb$ associated with the subspace $\NI'=+1$.

\subsection[S62]{The Partition Function 
in the $N$-Parity Projected HFB 
Approximation}

If one is only interested in 
the evaluation of the $N$-parity projected
grand partition function $\YI_\eta$ (obtained for $\ADh=\uni$), or
more generally of the characteristic function 
for conserved single-particle observables $\QDh_\ggg$
(obtained for $\ADh=\exp(-\sum_\ggg \xi_\ggg\QDh_\ggg)$
with $[\QDh_\ggg\,,\,\KDh]=0$),
it is not necessary to solve 
the coupled equations of Sect.~6.1 which are suited to
the evaluation of the most general characteristic function.

As in Sect.~5.4.2, we introduce a trial operator $\HDh_0$ of the
form (\ref{hd090}) which depends on the c-number
$h_0$ and the $2n\times2n$ matrix $\HI_0$; for 
the restricted variational spaces, we take 
\be\label{he170}
\DIh_0\uu=\PDh_\eta\,{\rm e}^{\displaystyle -u\HDh_0}\ ,\q
\AIh_0\uu={\rm e}^{\displaystyle -(\beta-u)\HDh_0}\q.
\ee
The matrices $\TId\uu$ and $\TIa\uu$ are therefore
equal to $\exp(-u\HI_0)$ and
$\exp(-(\bb-u)\HI_0)$, respectively, while the $u$-independent 
contraction matrices $\RI_0^{[ad]}$ and $\RI_0^{[da]}$
are expressed in terms of $\HI_0$ by
\be\label{he180}
\RI_0^{[ad]}\uu=\RI_0^{[da]}\uu=
\frac{1}{1+{\rm e}^{\bb\HI_0}}=\RIze\q.
\ee
We shall also need the specialization 
of the matrices $\RI_0^g$, defined in Eq.~(\ref{hd130}),
to the simple projection
group associated with $\PDh_\eta$~:
\be\label{he190}
\RI_0^0=\RIze\ ,\q
\RI_0^1=\frac{\RIze}{2\RIze-1}=\frac{1}{1-{\rm e}^{\bb\HI_0}}\q.
\ee

As in Sect.~5.4.2,
the c-number $h_0$ is given by Eq.~(\ref{hd160}) where
${\bar\RI}_\eta$ and $\EIzb_\eta$, defined by Eqs.~(\ref{hd140}), 
take now the explicit forms
\be\label{he200}
{\bar\RI}_\eta=\frac{1}{1+\eta\,r_0}
\pag \RIze+\eta\,r_0\,\RI_0^1 \pad\q,
\ee
\be\label{he210}
\EIzb_\eta=\frac{1}{1+\eta\,r_0}\pag
\EI\{\RIze\}+\eta\,r_0\,\EI\{\RI_0^1\}\pad\q.
\ee
In these equations the quantity $r_0=\YI_0^1/\YI_0^0$ is given by
\be\label{he220}
r_0=[\det \NI'\tanh\frac{\beta}{2}\HI_0]^{1/2}
=\prod_\lambda\tanh\frac{\beta}{2}\HI_{0\lambda}\q,
\ee
where $\NI'$ stands for the quasi-particle-number
operator and $\HI_{0\lambda}$ denote the eigenvalues of
the matrix $\HI_0$.
The sign of $r_0$ is obtained by continuity according to (\ref{he160}).

The matrix $\HI_0$ is the solution
of the self-consistent Eq.~(\ref{hd170})
which, after a suitable reorganization,
can be written 
\be\label{he230}
\HI_0\,(1-\eta\,r_0\,\coth^2\frac{\beta}{2}\HI_0)={\bar\HI}\q.
\ee
Using Eq.~(\ref{he190}) and Eq.~(\ref{hd180})
to define $\HI^0$ from
$\RIze\equiv\RI_0^0$ and $\HI^1$ from
$\RI_0^1$, the matrix $\bar\HI$ is given by
\be\label{he240}
{\bar\HI}=\HI^0-\eta\,r_0\,
\left(\coth\frac{\beta}{2}\HI_0\right)\,\HI^1\,
\left(\coth\frac{\beta}{2}\HI_0\right)\q.
\ee
In terms of 
the matrices $\RIze$ and $\RI_0^1$, 
the operator ${\bar\HI}$ can be expressed as
\be\label{he250}
\displaystyle
{\bar\HI}=\HI\{\RIze\}-\eta\,r_0\,
\left(\coth\frac{\beta}{2}\HI_0\right) 
\,\HI\{\RI_0^1 \}\,
\left(\coth\frac{\beta}{2}\HI_0\right)
- 2\,\eta\,r_0\,\ksu\,\coth\frac{\beta}{2}\HI_0
\ ,
\ee
in which the quantity $\ksu$ 
(see Eq.~(\ref{hd190}) and note 
that $k^0=-\eta\,r_0\,\ksu$) is 
\be\label{he260}
\ksu=
\frac{1}{1+\eta\,r_0}\left(\usd\trd{
\frac{\HI_0}{\sinh\beta\HI_0}
}
+
\EI\{\RI_0^1\}-\EI\{\RIze\}\right)\q.
\ee

In this projected 
variational scheme, Eq.~(\ref{he230}) replaces the standard
HFB self-consistent equation $\HI_0=\HI\{\RIze\}$.
The logarithm of the 
{exact} $N$-parity projected grand-partition function
$\YI_\eta$ (Eq.~(\ref{he010})) is approximated by $\log\YB_\eta$,
the stationary
value of the functional (\ref{hd110}), as 
\be\label{he270}
\log\YB_\eta=
{\textstyle\usd}\trd{\log[1+{\rm e}^{-\beta\HI_0}}]
+\textstyle{\usd}\trd{[\beta\HI_0{\bar\RI}_\eta]}-\beta\EIzb_\eta
+\log[{\textstyle\usd}(1+\eta\,r_0)]\ .
\ee
Equivalently, this quantity can be written
as
\be\label{he280}
\log\YB_\eta=
\log\YI-\eta\,r_0\,\beta\,\ksu
+\log[{\textstyle\usd}(1+\eta\,r_0)]
\q,
\ee
where $\log\YI$ is the logarithm of the HFB partition function,
$\SI\{\RIze\}-\bb\EI\{\RIze\}$, which is
associated with the independent-quasi-particle contraction matrix
$\RIze$ (Eq.~(\ref{ga050})).

The entropy $\SI_\eta$ of the exact projected
state  is approximated by the specialization 
to the present case of the quantity $\SIzb_\eta$
given in Eq.~(\ref{hd250}):
\be\label{he290}
\SIzb_\eta=-{\textstyle\usd}
\trd{ [{\bar\RI}_\eta\,\log\RIze\,+\,(1-{\bar\RI}_\eta)\log(1-\RIze)] }
+\log[\textstyle{\usd}(1+\eta\,r_0)]\ .
\ee
After some rearrangement, $\SIzb_\eta$ takes the form
\be\label{he300}
\SIzb_\eta=\SI\{\RIze\}-
\frac{\eta\,r_0}{2(1+\eta\,r_0)}\,
\trd{\frac{\beta\HI_0}{\sinh\beta\HI_0}}
+\log[\textstyle{\usd}(1+\eta\,r_0)]\ ,
\ee
where $\SI\{\RI_0\}$ is the usual
HFB entropy given by Eq.(\ref{ga060}).
As stressed in Sect.~5.4.2, the
equations (\ref{he230}-\ref{he260})
satisfied by $\HI_0$
ensure that {the approximate self-consistent projected
entropy $\SIzb_\eta$ obeys the standard thermodynamic relations
(\ref{e262}) and (\ref{e263}) with $\FI_{\rm G}=-\bb^{-1}\log\YB_\eta$.}

Since our trial operators $\DIh\uu$ and $\AIh\uu$ satisfy
the validity conditions of Eq.~(\ref{e230}),
the density matrix
which optimizes the partition function can be used
to obtain the variational approximation for the expectation
value of {\it any} single-quasi-particle operator $\QDh$ of the
type (\ref{f120}).
For instance, the transposition of Eq.~(\ref{e230}) gives :
\be\label{he310}
\langle\QDh\rangle=\qi+\usd\trd{\QI{\bar\RI}_\eta}\q,
\ee
which generalizes the HFB formula (\ref{gb010}).
Moreover, since these operators $\QDh$ commute with $(-1){^\NDh}$,
their characteristic function can be found by means of
the formalism above by changing $\KDh$ into
$\KDh+\xi\QDh/\bb$, as indicated in Sect.~2.3.

Altogether, to obtain the variational approximation $\YB_\eta$
for the $N$-parity projected grand-partition function $\YI_\eta$ defined
by Eq.~(\ref{he010}), one first determines the matrix $\HI_0$  
(jointly with the contraction matrix
${\bar \RI}_\eta$) from the self-consistent 
equation (\ref{he230}) which contains (through $\bar\HI$) 
the c-numbers $r_0$ and $\ksu$, 
given by (\ref{he220}) and  (\ref{he260}), 
respectively. One then evaluates $\YB_\eta$
by means of Eq.~(\ref{he280}). 
Approximations for the
thermodynamic quantities 
such as the entropy (\ref{he290}-\ref{he300}),  
or for the expectation values (\ref{he310})
of single-particle observables, in particular of
$\NDh$,
involve the averaged contraction matrix ${\bar\RI}_\eta$
defined in (\ref{he200}); this matrix is positive {since, from 
Eqs.~(\ref{he180}), (\ref{he190}) and (\ref{he220}),
its diagonal eigenvalues are given by}
\be\label{he320}
{\bar\RI}_{\eta\lambda}=
\frac{1}{{\rm e}^{\bb\HI_{0\,\lambda}}+1}
\frac{\displaystyle
1-\eta\prod_{\mu\neq\lambda
}\,\tanh\frac{\bb}{2}\HI_{0\,\mu}
}{\displaystyle
1+\eta\prod_{\mu}\,\tanh\frac{\bb}{2}\HI_{0\,\mu}}\q.
\ee

According to the general discussion of Sect.~2.3, our
variational approximation for the fluctuation of $\NDh$ is given by
the thermodynamic relation
\be\label{he330}
\Delta\NDh^2=
\frac{1}{\bb}\frac{\partial\langle\NDh\rangle}{\partial\mu}=
\frac{1}{2\bb}\trd{\NI
\frac{\partial{\bar\RI}_\eta}{\partial\mu}}\q.
\ee
The derivative $\partial{\bar\RI}_\eta/\partial\mu$ is deduced from
$\partial\HI_0/\partial\mu$, itself evaluated from Eq.~(\ref{he230}).

For large systems, $r_0$ is exponentially small while $\ksu$ is
proportional to the volume,
as shown by Eqs.(\ref{he220}) and (\ref{he260}),
and these behaviours imply that
the projection {does not modify} the HFB results. 
However, for sufficiently small
systems and at sufficiently low temperature, the factors {entering} 
$r_0$ (Eq.~(\ref{he220}))
may be close to $\pm 1$, except near the Fermi surface ;
the terms involving $r_0$ may thus be significant compared to the
dominant terms, although $\vert r_0\vert$ is always less than 1.

The results of this Section could have been obtained 
directly from the formalism developed in Sect.~6.1 by 
using the specific $u$-dependence of $\TId$ and the commutation of
$\HI_0$ with $\bar\HI$.
The latter property, already established in the Remark of 
Sect.5.4.2 for a
more general case, 
can also be proven from 
Eqs.(\ref{he140}) and (\ref{he150}) once
the $u$-independence (\ref{he180}) of the contraction matrix $\RIc$ 
is acknowledged.

\subsection[S63]{The Projected BCS model : Generalities}

We now use the formalism of the previous Section by specializing to 
the BCS-type pairing. 
In this case,
the $n$ creation operators $a^\dagger_i$
can be arranged (possibly by means of a unitary transformation)
in a set of couples
$\{(\acbcsp\,,\,\acbcsb)\,,\, 1\le p\le n/2\}$ such that the Hamiltonian
(\ref{f270}) takes the form
\be\label{hf010}
\HDh=\sum_p\,\epbcsp(\acbcsp\adbcsp+\acbcsb\adbcsb)-
\sum_{pq}\,\Gbcspq\acbcsp\acbcsb\adbcsc\adbcsq\q.
\ee
This may encompass models for electrons in a superconductor 
($p$ then denotes the momentum and a spin pointing upwards,
while $\bar p$ denotes the opposite momentum and spin),
or for nucleons in a deformed nuclear shell (the momentum is then
replaced by the component of the angular momentum along
the principal axis of inertia).
For the sake of simplicity 
we have not introduced a magnetic field, which would lift the
degeneracy between the states $p$ and $\bar p$. We have also assumed
that the diagonal matrix elements
$G_{pp}$ {are zero} (a non-zero 
value of $G_{pp}$ entails
a small {state-dependent}
shift of the single-particle energy $\epsilon_p$,
which leads to more complicated formulas but does not modify
the conclusions).
Moreover, we
consider the usual case
where the quantities $\Gbcspq$ are real and symmetric
in the exchange $p\leftrightarrow q$.

(The reader who wishes to begin with this Section 6.3 without
having followed the derivations of Sects.~5, 6.1 and 6.2
can do so if he is ready to adopt the definitions
and accept the expressions and equations that we now recapitulate.
Our variational space was introduced in Sect.~6.2, Eq.~(\ref{he170}), 
where the projection $\PDh_\eta$ was given in 
Eq.~(\ref{he020}) and
where the trial operator $\HDh_0$ (see Eq.~(\ref{hd090}))
involved a c-number $h_0$ (which will play no r\^ole in 
the following Sections) and a $2n\times 2n$
hermitian matrix $\HI_0$.
The contraction matrices $\RIze$ and $\RI_0^1$ were defined 
in Eqs.~(\ref{he180}-\ref{he190}) from 
the matrix $\HI_0$ which is itself determined self-consistently by 
Eqs.~(\ref{he230}) and (\ref{he250}).
The matrix ${\bar\RI}_\eta$ and the scalar quantities $r_0$, $\EIzb_\eta$ and
$\ksu$ were given in
Eqs.~(\ref{he200}), (\ref{he220}), (\ref{he210})
and (\ref{he260}), respectively.
The approximate projected partition function $\YB_\eta$,
entropy $\SIzb_\eta$ and expectation values $\langle\QDh\rangle$
were given in Eqs.~(\ref{he280}), (\ref{he300})
and (\ref{he310}).)

\subsubsection[S631]{The Variational Space} 

Within the BCS scheme the operator $\HDh_0$
can be put into the familiar diagonal
quasi-particle form
\be\label{hf020}
\HDh_0=(h_0-\sum_p\ebcsp)+
\sum_p\,\ebcsp(\,\bcbcsp\bdbcsp+\bcbcsb\bdbcsb\,)\q,
\ee
by means of the canonical transformation
\be\label{hf030}
\left\{\ba{rl}
\bdbcsp=&\ubcsp\adbcsp+\vbcsp\acbcsb\q,\\
\bdbcsb=&\ubcsp\adbcsb-\vbcsp\acbcsp\q,
\ea\right.\q\q
\left\{\ba{rl}
\adbcsp=&\ubcsp\bdbcsp-\vbcsp\bcbcsb\q,\\
\adbcsb=&\ubcsp\bdbcsb+\vbcsp\bcbcsp\q.
\ea\right.
\ee
(Note that two different sign conventions are currently used
in the literature for this transformation.)
The parameters $\ubcsp$ and $\vbcsp$ can be chosen real and
non-negative, with $\ubcsp^2+\vbcsp^2=1$.
The variational quantities are thus $\ebcsp$, $\vbcsp$ and $h_0$.

All the $2n\times 2n$ matrices that
enter our formalism can then be decomposed into 
a number $n/2$ of $2\times 2$ symmetric matrices labelled 
by $p$ and corresponding
to the operators $\adbcsp$ and $\acbcsb$~, 
plus another set of matrices labelled by $\bar p$ which
differs from the previous one through a
change in sign of $\vbcsp$.
In particular the $p$-components of the matrix $\HI_0$ 
(\*Eq.~(\ref{hd090})) are
\be\label{hf040}
\HI_{0\,p}=\ebcsp\,U_p\q,
\ee
where $U_p$ is the real symmetric orthogonal matrix
\be\label{hf050}
U_p\equiv\left(
\ba{cc}
\ubcsp^2-\vbcsp^2 & 2\ubcsp\vbcsp\\
2\ubcsp\vbcsp & \vbcsp^2-\ubcsp^2
\ea
\right)\q.
\ee
Accordingly, the $p$-components of the contraction matrix 
$\RIze$ (Eq.~(\ref{he180})) read
\be\label{hf060}
\RI_{0\,p}=\frac{1}{{\rm Tr}\,\TDhd\bbb}
{\rm Tr}\,\TDhd\bbb\,\left(
\ba{cc}
\acbcsp\adbcsp&\adbcsb\adbcsp\\
\acbcsp\acbcsb&\adbcsb\acbcsb
\ea
\right)={\textstyle\usd}
(1-U_p\,\tbcsp)
\q,
\ee
where we have introduced for shorthand the notation
\be\label{hf070}
\tbcsp\equiv\tanh{\textstyle\usd}\bb\,\ebcsp\q.
\ee
Likewise, the matrix $\RI_0^1$ (Eq.(\ref{he190}))
has the $p$-components
\be\label{hf080}
\RI_{0\,p}^1={\textstyle\usd}
(1-U_p\,\qbcsp)
\q.
\ee

The energy $\EI\{\RIze\}$, associated with the operator
$\KDh=\HDh-\mu\NDh$ and evaluated
for (\ref{hf060}), is now given by
\be\label{hf090}
\displaystyle
\EI\{\RIze\}=\sum_p\,
(\epbcsp-\mu)[1-(\ubcsp^2-\vbcsp^2)\tbcsp]
-\sum_{pq}\,\Gbcspq\,\ubcsp\vbcsp\,\ubcsq\vbcsq\,
\tbcsp\,\tbcsq\ .
\ee
The corresponding matrix $\HI_p\{\RIze\}$
has the $p$-components
\be\label{hf100}
\HI_p\{\RIze\}=\left(
\ba{cc}
\epbcsp-\mu & \Delta_{p}^0\\
\Delta_{p}^0 & \mu-\epbcsp
\ea
\right)\q,
\ee
with
\be\label{hf110}
\Delta_{p}^0\equiv
\sum_{q}\,\Gbcspq\,
\ubcsq\vbcsq\,\tbcsq\q.
\ee
{By substituting} $\qbcsq$ to
$\tbcsq$ in Eqs. (\ref{hf090}) and
(\ref{hf100}) {we obtain the energy} $\EI\{\RI_0^1\}$
and the matrix $\HI\{\RI_0^1\}$. {In particular,} $\Delta_p^0$ 
{is replaced by}
\be\label{hf120}
\Delta_p^1\equiv
\sum_{q}\,\Gbcspq\,
\ubcsq\vbcsq\,\qbcsq\q.
\ee
The c-numbers $r_0$ (Eq.~\ref{he220})) and $\ksu$ (Eq.~(\ref{he260}))
are respectively equal to
\be\label{hf130}
r_0=\prod_p\,\tbcsp^2\q,
\ee
and
\be\label{hf140}
\ksu=\frac{1}{1+\eta\,r_0}\sum_p\,(\qbcsp-\tbcsp)\left\{\,
\ebcsp-(\epbcsp-\mu)(\ubcsp^2-\vbcsp^2)
-\ubcsp\vbcsp\,
(\Delta_p^0+\Delta_p^1)
\right\}\ .
\ee

\subsubsection[S632]{The Variational Equations} 

The trial parameters $\ebcsp$ and $\vbcsp$
must be determined from the self-consistent matrix
equations (\ref{he230}) and (\ref{he250}).
By means of the canonical transformation (\ref{hf030}), let us
write this equation in the basis where $\HI_0$ is diagonal.
From the diagonal element of Eq.~(\ref{he230}) we thus obtain:
\be\label{hf150}\ba{l}
\displaystyle
\ebcsp(1-\eta\,r_0\,{\tbcsp^{-2}})
+2\eta\,r_0\,\ksu\,\qbcsp=
\\
\q\q\displaystyle
(\ubcsp^2-\vbcsp^2)(\epbcsp-\mu)(1-\eta\,r_0\,{\tbcsp^{-2}})
+2\ubcsp\vbcsp
(\Delta_p^0-\eta\,r_0\,\Delta_p^1\,{\tbcsp^{-2}})\ ,
\ea\ee
and from the off-diagonal element :
\be\label{hf160}
\displaystyle
0=2\ubcsp\vbcsp(\epbcsp-\mu)
(1+\eta\,r_0\,{\tbcsp^{-2}})-
(\ubcsp^2-\vbcsp^2)
(\Delta_p^0+\eta\,r_0\,\Delta_p^1\,{\tbcsp^{-2}})\ .
\ee
Together with the preceding definitions of $\Delta_p^0$,
$\Delta_p^1$, $\tbcsp$, $r_0$ and $\ksu$, Eqs.~(\ref{hf150})
and (\ref{hf160}) are the self-consistent equations which 
determine the variational parameters $\ebcsp$, $\ubcsp$ and $\vbcsp$.

Let us further work out these equations. By introducing
the {\it weighted averaged pairing gap}
\be\label{hf170}
\Delta_{\eta\, p}\equiv\frac{\displaystyle
\Delta_p^0+\eta\,r_0\,\Delta_p^1\,{\tbcsp^{-2}}
}{\displaystyle
1+\eta\,r_0\,{\tbcsp^{-2}}
}\q,
\ee
and making use of the relation
$(\ubcsp^2-\vbcsp^2)^2+(2\ubcsp\vbcsp)^2=1$,
we can simplify Eq.~(\ref{hf160}) into
\be\label{hf180}
\displaystyle
\frac{2\ubcsp\vbcsp}
{\Delta_{\eta\, p}}
=
\frac{\ubcsp^2-\vbcsp^2}
{\epbcsp-\mu}
=
\frac{1}
{\ebcstp}\ ,
\ee
where $\ebcstp$ has the familiar BCS form
\be\label{hf190}
\ebcstp\equiv\sqrt{(\epbcsp-\mu)^2+\Delta_{\eta\, p}^2}\q.
\ee
Except for the definition of $\Delta_{\eta\, p}$, Eq.~(\ref{hf180}) is the same
as the usual BCS equation for the coefficients
$\ubcsp$, $\vbcsp$ of the canonical transformation (\ref{hf030}).
As regards Eq.~(\ref{hf150}), it appears as an extension of the BCS
relation 
$\ebcsp=(\ubcsp^2-\vbcsp^2)(\epbcsp-\mu)+2\ubcsp\vbcsp\Delta_{p}$ ;
actually, we can rewrite Eq.~(\ref{hf150}) as
\be\label{hf200}\ba{rl}
\displaystyle
\ebcsp&=(\ubcsp^2-\vbcsp^2)\,(\epbcsp-\mu)+
2\ubcsp\vbcsp[\Delta_{\eta\, p}-
\frac{\displaystyle
2\eta\,r_0}{\displaystyle\tbcsp^2-r_0^2\tbcsp^{-2}}
(\Delta_p^1-\Delta_p^0)]\\
&\q\q\q\q\q\q\q\q\q\q\q\displaystyle
-\frac{2\eta\,r_0\,\ksu}{\tbcsp-\eta\,r_0\,\qbcsp}\q,
\ea\ee
and hence, by using Eq.~(\ref{hf180}) to eliminate $\ubcsp$ and 
$\vbcsp$, as
\be\label{hf210}
\ebcsp=\ebcstp
-\frac{2\eta\,r_0\,\ksu}{\tbcsp-\eta\,r_0\,\qbcsp}
-\frac{2\eta\,r_0}{\tbcsp^2-r_0^2\tbcsp^{-2}}
\frac{\displaystyle
\Delta_{\eta\, p}(\Delta_p^1-\Delta_p^0)
}{\displaystyle
\ebcstp
}
\q.
\ee
These quantities $\ebcsp$ are the energies which appear in the diagonal
form of the density operator.
Eqs.~(\ref{hf190}) and (\ref{hf210}) show that
$\ebcsp$ has no longer the BCS form.
Indeed, several quantities, $\Delta_p^0$, $\Delta_p^1$ and
$\Delta_{\eta\, p}$, play here the r\^ole of pairing gaps.
We recover the standard BCS relation between $\ebcsp$ and $\Delta_{\eta\, p}$
when the system becomes sufficiently large, in which case $r_0$
and $r_0\,k^1$ become small. 
We can eliminate $\ubcsp$ and $\vbcsp$
from the definitions (\ref{hf110}), (\ref{hf120}) and (\ref{hf170})
of $\Delta_p^0$, $\Delta_p^1$ and 
$\Delta_{\eta\, p}$. These quantities thus satisfy the
equations
\be\label{hf220}
\Delta_p^1-\Delta_p^0=\usd\sum_q\,\Gbcspq\,
\frac{\displaystyle
\Delta_{\eta\, q}\,(\qbcsq-\tbcsq)
}{\displaystyle
\ebcstq
}\q,
\ee
\be\label{hf230}\displaystyle
\Delta_{\eta\, p}=\usd\sum_{q}\,\Gbcspq\,
\frac{\displaystyle
\Delta_{\eta\, q}\,\tbcsq
}{\displaystyle
\ebcstq
}\,\frac{\displaystyle
1+\eta\,r_0\,\tbcsp^{-2}\,\tbcsq^{-2}
}{\displaystyle
1+\eta\,r_0\,\tbcsp^{-2}
}\q.
\ee
Eq.~(\ref{hf230}) exhibits the projection
correction to the so-called BCS
gap equation, while Eq.~(\ref{hf220}) obviously  
has no BCS counterpart. Finally, 
by using 
(\ref{hf200}) and the identity 
$\sum_p\,\ubcsp\vbcsp(\Delta_p^1\,\tbcsp-\Delta_p^0\,\qbcsp)=0$,
we can write the expression (\ref{hf140}) for $\ksu$ 
in a more explicit form :
\be\label{hf240}\ba{l}\displaystyle
\ksu\left[
1+
\frac{\displaystyle
2\eta\,r_0 
}{\displaystyle
1+\eta\,r_0
}\sum_p
\frac{\displaystyle
\tbcsp^{-2}-1
}{\displaystyle
1-\eta\,r_0\,\tbcsp^{-2}
}\right]=\\
\q\q\q\q\displaystyle
-\,\sum_p\,
\frac{\displaystyle
\qbcsp-\tbcsp
}{\displaystyle 
2(1+\eta\,r_0)
}\frac{\displaystyle
1+\eta\,r_0\,\tbcsp^{-2}
}{\displaystyle
1-\eta\,r_0\,\tbcsp^{-2}
}\frac{\displaystyle
\Delta_{\eta\, p}\,(\Delta_p^1-\Delta_p^0)
}{\displaystyle 
\ebcstp
}\q.
\ea\ee

\subsubsection[S633]{Comments on the Equations} 

Altogether then, we have thus to solve the self-consistent equations
(\ref{hf210}) and (\ref{hf230}) for the two sets $\ebcsp$ and
$\Delta_{\eta\, p}$.
The quantity $\ebcsp$ is indirectly involved through 
$\tbcsp\equiv\tanh(\bb\ebcsp/2)$.
{The two c-numbers} $r_0$ and $\ksu$ are expressed by
(\ref{hf130}) and (\ref{hf240}), respectively. The difference
$\Delta_p^1-\Delta_p^0$, which enters (\ref{hf210}) and (\ref{hf240}),
can be eliminated by means of Eq.~(\ref{hf220}).
The coefficients $\ubcsp$ and $\vbcsp$ are given explicitly
by (\ref{hf180}). 

In Eq.~(\ref{hf230}) the last fraction, which takes care of
 the projection effects,
can also be written as 
$1+(\tbcsq^{-2}-1)f(\eta\,r_0\tbcsp^{-2})$ with
$f(x)=x/(1+x)$. 
The positive term $(\tbcsq^{-2}-1)$ differs significantly from zero
only when $\bb\ebcsq<1$. 
For an even-number projection ($\eta=1$), 
the function $f$ is positive and varies between 0 and 1/2 while
for an odd-number projection 
($-1<\eta r_0<0$) $f$ is negative and varies from $-\infty$
to zero. The following conclusions can therefore be drawn :
\begin{itemize}
\item in the sums of Eqs.~(\ref{hf230}-\ref{hf240}), the
modifications due to projection affect predominantly the terms
corresponding to the single-particle states near the Fermi surface.
\item  the even-number projected gap is
larger than the BCS one while the odd-number gap is smaller.
\item larger differences with BCS are expected for the projection
on an odd number of particles.  
\end{itemize}

Once Eqs.~(\ref{hf210}) and (\ref{hf230})
are solved, the thermodynamics 
is governed by the equations (\ref{he300}), (\ref{he200}) and 
(\ref{he280}) which give $\SIzb_\eta$, $\EIzb_\eta$ and $\YB_\eta$, respectively. 
To compute the approximation $\SIzb_\eta$ given by (\ref{he300})
to the exact number-parity projected entropy $\SI_\eta$, 
we take into account 
the twofold degeneracy of the eigenvalues 
$({\rm e}^{\pm\bb\ebcsp}+1)^{-1}$ of the contraction matrix $\RIze$.  
We find that $\SIzb_\eta$
differs from the independent quasi-particle entropy $\SI_{\rm BCS}$,
\be\label{hf250}
\SI_{\rm BCS}\equiv\SI\{\RIze\}
=\sum_p\,[2\log(2\cosh\frac{\beta}{2}\ebcsp)-
\bb\ebcsp\tanh\frac{\beta}{2}\ebcsp]\q,
\ee
according to
\be\label{hf260}
\displaystyle
\SIzb_\eta=\SI_{\rm BCS}
-\frac{\eta\,r_0}{1+\eta\,r_0}
\sum_p\,\bb\ebcsp\,(\qbcsp-\tbcsp)
+\log[{\textstyle\usd}(1+\eta\,r_0)]\q.
\ee

It results immediately from
(\ref{he310}), (\ref{hf060}) and 
(\ref{hf180}) that the expectation value of the
particle number is given in terms
of the chemical potential by
\be\label{hf270}\displaystyle
\langle\NDh\rangle=\sum_p\,\left[
1-\frac{\epbcsp-\mu}
{\ebcstp}\,
\frac{\tbcsp+\,\eta\,r_0\,\qbcsp}{1+\,\eta\,r_0}
\right]\q.
\ee
This expression is also the derivative {with respect to $\mu$
of $F_{\rm G}=-\log\YB_\eta/\bb=\EIzb_\eta-\SIzb_\eta/\bb$, which is our
approximation
to the logarithm of the exact $N$-parity projected grand-potential
$-\log\YI_\eta/\bb$}. The 
expectation value $\langle \HDh-\mu\NDh\rangle$ is obtained 
from Eqs.~(\ref{he210}) and (\ref{hf090}) as
\be\label{hf280}\ba{rl}\displaystyle
\langle \HDh-\mu\NDh\rangle\equiv
\EIzb_\eta=&\displaystyle
\sum_p\,(\epbcsp-\mu)
\left[
1-(\ubcsp^2-\vbcsp^2)\,
\frac{\displaystyle
\tbcsp+\eta\,r_0\,\qbcsp
}{\displaystyle
1+\eta\,r_0}
\right]\\
&\q\q\displaystyle
-\sum_{pq}
\Gbcspq\ubcsp\vbcsp\,\ubcsp\vbcsq\,
\frac{\displaystyle
\tbcsp\tbcsq+\eta\,r_0\,\qbcsp\qbcsq}
{\displaystyle
1+\eta\,r_0}\ ,
\ea\ee
while, using Eqs.~(\ref{hf170}), (\ref{hf180}), (\ref{hf270})
and (\ref{hf280}),
the energy $\langle \HDh\rangle$ is given by 
\be\label{hf290}
\langle \HDh\rangle
=\sum_p\left[\epbcsp
-\frac{\displaystyle
\epbcsp(\epbcsp-\mu)+{\textstyle\usd}\Delta_{\eta\, p}^2}
{\displaystyle
\ebcstp}\,
\frac{\displaystyle
\tbcsp+\eta\,r_0\qbcsp
}{\displaystyle
 1+\eta\,r_0}
\right]\q.
\ee
Here again, the deviation from the BCS energy is exhibited by the
two terms that contain $r_0$.

In accordance with the general properties of our variational scheme, 
the {\it thermodynamic relations} (\ref{e261}) 
and (\ref{e262}) are satisfied by the approximate
entropy  $\SIzb_\eta$.
The expectation values of the single-particle observables are
given by (\ref{he310}). The specific heat, 
as expected, is equal both to $-\beta{\rm d}\SIzb_\eta/{\rm d}\beta$ 
and to the derivative of $\langle\HDh\rangle$ with respect to the
temperature. In Eq.~(\ref{hf290}), the temperature
appears directly in $\tbcsp=\tanh(\bb\epbcsp/2)$,
and indirectly through $\ebcsp$, $\Delta_{\eta\, p}$, $r_0$ and also
through $\mu$ since $\langle\NDh\rangle$ should be kept fixed
in the derivative. 
Finally, the general arguments
at the end of Sect.6.2 show 
that the characteristic function for 
any single-quasi-particle operator commuting with $\KDh$, 
such as $\NDh$,
is variationally obtained by adding a source term to 
the operator $\KDh$.
We find therefore that the fluctuation satisfies
\be\label{hf300} 
\Delta N^2=\frac{1}{\bb}\,\frac{
\partial\langle\NDh\rangle}{\partial\mu}\q,
\ee 
where in Eq.(\ref{hf270}) the derivative should be taken both explicitly
and through $\ebcsp$, $\Delta_{\eta\, p}$ and $r_0$.

A somewhat similar extension of the BCS theory, also involving a
projection on the spaces with a given parity of the particle number, 
has been worked out in Ref.\cite{JSA94}.
The results agree with ours for the partition 
function $\YB_\eta$.
The comparison between our variational equations 
(\ref{hf150}-\ref{hf160}) and those of Ref.\cite{JSA94}
appears more difficult. In particular it is not clear to us why the
entropy found in Ref.~\cite{JSA94} does not satisfy, in contrast
to ours, the thermodynamic relations (\ref{e261}) and (\ref{e262}).
On the other hand, the
general discussion of Sect.~2 as well as Eq.~(\ref{hf300})
show that fluctuations and correlations calculated with our
method must definitely be different; for instance we find here that 
the fluctuations of the particle number tend to zero with the
temperature, as expected from general grounds.

\subsection[S64]{The Projected BCS Model : Limiting Cases} 

\subsubsection[S641]{Low Temperatures}

At zero temperature the BCS (and HFB) solutions 
of the ground and excited states are exact 
eigenstates of the
number-parity operator and
their eigenvalues $\pm 1$, depend on the number of 
quasi-particles added to the even BCS vacuum. 

Let us first recall the
low-temperature limits of some quantities and equations of interest.
For large values of $\beta$,  the usual BCS gap 
equation becomes 
\be\label{hi010}
\Delta_{{\rm BCS}\, p}=\frac{1}{2}\,\sum_{q}\,\Gbcspq\,
\frac{\Delta_{{\rm BCS}\,q}}{\ebcstq}\,
(1-2{\rm e}^{\displaystyle -\bb\ebcsq})
\q.
\ee
In Eq.~(\ref{hi010}) the terms proportional to $\exp{(-\bb\ebcsq)}$
lead to a decrease of the gaps versus temperature $T=1/\bb$.
In the same limit, the BCS entropy (\ref{hf250}) 
is approximated by 
\be\label{hi020}
\SI_{\rm BCS}=2\,\sum_p\,\bb\ebcsp\,
{\rm e}^{\displaystyle -\bb\ebcsp}\q.
\ee

In the physics of mesoscopic superconductors, one often considers 
a simplified situation in which the
twofold degenerate unperturbed energies $\epsilon_p$
are sufficiently dense near the Fermi surface to
allow the replacement of a sum over $p$ by an integral over 
$\epsilon\equiv\epsilon_p-\mu$
with a weight $w_{\rm F}$ equal to the level density.
One also assumes that the pairing matrix elements
are constant and equal to $\tilde G$ over an interval 
$\Lambda$ around the Fermi energy $\mu$ and that they
vanish outside.
The width of this interval, taken of the order of the Debye energy, is
assumed to be much larger than the pairing
gap. It must be finite, however, to ensure the
convergence of the integrals.     
The ordinary BCS gap equation at zero temperature 
\be\label{hi030}
\frac{2}{\tilde G}=\int_{-\Lambda/2}^{\Lambda/2}\,
\frac{w_{\rm F}\,{\rm d}\epsilon}{\sqrt{\epsilon^2+\Delta^2}}\q,
\ee
defines a $p$-independent quantity $\Delta$ that we shall take
below as a reference energy.
In this field of physics, the ``low-temperature''
regime actually refers to an intermediate range of temperatures
such that the conditions
\be\label{hi040}
\frac{1}{\Lambda}\ll\frac{1}{\Delta}\ll\beta\ll w_{\rm F}
\ee
are valid. Under these conditions 
Eqs.~(\ref{hi010}) and (\ref{hi020}) give,
after elimination of the cut-off $\Lambda$ by means of Eq.~(\ref{hi030}), 
analytical expressions for the ``low temperature'' dependence of the
BCS pairing gap~:
\be\label{hi050}
\Delta_{\rm BCS}(\beta)=\Delta\,
\left[1-\sqrt{\frac{2\pi}{\beta\Delta}}\,
{\rm e}^{\displaystyle -\bb\Delta}\right]\q,
\ee
and of the BCS entropy :
\be\label{hi060}
\SI_{\rm BCS}=2\,{w_{\rm F}}\Delta\,
\sqrt{\,2\pi\bb\Delta}\,{\rm e}^{\displaystyle -\bb\Delta}\q.
\ee

In the $N$-parity
projected extension,
the quantity $r_0$ (Eq.~(\ref{hf130}))
plays a central r\^ole. In the ``low-temperature'' limit, its value tends
in general towards 1 and one has
\be\label{hi070}
\log r_0=-4\,\sum_p\,
{\rm e}^{\displaystyle -\bb\ebcsp}\q,
\ee
up to terms equal to, or smaller than $\exp(-3\bb\Delta)$.
(However, in the case $w_{\rm F}\Delta\simeq 1$
that we shall numerically investigate in Section~6.5.4,
the condition (\ref{hi040}) cannot be satisfied 
and we shall see that $r_0$
vanishes with the temperature.) 

We shall check below (Sect.~6.5.2) 
that, in the limit $w_{\rm F}\Delta\gg 1$,
the pairing gap $\Delta_{\eta\, p}$ obtained with the $N$-parity projection
does not depend much on $p$ and that it is almost identical
with $\Delta_{\rm BCS}(\beta)$. This will be illustrated by the 
numerical results plotted in Fig.~2.
Moreover, we shall see that the
difference between $\ebcsp$ and
${\tilde e}(\epsilon)=\sqrt{(\epsilon_p-\mu)^2+\Delta_{\eta\, p}^2}
\simeq\sqrt{\epsilon^2+\Delta^2}$ is small.
Under the conditions (\ref{hi040}), the approximation
\be\label{hi080}
{\rm e}^{\displaystyle -\bb({e_p}-\Delta)}\simeq
{\rm e}^{\displaystyle -\frac{\beta\epsilon^2}{2\Delta}}\,
(1+\frac{\bb\epsilon^4}{8\Delta^3})\simeq
\sqrt{\frac{2\pi\Delta}{\beta}}\left[
\delta(\epsilon)\left(1+\frac{3}{8\bb\Delta}\right)+
\frac{\Delta}{2\beta}\,\delta^{\prime\prime}(\epsilon)\right]\q,
\ee
holds. Hence the expression 
(\ref{hi070}) becomes
\be\label{hi090}
\log\,r_0=-4\,{w_{\rm F}}\,
\sqrt{\frac{2\pi
\,\Delta}{\beta}}\,
{\rm e}^{\displaystyle -\beta\Delta}
\,\,\left(1+\frac{3}{8\bb\Delta}\right)
\q.
\ee

We can then
evaluate the {\it odd-even grand potential difference}
for given values of $T$ and $\mu$,
\be\label{hi100}
{\delta\FI_{\rm G}}\equiv-\frac{1}{\bb}\,(\log\YB_{-} -\log\YB_{+})\q,
\ee
with good accuracy in terms of
the BCS solution. From Eqs.~(\ref{he280}) and (\ref{hi090}), 
neglecting the contribution $r_0(k^1_-+k^1_+)$, we obtain
for $T$ in the range (\ref{hi040})
\be\label{hi110}
\frac{\delta\FI_{\rm G}}{\Delta}
\simeq-\frac{1}{\bb\Delta}\,\ln\frac{1-r_0}{1+r_0}\simeq
1-\frac{T}{2\Delta}\,
\log\left(8\pi({w_{\rm F}}\Delta)^2\,\frac{T}{\Delta}\right)\q.
\ee
Ref.~\cite{LJE93} reports on a
direct measurement of the difference $\delta\FI$ between
the {\it free energies} of neighbouring odd and even systems 
for small superconducting metallic islands
and shows that it agrees remarkably well with 
the approximation (\ref{hi110}). We shall return to this point,
and to
the difference between $\delta\FI_{\rm G}$
and $\delta\FI$ in Sect.~6.5.3.

To proceed further, and in particular to deal with very low temperatures
$\beta\gg w_{\rm F}$, for which (\ref{hi040}) no
longer holds, we consider separately 
the projections on an even or odd number of particles. 
Let us begin with the even case.
 
\newpage
\noindent{\it Even Particle Number}

Using (\ref{hi070}), the expansion of Eq.~(\ref{hf230}) yields:
\be\label{hi120}
\Delta_{+\, p}=\frac{1}{2}\,\sum_{q\,(q\ne p)}\,\Gbcspq
\frac{\Delta_{+\, q}}{\ebcstq}\,
\left(1-2{\rm e}^{\displaystyle -2\bb\ebcsq}-
4\sum_{r\,(r\ne p,q)}
{\rm e}^{\displaystyle -\bb(\ebcsq+{e}_r)}\right)
\q.
\ee
In the low-temperature limit $\bb/w_{\rm F}\gg 1$,
Eqs.~(\ref{hi010}) and (\ref{hi120})
are identical. In Eq.~(\ref{hi120}),
the terms which determine the low-temperature
behaviour of the gaps are negative. However, they
are now of the order of, or smaller than,
$\exp{(-2\bb\Delta)}$ while they were of the order of
$\exp{(-\bb\Delta)}$ for BCS.
This reflects the fact that any excitation with fixed $N$-parity requires
the creation of at least two quasi-particles and therefore 
an energy larger than $2\Delta$ (instead of $\Delta$ for BCS).
Therefore the decrease of the pairing gaps with 
increasing temperature
will be slower than for the BCS case.
The low temperature behaviour of the
entropy (\ref{hf260}) is now given by
\be\label{hi130}
\SIzb_+=2\,\sum_p\,\bb\ebcsp\,
{\rm e}^{\displaystyle -2\bb\ebcsp}+
2\,\sum_{p\,q\,(p\ne q)}\,\bb(\ebcsp+\ebcsq)\,
{\rm e}^{\displaystyle -\bb(\ebcsp+\ebcsq)}
\q.
\ee
It is also governed by terms smaller than $\exp{(-2\bb\Delta)}$.
Under the conditions (\ref{hi040}), the gap equation (\ref{hi120})
provides an analytical expression for the low-temperature dependence
of the positive-parity-number gap $\Delta_+(\beta)$ :
\be\label{hi140}
\Delta_+(\beta)=\Delta\,\left[1-4\pi\,\frac{{w_{\rm F}}}{\bb}\,
{\rm e}^{\displaystyle -2\bb\Delta}\right]\q.
\ee
The extrapolation to zero temperature of $\Delta_+$ 
is therefore equal to the
BCS gap $\Delta$. 
Under the same conditions, one can can rewrite the
expression (\ref{hi130}) as
\be\label{hi150}
\SIzb_+=8\pi\,({w_{\rm F}}\Delta)^2\,
{\rm e}^{\displaystyle -2\bb\Delta}\q.
\ee 
The BCS entropy (\ref{hi060})
is therefore larger
than $\SIzb_+$. Indeed, the disorder 
decreases when odd-particle components are removed.

Terms proportional to
$\exp{(-2\bb\Delta)}$ occur in other thermodynamic 
quantities. For instance, for the
average particle number, the limit of large $\bb$  is
\be\label{hi160}\displaystyle
\langle\NDh\rangle=\sum_p\,\left[\,
1-\frac{\epbcsp-\mu}
{\ebcstp}\,\left(1-2
{\rm e}^{\displaystyle -2\bb\ebcsp}
-4\sum_{q\,(q\ne p)}\,
{\rm e}^{\displaystyle -\bb(\ebcsp+\ebcsq)}
\right)\,\right]
\q.
\ee

The analysis of Eqs.~(\ref{hf210}), (\ref{hf220}) and (\ref{hf240})
shows that, in the large $\beta$ limit, 
the energy $\ebcsp$ of a state in the vicinity
of the Fermi surface is smaller than its BCS counterpart.
The correction depends on the 
two degenerate quasi-particle states
with lowest energy. Hereafter we shall
denote $\Delta_0$ and $\ebcstz$ the corresponding 
gap and energy.
Assuming moreover that all the
matrix elements $\Gbcspq$ ($p\ne q$) have the
same value ${\tilde G}$, one finds
\be\label{hi170}
\ebcsp=\ebcstp-\frac{\tilde G}{6}\,\frac{\Delta_{+\, 0}}{\ebcstz}
\,\left\{
(1-\delta_{p\,0})\,(3\frac{\Delta_{+\,p}}
{\ebcstp}-\frac{\Delta_{+\,0}}{\ebcstz})
+\delta_{p\,0}\,\frac{\Delta_{+\,0}}{\ebcstz}
\right\}\q.
\ee
Actually, in the numerical analysis of Sect.~6.5, 
we shall find that  
$\ebcsp$ and $\ebcstp$ differ by only few percent. 

\noindent{\it Odd Particle Number}

Let us now study the case of
an odd system. It is useful to compare our results with
those obtained from two different types of BCS approximation. 
The first one (hereafter labeled BCS) corresponds to
the standard equation with the usual constraint on
the odd {\it average} number of particles.
The second one (labeled BCS1) corresponds to
a situation in which there is {\it exactly one particle} 
on the twofold-degenerate individual level 
whose energy $\epsilon$
is closest to the Fermi energy $\mu$.
If we label this level with the index $p=0$,
the BCS1 solution is obtained by effecting 
in all sums the substitution 
$(u_0,v_0)\rightarrow(v_0,-u_0)$ for {\it only one}
of the two quasi-particles states $0$ or $\bar0$ 
associated with $\epsilon_0$.
Under this transformation the
product $u_0v_0$ changes sign and
the contributions of these two quasi-particles states
to the gap 
cancel exactly. Thus for $p\ne 0$,
the BCS1 gap equation at arbitrary temperatures is
\be\label{hi180}
\Delta_{{\rm BCS1}\, p}=\usd\sum_{q\,(q\ne 0)}\,\Gbcspq\,
\frac{ \Delta_{{\rm BCS1}\,q}\,\tbcsp}{\ebcstq}
\q.
\ee
The only difference from the usual BCS equation
is the absence of the $q=0$ term in the sum.
However, through self-consistency, this affects
the value of the gaps.
(A constraint on the odd value of the 
average total number of particles, which includes 
the single particle occupying the level $0$,
must of course be enforced.) 
In the BCS1 solution, the particle which occupies the 
state $p=0$ (or ${\bar p}={\bar 0}$)
forbids the formation of a pair
$(0,\,{\bar 0})$.
Within the single-particle space from which 
the levels $0$ and $\bar 0$
are omitted, the variational pairing equations must then 
be solved for the (even) system with one fewer particle.
We shall see (Fig.~1) that this results in a sizeable difference
between BCS and BCS1.

Returning to the odd-number projection, we have to
treat separately the gaps $\Delta_{-\,p}$ for the quasi-particles 
$p\ne 0$ and the gap $\Delta_{-\,0}$ for $p=0$.
Let us first consider the case $p\neq 0$. 
At low temperature both the numerator and the denominator 
in the fraction in the right hand side of Eq.~(\ref{hf230})
vanish as $\exp{(-\bb e_0)}$.
After dividing out by this common factor, we
obtain the well-conditioned equation:   
\be\label{hi190}
\Delta_{-\,p}=
\frac{1}{2}\,\sum_{q\,(q\ne 0)}\,\left\{
\Gbcspq\,\frac{\Delta_{-\,q}}{\ebcstq}
+\left(
G_{p0}\frac{\Delta_{-\,0}}{\ebcstz}-\Gbcspq\frac{\Delta_{-\,q}}{\ebcstq}
\right)
{\rm e}^{\displaystyle -\bb(\ebcsq-e_0)}
\right\}
\q.
\ee 
The zero-temperature limit of Eq.~(\ref{hi190})
differs from that of BCS, but is
identical with the limit of (\ref{hi180}). 
This implies that, as
$\bb\rightarrow\infty$,
Eq.~(\ref{hf230}) automatically singles out the level
with the lowest quasi-particle energy and
generates the odd-particle-number
ground state given by BCS1.
The low-temperature correction to the gap 
$\Delta_{-\,p}$ is positive
and depends on terms which are proportional to 
an exponential of the
difference between a quasi-particle energy $e_q$ and $e_0$.
Therefore, starting from the value provided by the BCS1 equation
for $\bb\rightarrow\infty$, 
one expects the gap to first {\it increase} 
with temperature as $\exp{(-\bb/ 2{w_{\rm F}}^2 \Delta)}$.

The case $p=0$ is slightly more
complicated. One finds that the temperature corrections 
involve no longer
$e_0$ but the second smallest quasi-particle energy
$e_1$, the corresponding gap $\Delta_{-\,1}$ and degeneracy $g_1$ :
\be\label{hi200}\ba{rl}
\Delta_{-\,0}=&\displaystyle
-\frac{1}{2}\,G_{01}\,\frac{\Delta_{-\,1}}{{\tilde e}_1}\\
&\q\displaystyle
+\frac{1}{2}\,\sum_{q\,(q\ne 0)}\left\{
\,G_{0q}\,\frac{\Delta_{-\,q}}{\ebcstq}
-\frac{1}{g_1}\,\left(
G_{0q}\,\frac{\Delta_{-\,q}}{\ebcstq}-G_{01}\,\frac{\Delta_{-\,1}}{{\tilde e}_1}
\right)
{\rm e}^{\displaystyle -\bb(\ebcsq-e_1)}
\right\}
\q.
\ea\ee
Despite their formally different expressions, we shall find 
in the numerical application performed in Sect.~6.5
that the values of $\Delta_{-\,0}$ and $\Delta_{-\, p}$
are equal within a few percent.  
In nuclear physics, one often encounters situations
where $1/\Delta$, $\beta$ and ${w_{\rm F}}$ are of the same
order of magnitude.
Then, the {\it very low-temperature entropy} (\ref{hf260}) behaves as  
\be\label{hi210}
\SIzb_-=\log2+\sum_{p\,(p\ne 0)}\,\bb\,(\ebcsp-e_0)\,
{\rm e}^{\displaystyle -\bb(\ebcsp-e_0)}
\q.
\ee
When $\bb\rightarrow\infty$, the entropy $\SIzb_-$ is equal to $\log2$, 
a value which reflects the twofold degeneracy of the
ground state, and to which we shall return
in Sect.~6.5.4.

Under the conditions (\ref{hi040}), 
using Eqs.~(\ref{hi080}) and (\ref{hi090}),
one can reformulate Eq.~(\ref{hf230}) as
an equation for the
($p$-independent) odd-number gap $\Delta_-(\beta)$ :
\be\label{hi220}
\frac{2}{\tilde G}=\int_{-\Lambda/2}^{\Lambda/2}\,
\frac{{w_{\rm F}\,{\rm d}\epsilon}}{\sqrt{\epsilon^2+\Delta_-(\beta)^2}}
-\frac{1}{\Delta_-(\beta)}
+\frac{1}{2\beta\Delta_-(\beta)^2}
\q.
\ee
The limit of this equation, in the regime (\ref{hi040}), yields
\be\label{hi230}
\Delta_-(\beta)=\Delta\,\left[\,
1-\frac{1}{2{w_{\rm F}}\Delta}+\frac{1}{4\beta{w_{\rm F}}\Delta^2}
\,\right]\q.
\ee
At zero temperature the odd-$N$ gap $\Delta_-$ is therefore
smaller than the BCS gap by the quantity $1/2{w_{\rm F}}$.
This point has already been 
mentioned in connection with the discussion
of the low-temperature
gap equation (\ref{hi190}) and its relation with the BCS1
Eq.~(\ref{hi180}).
In the temperature range defined by the
conditions (\ref{hi040}),
the gap grows linearly with temperature. We have already noted 
that it grows as $\exp(-\beta/2{w_{\rm F}}^2\Delta)$
when $\beta$ is larger than ${w_{\rm F}}$, 
or of the same order.
The odd-$N$ projected thermodynamic quantities 
have a larger temperature variation than 
the corresponding BCS ones. 
This results from
the fact that excitation energies 
are then of the order ${w_{\rm F}}^{-1}$ and
therefore much smaller than $\Delta\,$.

For the entropy (\ref{hf260}) and for {\it low temperatures 
satisfying the conditions} (\ref{hi040}), we obtain~:
\be\label{hi240}
\SIzb_-=\log2 +\frac{1}{2}+
\usd\log \left(\frac{2\pi\,{w_{\rm F}}^2\Delta}{\beta} \right)\q.
\ee
This is simply the expression of the {\it Sakur-Tetrode
entropy} \cite{Bal91a} {\it for one classical particle} of mass $\Delta$
with an internal twofold degeneracy,
enclosed in a one-dimensional box of length
$2\pi\hbar\,{w_{\rm F}}$.
Such a result is natural since the elementary
excitations concern only a single
unpaired particle, with energy
$e_p\simeq\Delta+\epsilon^2/2\Delta$;
if we identify $\epsilon=\epsilon_p-\mu$
with the momentum of a particle in a box,
the identification of the level densities leads to
the size $2\pi\hbar\,{w_{\rm F}}$ for this box.
While elementary excitations of even-$N$ systems require energies
of at least $2\Delta$, the effective spectrum for odd-$N$ systems is
nearly continuous, 
with  the form $m^2/2w_{\rm F}^2\Delta$ where $m$ is an
integer. This occurence of
{\it gapless elementary excitations} is consistent
with recent experiments on the conductance of odd-$N$ ultrasmall
Aluminium grains\cite{RBT97}.
The analogy with the Sakur-Tetrode entropy shows also
that in the variational projected solution, contrary to the
BCS1 one, the unpaired particle does not remain on
the single-particle level
of energy $\epsilon_0=\mu$. The projected solution is a
coherent superposition of configurations in which this particle
explores all single-particle levels consistent
with the temperature. 
The fact that we recover
a formula of classical thermodynamics is another indication
that, strictly speaking, the regime associated 
with the conditions (\ref{hi040}) is  
not a low temperature regime.
Actually, (\ref{hi040}) implies that the temperature is sufficiently
large so that ${w_{\rm F}}^2\Delta/\bb\gg 1$, which ensures the 
positivity of the entropy $\SIzb_-$ as given by Eq.~(\ref{hi240}).

For the {\it average number of particles},
the low temperature limit of Eq.~(\ref{hf270}) yields 
\be\label{hi250}
\langle\NDh\rangle=1+\sum_{p\,(p\ne 0)}\,
\left[\,1-\frac{\epsilon_p-\mu}{\ebcstp}\left(1-2
{\rm e}^{\displaystyle -\bb(\ebcsp-e_0)}\right)
\,\right]\q.
\ee
In contrast to the corresponding BCS formula, the
temperature correction depends also on the difference
$\ebcsp-e_0$.

Using Eqs.~(\ref{hf210}), (\ref{hf220}) and (\ref{hf240})
one finds, in the limit
$\bb/w_{\rm F}\gg 1$, that $\ebcsp$ is larger than
$\ebcstp$. Taking again all the
matrix elements $\Gbcspq$ ($p\ne q$) equal to the
same value $\tilde G$, one has
\be\label{hi260}
\ebcsp=\ebcstp+\frac{\tilde G}{2}\,\left\{
(1-\delta_{p\,0})\,\frac{\Delta_{-\,p}}{\ebcstp}
+\delta_{p\,0}\,\frac{\Delta_{-\,1}}{{\tilde e}_1}\right\}
\q.
\ee
To derive this formula, we have taken $g_1=2$ as
in the numerical examples presented in Sect.~6.5. 
In these applications,
we shall find that the magnitude of $\ebcsp-\ebcstp$ is small 
(a few percent). 

The inequalities $\SIzb_->\SI_{\rm BCS}>\SIzb_+$
implied by Eqs.~(\ref{hi060}), (\ref{hi150}),  
(\ref{hi210}) and (\ref{hi240}) are easily understood. 
When the system is 
constrained to have an even particle number, 
we have seen that its lowest excited states
are obtained from the ground state by 
the creation of two quasi-particles, which requires at least an
energy $2\Delta$. For an odd system, the excitation of the
(gapless) unpaired last particle is much easier.
At non-zero temperature the BCS state 
is a weighted sum over even and odd components.
The concavity of entropy implies that
$\SI_{\rm BCS}$ is larger than the smallest 
of the quantities $\SIzb_+$ and $\SIzb_-$, that is $\SIzb_+$. 
The inequality $\SI_{\rm BCS}<\SIzb_-$ is made possible by
the very small weight of the odd components entering 
the BCS state at low temperatures.

The smallness of this weight also explains why, 
as already noted in Sect.~6.3.3, the odd-$N$ projection
affects the BCS results more than the even one.
Indeed, we have already met with several qualitative differences
between the properties of odd-systems and their BCS description, 
while BCS explains fairly well the even systems. This point will
also be exhibited by the numerical applications of Sect.~6.5, 
especially on Figs.~1 and 6 which show the gaps at the
Fermi surface as a function of temperature.

\subsubsection[S642]{High Temperatures}

High temperatures relevant for superconductivity
are of the order of the critical temperature, that is
$T_c=0.57\Delta$. General conclusions on the behaviour
of the equations can only be 
obtained when the density of single-particle
states at the Fermi surface
is significantly larger than one, which entails 
${w_{\rm F}}\Delta\gg 1$.
In this limit, the quantity 
$\prod_{r\neq p,\,q}\,t_r^{2}\,$
tends towards 0 for any value of $p$ and $q$. Then, 
Eq.~(\ref{hf230})
is equivalent to :
\be\label{hi270}
\Delta_{\eta\, p}=\frac{1}{2}\,\sum_{q\,(q\ne p)}\,\Gbcspq
\frac{\Delta_{\eta\, q}}{{\tilde e}_q}\tbcsq\,
\left\{1+\eta\,(1-\tbcsq^2)\prod_{r\neq p,\,q}\,t_r^{2}\right\}
\q.
\ee
This differs from the BCS equation by the factor within the braces,
which deviates from 1 by a vanishingly small correction.
The projected gap values converge (from above when $\eta=1$ 
or from below when $\eta=-1$) towards the BCS values as one reaches
the critical temperature. In particular, the odd solution
converges towards the standard BCS solution and {\it not} towards
BCS1.

In Sect.~6.5.4, we present a numerical case
which does not pertain to the limit ${w_F}\Delta\gg 1$
and yields qualitatively different results.

\subsubsection[S643]{Large Systems}

In large systems, at non-zero temperatures, 
the product $r_0$ decreases as the inverse 
of an exponential of the volume of the system.
Then, an iterative solution of 
Eqs.~(\ref{hf210}), (\ref{hf230}), (\ref{hf290}) and (\ref{hf240}) 
is attainable by starting from the BCS solution. 
Since in several places $r_0$ appears multiplied by 
$\coth^2(\bb\ebcsp/2)$, the effect
of the projection is surmised to be especially sizeable near the
Fermi surface. 

In this large system limit, 
the approximate projected entropy (\ref{hf260}),
for both values of $\eta$,
differs from the BCS entropy by a constant:
\be\label{hi280}
{\SIzb}_\eta=\SI_{\rm BCS}-\log2\q.
\ee
This result, shows that in large systems (for a given
 temperature) the numbers of states
with odd and even $N$-parity
are equal.

The quantity $\ksu$ increases linearly with the volume,
so that $e_p-{\tilde e}_p$
is expected to be dominated by the 
second term in the r.h.s. of Eq.~(\ref{hf210}). 
Then, at the lowest non-trivial order, the quasi-particle energy
can be expressed as 
\be\label{hi290}
\ebcsp\simeq\ebcstp
+\frac{\displaystyle \eta\,r_0\,\qbcsp
}{\displaystyle 2}\,\sum_{q\,q'}\,G_{q\,q'}
\frac{\displaystyle
\Delta_{\eta\, q}\,(\qbcsq-\tbcsq)
}{\displaystyle
\ebcstq
}
\frac{\displaystyle
\Delta_{\eta\, q'}\,(t_{q'}^{-1}-t_{q'})
}{\displaystyle
{\tilde e}_{q'}
}\q.
\ee
The self-consistency and the constraint on the particle-number
moreover shift $\mu$ and $\Delta_{\eta\, p}$ away from their BCS values.

\subsection[S65]{The Projected BCS Model : A Numerical Illustration}

To evaluate the corrections to BCS introduced by the 
number-parity projection,
we have performed three schematic calculations that exemplify
situations encountered in the physics of (i) superdeformed
nuclear states around mass $A=190$, 
(ii) superconducting 
mesoscopic metallic islands and 
(iii) ultrasmall metallic grains. 

\subsubsection[S651]{Superdeformed Heavy Nuclei}
We first consider an application to a field of nuclear structure physics
which, recently, has been the subject 
of an intense experimental exploration.
Superdeformed nuclear states are generally
populated via fusion reactions between two heavy ions.
Once the colliding nuclei 
have fused and a few neutrons have evaporated, 
the compound nucleus decays with a small probability (a few percents)
into superdeformed
metastable configurations via the emission of 
statistical $\gamma$-rays.
In heavy nuclei, the angular momentum dependence
of the moment of inertia observed in superdeformed bands gives 
strong evidence of the presence of pairing correlations.
Most of the presently available data concern 
the properties of superdeformed ground states or of the lowest excitations.
However, it is expected that
the next generation of $\gamma$-arrays, in Europe and in the US, will
soon allow investigations of the statistical behaviour of 
the spectrum and of the properties of heated nuclei.

In view of the general scope of the present work, rather than 
studying a realistic nuclear configuration 
(see for instance Ref.\cite{GFB94}) that
would require the solution of the full extended HFB equations
written in Sect.~6.2, 
we prefer to investigate a 
simple model in which the nuclear context
outlined above is used as a guidance for
selecting the relevant parameter values.
Thus we consider twofold degenerate single-particle levels
of energies $\epbcsp$, regularly spaced
with a density ${w_{\rm F}}=$2MeV$^{-1}$. Our
active pairing space includes 21 levels and 
extends 5 MeV above and 5 MeV below the Fermi surface,
so that the cut-off $\Lambda$ equals 10 MeV.
The matrix elements of the interaction are taken to have the form
$\Gbcspq={\tilde G}\,(1-\delta_{pq})$ with
${\tilde G}=0.23$MeV. For this value of $\tilde G$,
the BCS pairing gap $\Delta$ at zero temperature 
is 1MeV, consistent with the data for the even heavy nuclei.
In fact, it is partly because $\Delta$ is close to 1MeV
in medium and heavy nuclei that
the MeV turns out to be a natural energy unit
in nuclear dynamics despite the fact that binding energies
are several orders of magnitude larger.

For these values of the parameters $w_{\rm F}$, $\Lambda$
and $\tilde G$ (or equivalently $\Delta$), we have solved
the $N$-parity projected coupled equations (\ref{hf210}-\ref{hf240})
with $\eta=\pm 1$ as function of the temperature.
This provides us with the projected pairing gap $\Delta_{\eta\, p}$
defined by Eq.~(\ref{hf170}). Although our interaction is very simple,
it leads to state-dependent gaps
for the BCS as well as for the projected cases. 
The numerics, however, shows that this state 
dependence is weak:
we find that at all temperatures the
variation of the gap $\Delta_{\eta\, p}$ versus $p$ never exceeds 9\%. 
We thus content ourselves
with plots of $\Delta_{\rm F}$, that is, of the value
of the gap associated with the quasi-particle having the 
lowest energy in either the even or odd system.

The quantity $\Delta$ which serves as a reference energy in all
the figures of the present Section, 
both for the temperature in abscissa
and for the gap $\Delta_{\rm F}$ in ordinates, is likewise the 
zero-temperature BCS gap associated with the lowest energy
quasi-particle.

We first consider (upper part of Fig.~1) the {\it even-number}
case with $\langle\NDh\rangle=20$.
(Here, $\langle\NDh\rangle$
should not to be confused with the numbers of neutrons
and protons which, for a heavy  nucleus, are of the order of 120
and 80, respectively.
$\langle\NDh\rangle$ corresponds to the particles
within the active pairing space; it does not take into account
the nucleons of the fully occupied levels whose energies are smaller
than $\mu-\Lambda/2$.)
We have plotted the temperature dependences of the BCS gap 
and of the projected gap $\Delta_{\rm F}$.
As expected from the discussion of Eq.~(\ref{hf230}) in Sect.~6.3.3
the even-$N$ projected gap is 
always larger than the BCS one. The two curves meet
at $T=0$ and also at the critical temperature $T_{\rm c}=0.57\Delta$,
consistent with the analysis of Sect.~6.4.2.
\begin{figure}[hp]\label{F1}
\begin{center}
\includegraphics*[scale=0.6,angle=-90.]{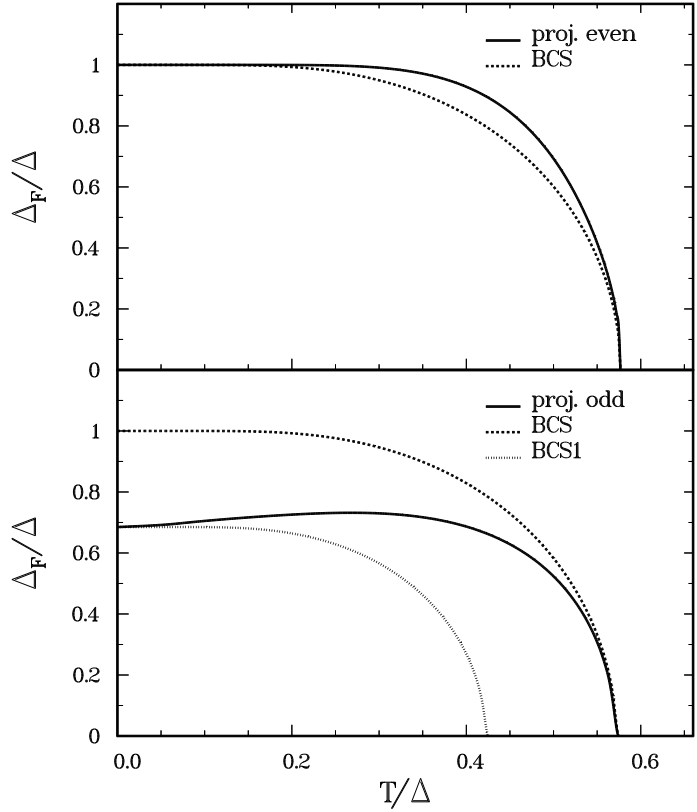}
\caption[Fig1]{
Comparison of BCS and number-parity projected gaps at the Fermi
surface versus temperature for ${w_{\rm F}}\Delta= 2$,
${w_{\rm F}}\Lambda= 20$ and $\langle\NDh\rangle=20$ or 21.
These parameter values are illustrative of the physics of heavy nuclei.
The upper (respectively lower) part of
the figure corresponds to an even (respectively odd) average number of
particles. With respect to BCS, the pairing gap is enhanced 
by even-$N$ projection and
(more strongly) reduced by odd-$N$ projection, except
near $T_{\rm c}$. For the BCS1 curve, see Sect.~6.4.1 and 6.5.1.
The odd-$N$ projected gap increases with $T$ up to 0.4$\Delta$.
}
\end{center}
\end{figure}

The lower part of Fig.~1 presents results for the
{\it odd-number} case with $\langle\NDh\rangle$=21.
It compares the temperature-dependent gaps obtained with our
odd-$N$ projected method, with
BCS and with BCS1. As explained in
Sect.~6.4.1, in the BCS1
option the level $p=0$ (with $\epsilon_0=\mu=0$)
is always occupied by exactly one nucleon and therefore not
available for pairing.
We have therefore
chosen to show in all three cases the gap $\Delta_{\rm F}$
associated with the lowest-energy quasi-particle
whose index is {\it distinct from} $0$.
The odd-$N$ projected curve always lies
between the BCS and BCS1 ones.
In agreement with the discussion of Sect.~6.4.2,
the projected curve joins
BCS1 at $T=0$
and BCS at the critical temperature.
The failure of BCS at low temperature
is associated with the small weight
of the odd-$N$ components in the BCS state.
The BCS1 approximation, by enforcing the odd parity at zero 
temperature, approaches at low temperature the projected state 
better than does BCS.
On the other hand, when the temperature increases, 
the even and odd components tend to have equal weights, 
while the differences between even and odd systems fade out.
In that limit the BCS1 approximation becomes inadequate
since it prevents the formation of
a pair in the level $\epsilon_0$.
This inhibits a correct description of 
pairing correlations which, at high temperature,  
take place coherently on all levels in 
the energy range of the gap for the odd as well as for the even systems.

When the temperature grows, the odd-$N$ projected gap 
{\it begins to increase} as expected from (\ref{hi190}-\ref{hi230}), 
then reaches a maximum,
decreases and converges asymptotically towards the BCS curve
near the critical temperature. 
At low temperatures, there is a significant difference between the 
usual BCS gap and $\Delta_{\rm F}$. 
In nuclear physics, this difference 
should be taken into account when mass differences 
between odd and even neighbour
nuclei are used to extract pairing gap values.

In the numerical example of Fig.~1, the condition 
${w_{\rm F}}\Delta\gg 1$ is not satisfied.
It is therefore not surprising that the results 
do not comply with Eqs.~(\ref{hi140}) and (\ref{hi230})
derived under the conditions (\ref{hi040}).

When gaps and temperatures
are measured in $\Delta$ units, 
the BCS curve is universal up to small differences
resulting from the details of the spectrum.
For instance, the critical temperature always comes out equal 
(with an accuracy better than 1\%) to the value 
$0.57\Delta$ which is derived for a
continuous spectrum with constant level density.
In contrast, as anticipated in Sect.~6.4, 
the number-parity projected
curves show more diversity.
In our model, their characteristics 
are mostly determined by 
the product ${w_{\rm F}}\Delta$ 
and to a lesser degree by the product $w_{\rm F}\Lambda$ and the
ratio $\Lambda/\Delta$.
The quantity ${w_{\rm F}}\Delta$, ``the number of Cooper pairs'',
is the number 
of twofold degenerate single-particle levels in an
energy interval equal to the gap, while $w_{\rm F}\Lambda$
is the number of levels in the active pairing space.
In the nuclear physics case of Fig.~1
we have taken ${w_F}\Delta= 2$ and $\Lambda/\Delta= 10$,
the number $w_{\rm F}\Lambda$ of levels being 21.  
The data corresponding to condensed matter physics
are expected to be quite different.

\subsubsection[652]{Mesoscopic Metallic Islands}

For superconducting materials, $\Lambda$ is of the order of the
Debye energy while $\Delta$ is of the order of the critical temperature,
so that $\Lambda/\Delta$ is larger than above.
On the other hand, depending
on the size of the metallic islands or grains, such as
those considered
in recent experiments\cite{LJE92,RBT95,DEU94,BRT96}, 
the product ${w_{\rm F}}\Delta$ can vary between
0.5 and several thousands.

Let us first consider the case
${w_{\rm F}}\Delta\gg 1$,
which corresponds to most of the
experimental studies performed on
superconducting metallic islands.
In this limit the effects
on the pairing gap $\Delta_{\rm F}$
of the number-parity projection are weak.
In order to be able
to visualize them, we have considered 
in Fig.~2 not too large a
value of ${w_{\rm F}}\Delta$, taking ${w_{\rm F}}\Delta=10$ with
$\Lambda/\Delta=20$ and hence $w_{\rm F}{\tilde G}=0.34$.
Thus, the active space comprises 201 twofold-degenerate levels.
The average particle numbers of the even and odd systems are
200 and 201, respectively.
The BCS and even-$N$ projected curves for $\Delta_{\rm F}$
are almost indistinguishable.
In agreement with the discussion of the projected gap
equation in Sect.~6.3.3, the
odd-$N$ projected gap differs from the BCS value
more than the even-$N$. 
At low temperature, it
increases and is represented with good accuracy
by the linear $T$-dependence of Eq.~(\ref{hi230}).
\begin{figure}[hp]\label{F2}
\begin{center}
\includegraphics*[scale=0.6,angle=-90.]{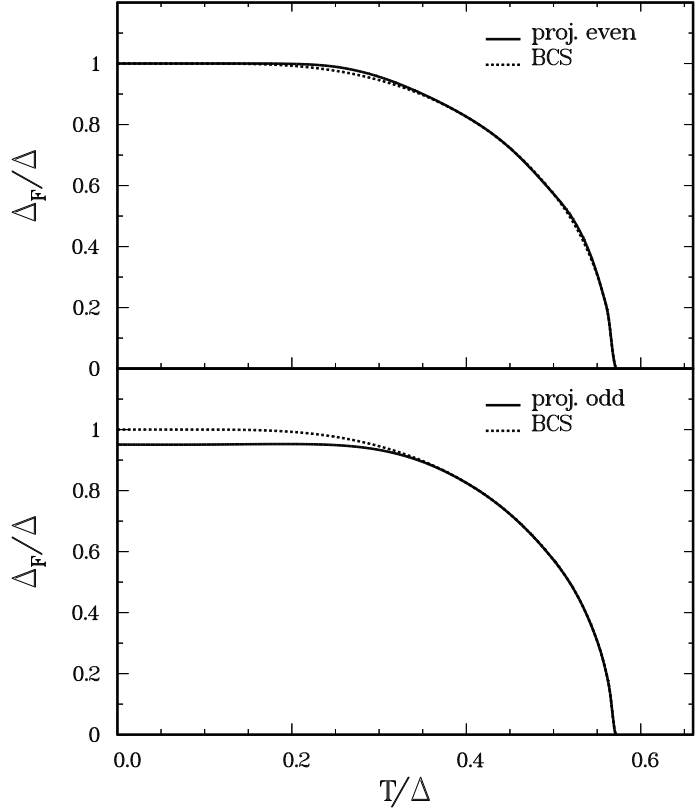}
\caption[Fig2]{Comparison of BCS and number-parity
projected gaps at the Fermi
surface versus temperature for $w_{\rm F}\Delta= 10$,
$w_{\rm F}\Lambda=200$ and $\langle\NDh\rangle=200$
(upper part) or 201 (lower part).
Such parameter values are illustrative
of superconducting mesoscopic islands.
The effects are qualitatively the same as in Fig.~1 but weaker. Even-odd
effects disappear above $T_\times=0.3\Delta$.
}
\end{center}
\end{figure}

In Fig.~3 we have plotted the {\it entropy of the odd system}
calculated in four different ways : with or without (i.e. ${\tilde G}=0$)
pairing interaction and with or without projection
on number parity.
The odd-$N$ projected entropy $\SIzb_-$
tends to $\log 2$ when $T\rightarrow 0$;
above the critical temperature it joins the non interacting
odd-$N$ projected entropy.
In contrast, the BCS entropy vanishes at small temperature;
above the critical temperature,
it joins the grand-canonical entropy of the non-interacting
system.
For $T$ larger than $0.35\Delta$ the odd-$N$ projection
lowers the BCS entropy, according to $\SI_{\rm BCS}-\SIzb_-=\log2$.
In contrast, at small $T$, one has
$\SIzb_-\gg\SI_{\rm BCS}$ and the projected entropy $\SIzb_-$
approaches the Sakur-Tetrode regime (Eq.~(\ref{hi240})).
Although we are working with a discrete spectrum, for
$T/\Delta\ge 0.1$,
the grand-canonical
unprojected entropy of the non-interacting system  
is almost undistinguishable from the 
expression 
\be\label{hj010}
S=\frac{2\pi^2}{3} w_{\rm F}\,T\quad,
\ee
obtained by assuming
a continuous spectrum with constant level density.
The only effect of the odd-$N$ projection 
is again a lowering of the entropy by $\log2$.
\begin{figure}[hp]\label{F3}
\begin{center}
\includegraphics*[scale=0.5,angle=-90.]{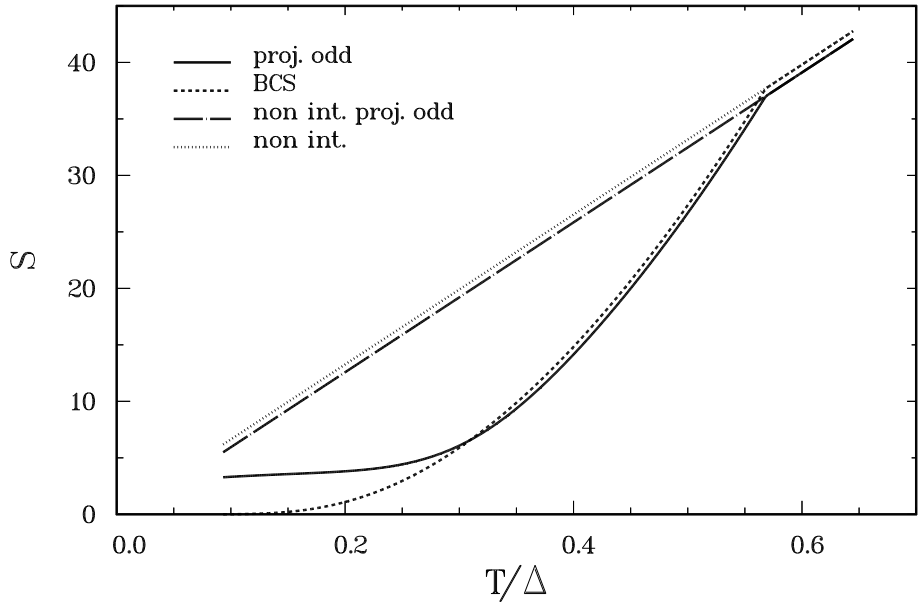}
\caption[Fig3]{
Entropy of an odd-number system for $w_F\Delta=10$,
$w_{\rm F}\Lambda=200$ and $\langle\NDh\rangle=201$.
The solid curve corresponds to the variational solution with
projection on an odd number of particles.
Around $T=0.1\Delta$ it has a Sakur-Tetrode form and it tends to
$\log 2$ for $T\rightarrow 0$.
The dashed curve (BCS) is
obtained with the standard finite-temperature BCS formalism.
The dashed-dotted and dotted curves
are obtained by dropping the interactions.
The latter unprojected non-interacting case coincides 
with the entropy of the Fermi gas.
}
\end{center}
\end{figure}

In Fig.~4 we show the {\it specific heat}
$C_{\rm V}=
{\rm d}\langle\HDh\rangle/{\rm d}T=
T{\rm d}S/{\rm d}T$
{\it for the odd system}.
With or without projection, above the
critical temperature $T_{\rm c}=0.57\Delta$,
the specific heat of the interacting system
follows the linear form (\ref{hj010}).
Both the critical temperature and the magnitude of
the discontinuity of $C_{\rm V}$ are the
same for BCS and odd-$N$ projected curves.
The low temperature behaviour, however, is
affected by the projection.
The enlarged scale inset in Fig.~4
shows that the BCS specific heat drops to zero
at low temperature, while the odd-$N$ projected curve
approaches the value $\usd$ expected for a single classical
particle moving in a one-dimensional
box (see Sect. 6.4.1). As mentioned in Sect.~6.3.3, the odd-$N$ specific heat
$C_{\rm V}$ takes this value $\usd$
only over the intermediate range of temperatures
where the quasi-particle spectrum $e_p$ is well
approximated by a quadratic function of
$\epsilon_p-\mu$. In Sect.~6.5.4
we shall investigate in more details its zero-temperature
limit. In the absence of
pairing interaction $C_{\rm V}$ does not depend on
whether odd-$N$ projection is or is not effected,
and it is again given by the linear function (\ref{hj010}).
\begin{figure}[hp]\label{F4}
\begin{center}
\includegraphics*[scale=0.5,angle=-90.]{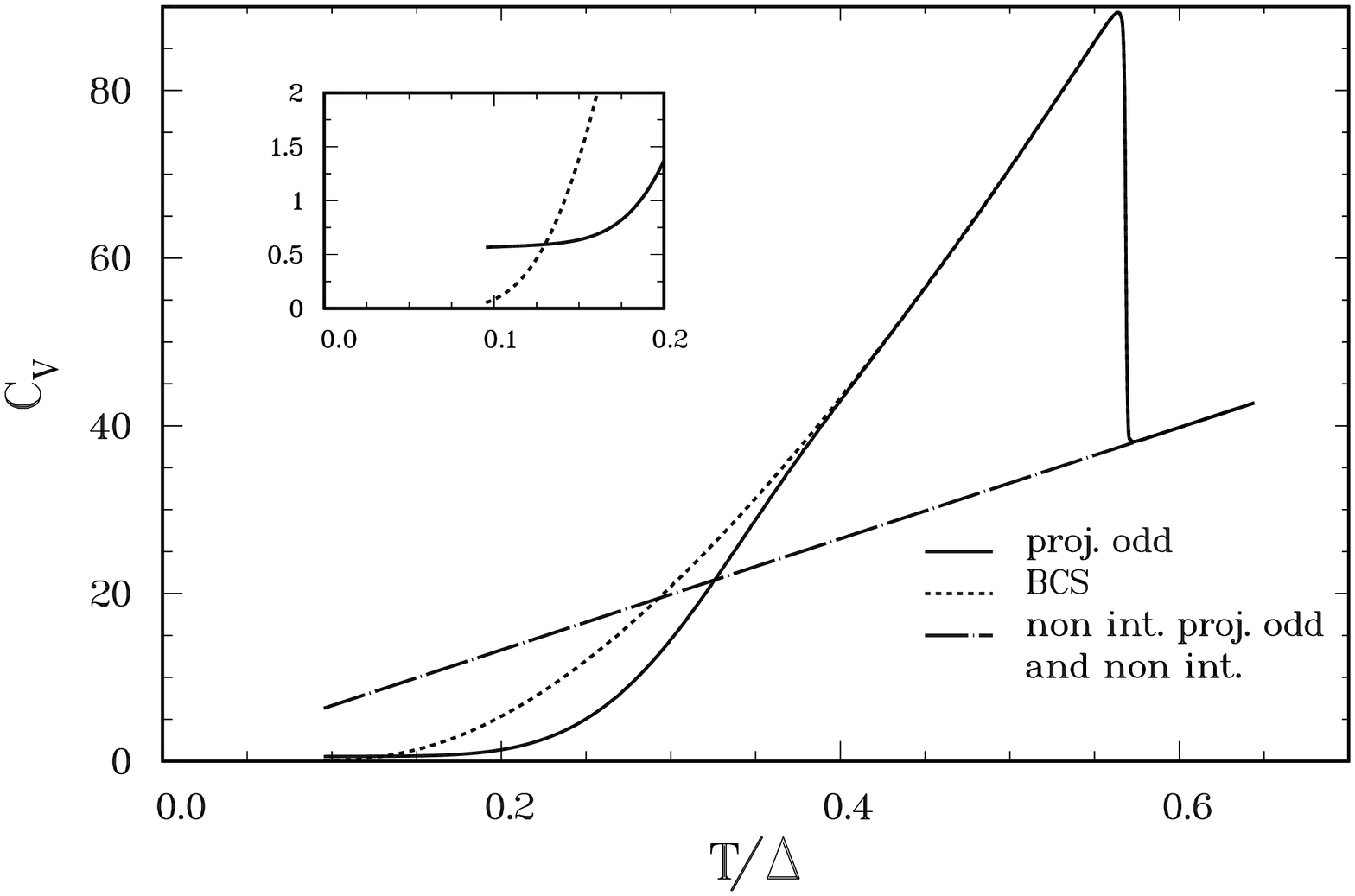}
\caption[Fig4]{
Specific heat of an odd-number system for $w_F\Delta=10$,
$w_{\rm F}\Lambda=200$ and $\langle\NDh\rangle=201$.
Symbols and abbreviations are the same as in Fig.~3.
The solid curve reaches the classical value for one
degree of freedom near $T=0.1\Delta$ while the
BCS value tends to zero, as exhibited by the inset.
The projected and unprojected curves with no interaction 
are identical and have the linear behaviour of the Fermi gas.
}
\end{center}
\end{figure}

\subsubsection[653]{The Free-Energy Difference}

We now consider a quantity which is experimentally
accessible: the difference $\delta F$ between the free energies
of the odd and even systems\cite{LJE93}.
Fig.~5 shows the variation with temperature of this quantity,
for the same parameters 
$w_{\rm F}\Delta=10$, $\Lambda/\Delta=20$ as in Figs.~2-4 (we have
taken as our unit scale $\Delta$ the value  
for the odd system $\langle\NDh\rangle$=201, which differs from
the even system value by less than 0.1\%).

In the BCS case, for a finite system in grand-canonical equilibrium,
the free energy $F_{\rm BCS}(\langle\NDh\rangle)$ is a function
of the expectation value $\langle\NDh\rangle$
which may vary continuously from $\langle\NDh\rangle=N_0\equiv 200$
to $\langle\NDh\rangle=N_0+1=201$. On the other hand,
$\partial F_{\rm BCS}/\partial\langle\NDh\rangle
=\mu(\langle\NDh\rangle)$
is the BCS chemical potential as function of
$\langle\NDh\rangle$ at the considered temperature.
Accordingly, we can obtain the difference
$\delta F_{\rm BCS}=
F_{\rm BCS}(N_0+1)-F_{\rm BCS}(N_0)$
by integration of $\mu(\langle\NDh\rangle)$ between $N_0$
and $N_0+1$. In the present model with $N_0+1$ equidistant
twofold degenerate single-particle levels,
for symmetry reasons, $\mu(N_0+1)$ is equal
to the energy of the level at mid-spectrum
(level 101) which we take equal to zero.
The Fermi level $\mu(\langle\NDh\rangle)$
lies halfway between two consecutive levels for even
$\langle\NDh\rangle$, and coincides with a level
for odd $\langle\NDh\rangle$ provided
the occupation number is close to 1 (resp.0) at the bottom (resp. top)
of the active space. Moreover, when $w_{\rm F}T\gg 1$
or $w_{\rm F}\Delta\gg 1$,
$\mu(\langle\NDh\rangle)$ is expected to vary smoothly, without
oscillations. Hence we find
\be\label{hj020}
\mu(\langle\NDh\rangle)\simeq\frac{1}{2w_{\rm F}}
(\langle\NDh\rangle-N_0-1)\q,
\ee
from which we derive the BCS difference of free energies:
\be\label{hj030}
\delta F_{\rm BCS}=
\int_{N_0}^{N_0+1}\mu(\langle\NDh\rangle){\rm d}\langle\NDh\rangle
\simeq-\frac{1}{4 w_F}\quad.
\ee
At this level of approximation one finds that
$\delta F_{\rm BCS}$ is temperature independent, a
result supported by the numerical evaluation, as illustrated by
Fig.~5. One observes moreover on Fig.~5 that for temperatures above 
0.4$\Delta$ the projected and BCS curves are identical.
\begin{figure}[hp]\label{F5}
\begin{center}
\includegraphics*[scale=0.5,angle=-90.]{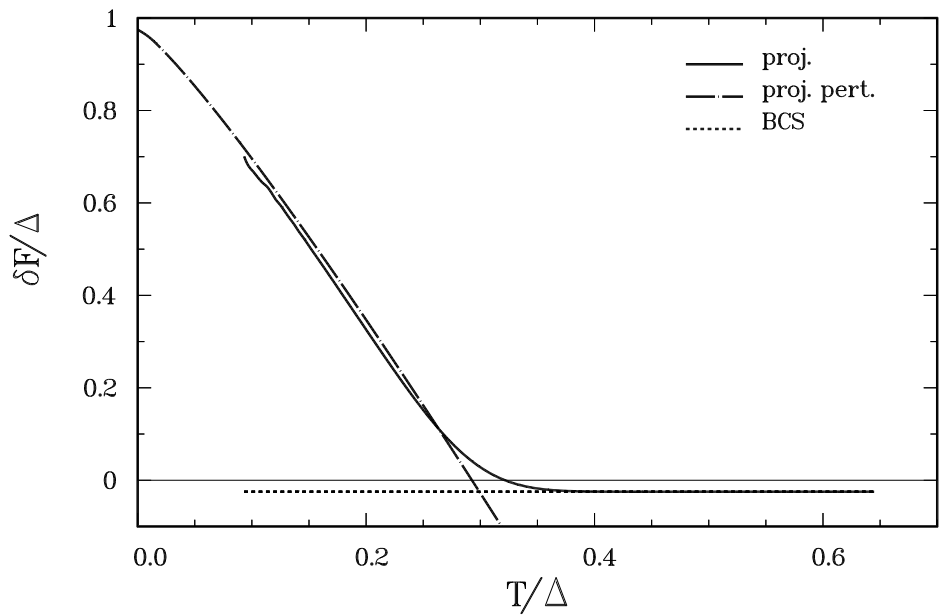}
\caption[Fig5]{
Difference between the free energies
of neighbouring odd $(\langle\NDh\rangle=201)$ and
even $(\langle\NDh\rangle=200)$ systems
as function of temperature for the same parameter values
$w_{\rm F}\Delta=10$ and
$w_{\rm F}\Lambda=200$ as for Figs.~2-4.
The solid and dashed curves correspond to self-consistent
calculations with and without projection on number-parity,
respectively.
Above the crossover temperature $T_\times=0.3\Delta$, the
even-odd effect disappears, as for $\Delta_{\rm F}$ in Fig.~2.
Below $T_\times$ the projected difference $\delta F$ decreases
nearly linearly. It lies close to the
dashed-dotted curve which displays the 
low-temperature perturbative approximation
(\ref{hj040}).
}
\end{center}
\end{figure}

At lower temperatures,
the projected free-energy difference
$\delta F\equiv F_-(N_0+1)-F_+(N_0)$ is seen to
decrease almost linearly from
the zero-temperature value which, by extrapolation,
can be estimated to lie close to $\Delta$.
Let us see how this behaviour, found numerically, can be 
understood from the approximation (\ref{hi110}) for the
grand potential difference $\delta F_{\rm G}$,
derived in the
low-temperature regime (\ref{hi040}) that is relevant here.
To obtain Eq.~(\ref{hi110}) we approximated the solution 
of the full self-consistent equations
by a perturbative treatment starting from the BCS solution.
In order to relate $\delta F$ to
$\delta F_{\rm G}=F_{{\rm G}-}(\mu)-F_{{\rm G}+}(\mu)$,
we note that the derivative with respect to $\mu$
of the approximate expression (\ref{hi110}) is practically zero.
(Indeed $\partial\Delta/\partial\mu\vert_{\mu=0}=0$ for
symmetry reasons.) This means that the relation between
$\langle\NDh\rangle$ and $\mu$ is the same for 
the even-$N$ and odd-$N$ projected solutions.
Hence this relation is also the same as for
BCS (Eq. (\ref{hj020})). Since $F=F_{\rm G}+\mu\langle\NDh\rangle$
and since $\mu$ vanishes for $\langle\NDh\rangle=N_0+1$,
the difference $\delta F_{\rm G}$ coincides with 
$F_-(N_0+1)-F_+(N_0+1)$.
The extra contribution $F_+(N_0+1)-F_+(N_0)$ to $\delta F$
is evaluated by integration of 
$\partial F_+/\partial\langle\NDh\rangle=\mu(\langle\NDh\rangle)$
as in Eq.~(\ref{hj030}). We find:
\be\label{hj040}
\delta F=\delta F_{\rm G}-\frac{1}{4 w_F}\simeq
\Delta-\frac{1}{4 w_{\rm F}}-\frac{T}{2}
\log(8\pi\Delta\,w_{\rm F}^2\,T)
\quad.
\ee
This expression is represented on Fig.~5 
by the curve labelled ``proj.pert.''.
It is very close to the full solution of the projected equations up
to $T=0.3\Delta$. 

Thus, our results agree with
the observation\cite{LJE93} that, when the spacing
between single-particle levels is small
compared to the gap ($w_F\Delta\gg 1$),
{\it number-parity effects
disappear above some temperature} $T_\times$ which lies
below the critical temperature.
Actually, Fig.~5 shows that for $w_{\rm F}\Delta=10$
a {\it sharp crossover} takes place around $T_\times =0.3\Delta$
(that is, around 0.5 $T_{\rm c}$), between the
$N$-parity dependent and independent regimes.
Below this value, projection strongly modifies $\delta F$,
while parity effects disappear above. Figs.~2, 3 and 4
exhibit the identity of the various BCS, even-$N$
and odd-$N$ projected results above $T=0.4\Delta$. 

More generally, for arbitrary values
of $w_{\rm F}\Delta$ much larger than 1, the 
difference $\delta F$ is given at
low temperature by the
approximation (\ref{hj040}), at higher temperature by 
the value $-1/(4w_{\rm F})$ 
associated with the same trivial parity effect than for a
non-interacting system or for the BCS approximation.
The crossover between the two regimes takes
place around the solution of the equation
\be\label{hj050}
\frac{T_\times}{\Delta}
\simeq\frac{1}
{
\log w_{\rm F}\Delta
\sqrt{\displaystyle
\frac{8\pi\,T_\times}{\Delta}
}
}\quad.
\ee
This crossover temperature is also visible on Fig.~2, which shows that
$\Delta_{\rm F}(T)$ is the same for BCS and for both the projected cases
in the range $T_{\times}<T<T_{\rm c}$~, where $T_\times\simeq 0.3\Delta$
for $w_{\rm F}\Delta=10$.

When the size of the system decreases, $w_{\rm F}\Delta$
also decreases. The crossover temperature
$T_\times$ thus rises; from Eq.~(\ref{hj050}) 
one finds that it reaches
the critical temperature $T_{\rm c}=0.57\Delta$ for 
$w_{\rm F}\Delta\simeq 1.5$.
Thus, if the distances between levels become comparable to the gap,
one can expect that odd-even effects will take place over the
entire range of temperatures where pairing occurs.
Physical systems which satisfy this condition
seem presently available since one can
prepare and study ultrasmall
aluminium grains\cite{RBT95,BRT96}.

\subsubsection[654]{Ultrasmall Metallic Grains}

We have chosen, as an example of modelisation of ultrasmall
metallic grains, the parameter values  
${w_{\rm F}}\Delta\simeq 1.1$ and 
${w_{\rm F}}\Lambda=100$. This is obtained for
${w_{\rm F}}{\tilde G}=0.23$, and for
an active space having 101
twofold degenerate levels. 
For the odd (resp. even) system
we take $\langle\NDh\rangle=101$ (resp. 100).
In Fig.~6 we present the
gap $\Delta_{\rm F}$ as function of the temperature
for  odd and even systems.
The critical temperature associated with the 
even-$N$ projection comes out larger than the BCS value.
Even more striking is the behaviour of
the odd-$N$ projected curve. Then the gap is zero
at $T=0$ but takes a non-zero value
above some {\it first transition temperature up to a second one}, 
lower than the BCS critical temperature.
This peculiar behaviour can be explained
as follows.
At $T=0$, the unpaired particle occupies
one of the two single-particle states $\epsilon_0=\mu$
with a probability exactly equal to one. This blocks the formation
of pairs utilizing the two degenerate states with energy $\epsilon_0$
and creates an effective single-particle spectrum 
with a distance between the levels at the Fermi surface
which is twice as
large as $\Delta$.
Consequently, pairing is inhibited. 
When the temperature grows, the bachelor particle
statistically populates other single-particle states, as indicated
in the discussion above on the entropy (\ref{hi240}). This 
allows pairing correlations on the strategic level at the
Fermi surface ($\epsilon_0=\mu$), provided the
temperature is not too low. One thus 
observes a {\it reentrance effect,
with two transition temperatures} :
one at which the pairing switches on  and, 
at a higher temperature, the usual one at which it switches off.
\begin{figure}[hp]\label{F6}
\begin{center}
\includegraphics*[scale=0.6,angle=-90.]{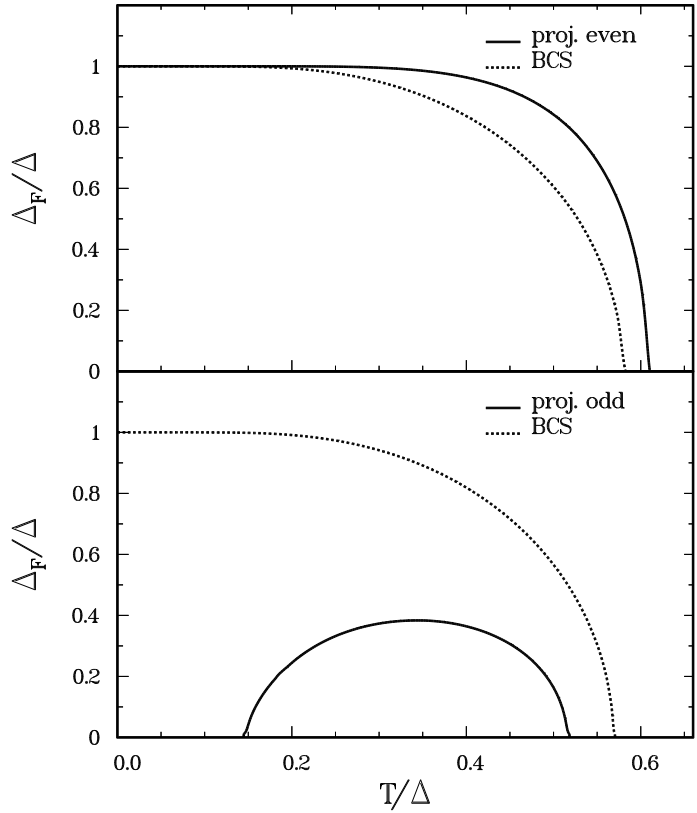}
\caption[Fig6]{Comparison of BCS and number-parity
projected gaps at the Fermi
surface versus temperature
for ${w_{\rm F}}\Delta\approx 1.1$,
${w_{\rm F}}\Lambda= 100$, $\langle\NDh\rangle=100$ (upper part)
or 101 (lower part of the figure).
Such parameters are illustrative of extremely small
superconducting grains. The critical temperature is raised by even-$N$
projection, lowered by odd-$N$ projection.
In the latter case, pairing disappears below a new, lower critical
temperature $0.15\Delta$, thus displaying a reentrance phenomenon.
}
\end{center}
\end{figure}

The {\it entropy of the even system} $\langle\NDh\rangle$=100 
is shown in Fig.~7.  
Projection strongly reduces the entropy below
the critical point. This projection effect
is seen to be more important for the interacting system.
At small $T$, the four entropies vanish exponentially.
As expected the decrease rate of the projected entropy is the fastest.
For $T>0.3\Delta$, the grand-canonical entropy
of the non-interacting system agrees with the linear
form of Eq.~(\ref{hj010}).
It bends down below $T=0.3\Delta$ under
the effect of the spectrum discreteness;
in the domain $T>0.3\Delta$, projection again induces only
a lowering by $\log2$.
\begin{figure}[hp]\label{F7}
\begin{center}
\includegraphics*[scale=0.5,angle=-90.]{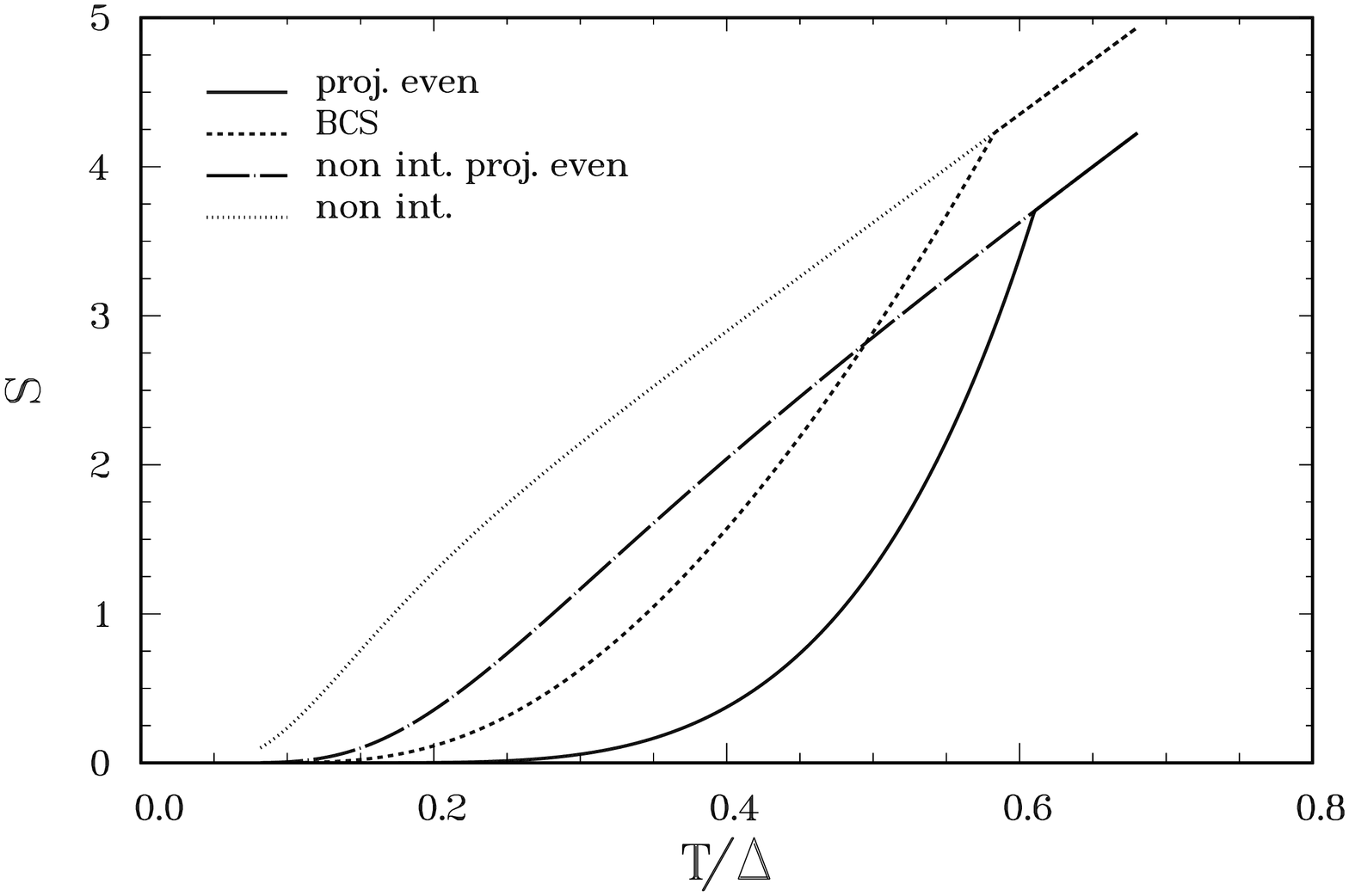}
\caption[Fig7]{Entropy of an even-number system 
for $w_{\rm F}\Delta\approx 1.1$,
${w_{\rm F}}\Lambda= 100$, $\langle\NDh\rangle=100$.
The symbols and notations are the same as in Fig.~3 .
Projection strongly reduces the entropy.
}
\end{center}
\end{figure}

Fig.~8 displays
the {\it entropy of the odd system} $\langle\NDh\rangle$=101.
When the pairing interactions are present,
the BCS formalism, which only constrains
the average number of particles, yields the result 0 instead
of the right zero-temperature limit $\log2$.
This correct limit is obtained by
the odd-$N$ projection.
When there is no interaction, for $T>0.3\Delta$,
the grand canonical value (dotted curve)
is again described by Eq.~(\ref{hj010}).
However the entropy tends to $\log4$ at small $T$. This comes from
the fact that, at $T=0$, the Fermi energy {\it exactly} coincides
with the energy of the last occupied level whose average occupation
is equal to 1/2.
When this value is inserted in
the standard formula
for the grand-canonical entropy of a non-interacting fermion system
(while taking into account
the twofold level degeneracy),
one obtains $\log4$ instead of the
expected value $\log2$.
Projection on odd-$N$ parity decreases the entropy by
$\log2$ and restores the correct limit.
Note that this is not a trivial result since
the exact particle number
is not restored by the projection; only its parity is.
On Fig.~8 the reentrance phenomenon manifests itself 
by the fact that the
entropy of the interacting system is larger than that
of the non-interacting system 
over the temperature interval $0.14<T/\Delta<0.34$;
otherwise the effect of interactions is rather small.
\begin{figure}[hp]\label{F8}
\begin{center}
\includegraphics*[scale=0.5,angle=-90.]{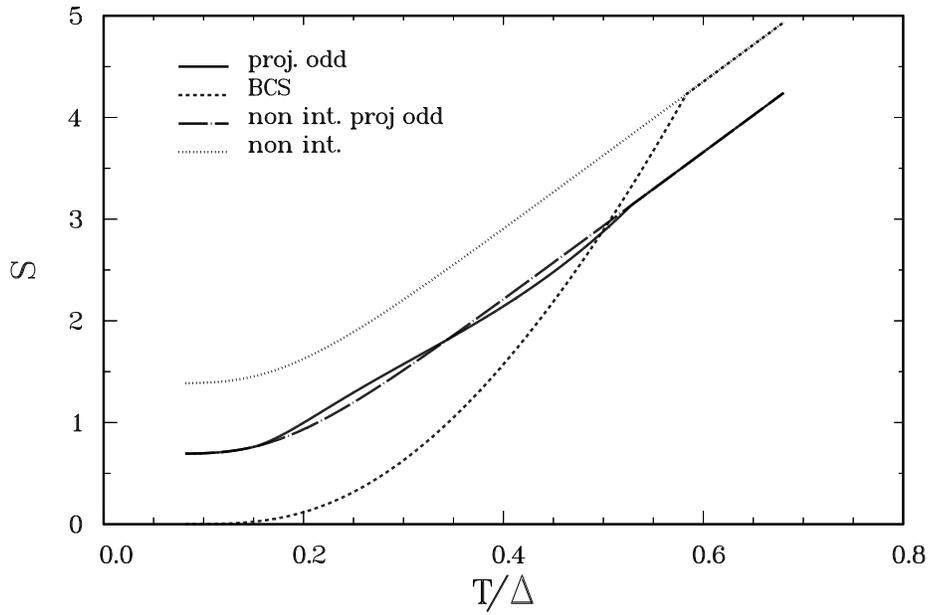}
\caption[Fig8]{Entropy of an odd-number system
for $\langle\NDh\rangle=101$
and the same parameters $w_{\rm F}\Delta\approx 1.1$,
${w_{\rm F}}\Lambda= 100$ as in Figs.~6 and 7.
The symbols and notations are the same as in Fig.~3.
The projection lowers the entropy above $T_{\rm c}$,
but raises it at lower temperatures with the limit
$\log 2$ at $T\rightarrow 0$.
}
\end{center}
\end{figure}

The original features associated
with values of $w_F\Delta$ close to one are even more
visible on the {\it difference of the specific heats}
of neighbouring odd and even systems.
In Fig.~9 one sees that the projected curve shows
discontinuities for the even-system critical temperature,
and also for the two odd-system
critical temperatures. The comparison with the
same curve with $w_{\rm F}\Delta=10$
shown in the inset of Fig.~9, which corresponds to the second derivative 
of the curves on Fig.~5, displays the
striking qualitative change arising from 
the variation of $w_{\rm F}\Delta$
over one order of magnitude.
\begin{figure}[hp]\label{F9}
\begin{center}
\includegraphics*[scale=0.5,angle=-90.]{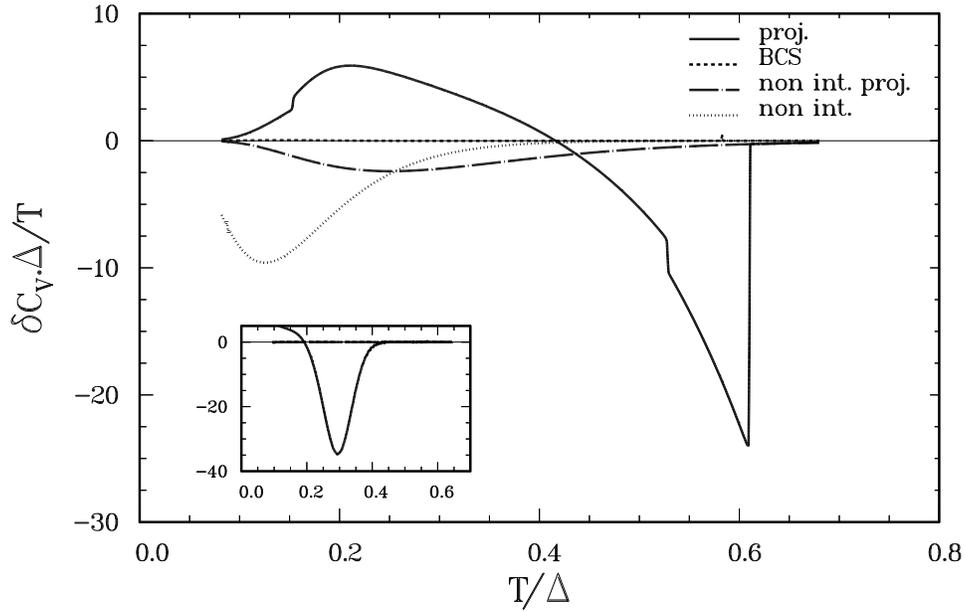}
\caption[Fig9]{Difference $\delta C_{\rm V}$
between the specific heats
of neighbour odd and even systems
$\langle\NDh\rangle=101$ and $\langle\NDh\rangle=100$
(multiplied by the
ratio of the zero-temperature BCS gap $\Delta$ over $T$)
as function of temperature for the same parameters
$w_{\rm F}\Delta\approx1.1$, 
${w_{\rm F}}\Lambda= 100$ as in Fig.~6-8.
The solid and dashed curves correspond to self-consistent
calculations with and without projection on 
particle-number parity, respectively.
The critical temperature for the even system and the two critical
temperatures for the odd system show up as discontinuities.
The curve differs qualitatively from that of the inset, which
for $\langle\NDh\rangle=200$ and $\langle\NDh\rangle=201$ 
corresponds to the parameters $w_F\Delta=10$, ${w_{\rm F}}\Lambda= 200$ 
of Figs.~2-5.
In this inset, the dashed, dashed-dotted and
dotted curves are indistinghishable from
the temperature axis. 
}
\end{center}
\end{figure}

\subsubsection[654]{Comments}

From our numerical analysis,
the pairing properties of heavy nuclei (Sect.~6.5.1)
appear to be intermediate 
between that of  metallic mesoscopic 
islands (Sect.~6.5.2) 
and that of ultrasmall aluminium grains (Sect.~6.5.4). 

In both the BCS and projected results, the sharp transition from 
the superconducting to the normal solution 
arises from the mean-field nature of the approximations.
The transitions should be smoothed by a projection onto
a well-defined particle number, that we have not performed.
This projection would probably have
a negligible influence on the results obtained
for mesoscopic metallic islands where 
$\langle\NDh\rangle$ is large and hence has small relative
fluctuations. In nuclei, on the other hand,
the smoothing of the superconducting to normal phase transition
due to the finite number of neutrons and protons
is experimentally observed in the dependence
on angular momentum of the moments of inertia. 
The same phase-transition smoothing effect
should also  be observable in metallic grains 
when they become sufficiently small, through
experiments on temperature-dependent properties
or on magnetic response.
The general equations derived in Sect.~5.2 and 5.3
could be helpful to perform the required projection. They
could also help to account for fluctuation effects, such as those
invoked in Ref.\cite{MLa97}.
\newpage

\section[S7]{Summary and Perspectives}

\setcounter{equation}{0}

A general variational method for evaluating
the properties of finite systems at thermodynamic equilibrium has
been proposed and applied
to finite fermion systems with pairing
correlations. 
Extensions of the standard mean-field theories were thus obtained for
two types of problems:
1°) the variational evaluation not only of
thermodynamic quantities, but also of expectation values,
fluctuations or correlations between observables and more generally
of characteristic functions; 2°) the 
restoration of some constraints on the physical space of states 
or/and of some broken symmetries violated by the
variational Ans\"atze. 

To reach this double goal, we relied on 
an action-like functional (\ref{e040}) which
is designed to yield as its stationary value 
the relevant characteristic function. Two dual trial quantities
enter this functional. The operator $\DIh\uu$ 
is akin to a statistical operator evolving
from $\uni$ to the actual density operator 
\be\label{ia010}
\DDh={\rm e}^{-\beta\KDh}\q
\ee
as $u$ goes from $0$ to $\beta$,
while the
operator $\AIh\uu$  is akin to the observable $\ADh$ defined
in Eq.~(\ref{I060}). Such a
doubling of the number of variational degrees of
freedom is distinctive of a general class of
variational principles built to optimize some quantity
of interest defined by given constraints.
Here the constraint is the (symmetrized) Bloch equation
(\ref{e030}) satisfied by $\DIh\uu$, while $\AIh\uu$ appears as
a Lagrangian multiplier associated with this constraint.
For unrestricted trial spaces,
the stationarity of the functional (\ref{e040})
expresses the Bloch equation (\ref{e030}) for $\DIh\uu$; in
addition, it yields the equation (\ref{e100}) for $\AIh\uu$ 
which is the Bloch analogue 
of a (backward) Heisenberg equation.
These two equations are decoupled and duplicate
each other.
However, when the trial
spaces are restricted (as is unavoidable for any application), 
the two equations generally become coupled, with mixed 
boundary conditions involving both ends of the integration range
$[0,\,\beta]$. 
The fundamental thermodynamic relations are preserved
by the variational approximations 
under rather general conditions on the choice of the trial spaces (Sect.~2.2);
these conditions are satisfied by the Ans\"atze that have 
been used throughout the present article. 
For conserved observables, the characteristic
functions, and therefore all the cumulants, 
are obtained from the approximate 
partition function via a simple shift of the operator
$\KDh$ (Sect.~2.3). 
Moreover, in the limit when the 
sources $\xi_\gamma$ entering the characteristic function
vanish, and provided the
trial spaces satisfy two additional conditions stated in
Sect.~2.4, our variational principle reduces to the standard 
maximization principle for thermodynamic potentials.

Beginning with Sect.~3, we used the functional (\ref{e040}) to 
study systems of fermions in which 
pairing correlations are effective. 
For the variational operators $\DIh\uu$ and 
$\AIh\uu$ we choose the trial forms
\be\label{ia020}
\TDhd\uu\equiv\exp( -\lid\uu -
{\scriptstyle \usd}\gbt\LId\uu\gbd )\q,
\ee
\be\label{ia030}
\TDha\uu\equiv\exp( -\lia\uu -
{\scriptstyle \usd}\gbt\LIa\uu\gbd )\q,
\ee 
where $\gbd$ denotes the $2n$ operators of creation
$\alpha^\dagger_\lambda$ or annihilation $\alpha_\lambda$;
the trial quantities are the scalars $\lid$ and $\lia$
and the $2n\times 2n$ matrices $\LId$ and $\LIa$.
The generalized Wick theorem then allows 
to write explicitly the resulting functional
as Eq.~(\ref{f220}). The 
stationarity conditions are expressed by the set
of coupled equations~(\ref{f370}),
(\ref{f430}), (\ref{f460}) and 
(\ref{f470}) which constitute an extension of the 
Hartree-Fock-Bogoliubov approximation. 
The equation~(\ref{f370}) proceeds forwards with respect 
to the imaginary time $u$ while Eq.~(\ref{f430}) 
evolves backwards, since the boundary conditions
are fixed at $u=0$ for the former and at $u=\bb$ for the latter.

In Sect.~4 thermodynamic quantities, fluctuations and
correlations have been evaluated via
 the expansion of the coupled
equations in powers of the sources $\xi_\gamma$.
The thermodynamic quantities are given by the zeroth order term in the
expansion while 
expectation values $\langle\QDh_\ggg\rangle$ of
single-quasi-particle observables 
\be\label{ia040}
\QDh_\gamma=q_\gamma+\usd\gbt\QI_\gamma\gbd
\ee
are given by the first order. 
In both cases, one recovers (Sect.~4.1) the standard HFB results 
in agreement with the discussion of Sect.~2.
 
The fluctuations and correlations 
$\Cgd\equiv
\usd\langle\QDh_\gamma\QDh_\delta+\QDh_\delta\QDh_\gamma\rangle
-\langle\QDh_\gamma\rangle\langle\QDh_\delta\rangle$
are given by the second order.
In the case where the two observables $\QDh_\ggg$ and
$\QDh_\delta$ commute with the operator $\KDh$,
we obtain (Eq.~(\ref{gd120}))
\be\label{ia050}
\Cgd=\frac{1}{\bb}\,\QI_\ggg:\FLl_0^{-1}:\QI_\delta\q.
\ee
The $2n\times 2n$ matrices
$\QI_\gamma$ and $\QI_\delta$ defined in (\ref{ia040})
are regarded
in Eq.~(\ref{ia050}) as vectors in the
Liouville space, and the colon symbol
stands as a scalar product in this space, or equivalently as half
the trace ($\trd{}$) in the original space of the
$2n\times 2n$ matrices (Sect.~4.2.2 and Eq.(\ref{aa010})).
In the Liouville space, the stability matrix $\FLl_0$ is
defined in Eq.~(\ref{gd050}) as the second
derivative with respect to the
reduced density $\RI$ of the HFB grand potential $\FI_{\rm G}$ around
its minimum, the HFB solution $\RIze$. If the observables
$\QDh_\ggg$ and $\QDh_\delta$ do not commute
with $\KDh$, Eq.~(\ref{ia050})
expresses their Kubo correlation (Sect.~4.3.1). It then appears as
a generalization to the non-commutative case of the Ornstein-Zernike
formula, here derived from a general variational principle.

The ordinary correlation of two
observables $\QDh_\ggg$ and $\QDh_\delta$
that do not commute with $\KDh$
is covered by the more general formula (Eq.~(\ref{gg260})) 
\be\label{ia060}
\Cgd={\textstyle\usd}
\QI_\ggg:(\KLl_0\coth{\textstyle\usd}\bb\,
\KLl_0)\FLl_0^{-1}:\QI_\delta\q.
\ee
The matrix $\KLl_0$ 
is the usual RPA kernel written 
in the Liouville space (Eq.~(\ref{gg100}) or (\ref{ac010})).
The HFB stability matrix $\FLl_0$ is related to the dynamical
RPA matrix $\KLl_0$ through  
Eq.~(\ref{gg120}), or (\ref{ac080}).

Notwithstanding the independent-quasi-particle nature
of the operators $\DIh\uu$ and $\AIh\uu$,
we have obtained non-trivial approximate expressions 
for the correlations.
In particular, formulas (\ref{ia050}) or (\ref{ia060}) incorporate
long-range effects. This was made possible
because, through the optimization of the functional (\ref{e040}), 
the variational
quantities $\DIh\uu$ and $\AIh\uu$ 
{\it acquire a dependence on the sources $\xi_\ggg$}.

In spite of the coupling inherent to the stationarity
equations (\ref{f370}) and (\ref{f430}), we have found the
explicit expressions (\ref{ia050}) and (\ref{ia060})
for the correlations. 
Moreover, we have found a simple $u$-dependence for the solutions
of these equations to first order in the sources.
Despite the compactness of the final results (\ref{ia050}) or (\ref{ia060}),
their derivation was complicated 
by the existence of zero eigenvalues in the kernel
$\FLl_0$, due to some broken invariance,
and in the kernel $\KLl_0$ (Sect.~4.3.2).
Indeed, for the formulas (\ref{ia050}) and 
(\ref{ia060}) to be meaningful, 
one must specify how to handle these 
zero eigenvalues. 
We showed in Sects.~4.2.3 and 4.3.2 that the precise meaning 
to be given to these formulas depends on
the commutation, 
or non-commutation, of the observables $\QDh_\gamma$ and
$\QDh_\delta$ with the conserved single-particle operator 
($\NDh$ in our example) associated 
with the broken invariance.

Analyzed in perturbation theory, the 
formulas (\ref{ia050}) and (\ref{ia060}) were shown to correspond 
to a summation of {\it bubble diagrams} (Sect.~4.3.3). 
Let us stress that the RPA kernel comes out
from our variational approach without any additional 
assumption beyond the choice 
(\ref{ia020}), (\ref{ia030}) of the trial Ans\"atze.
In this way, the {\it RPA acquires a genuine variational status}
(albeit not in the Rayleigh-Ritz sense). 

Sects.~5 and 6 are devoted to the second 
type of problems posed in the
Introduction, namely to those situations
which require projections. 
Though the operator $\KDh$,
and hence the {exact} density operator (\ref{ia010}), 
commute with any conserved quantity, the corresponding
invariance can be broken by the variational approximation 
if the commutation is not preserved by the trial 
operator $\DIh\uu$. In this case, a projection $\PDh$ over an eigenspace of
the conserved quantity is 
required on both sides of $\DIh$
to restore the symmetry. 
On the other hand, even when the invariance is not broken, 
a projection is needed 
if the trial operators act in a space 
which is wider than the physical Hilbert space $\HI_1$.
(For instance we consider below the problem of extending the
thermal Hartree-Fock approximation to a system with a well defined
particle number.) Sects.~5 and 6 
cope with both situations. 

In the first case where the invariance is broken,
we replace the trial expression (\ref{ia020}) for the
density operator by
\be\label{ia070}
\DIh\uu=\PDh\TDhd\uu\PDh\q,
\ee
where $\TDhd\uu$ still has the independent quasi-particle form
(\ref{ia020}); for $\AIh\uu$ we keep the Ansatz (\ref{ia030}).
In order to handle (\ref{ia070}) we take advantage of the fact that
the projection $\PDh$ can be written as a sum of group
elements $\TDhgu$ which have an independent quasi-particle form,
so that ({\ref{ia070}) is also a sum of terms of the
$\TDh$-type.
The resulting functional 
(Eq.~(\ref{hb050})) and 
coupled stationarity equations ((\ref{hc010}), (\ref{hc020}),
(\ref{hc060}) and (\ref{hc070})) 
are written in Sects.~5.2 and 5.3.
Although these coupled equations are noticeably 
more complicated than
the unprojected ones, the number of variational parameters,
and therefore the dimension of the numerical problem,
do not increase.

The case of unbroken $\PDh$-invariance,
where the trial operators $\TDhd\uu$ and
$\TDha\uu$ commute with the group elements $\TDhgu$, is
discussed in Sect.~5.4.
Since $\PDh$ now commutes with $\TDhd\uu$, the Ansatz (\ref{ia070})
reduces to
\be\label{ia080}
\DIh\uu=\PDh\TDhd\uu=\TDhd\uu\PDh\q,
\ee
which entails several simplifications in the 
action-like functional (Eq.~(\ref{hd050})) and in the 
stationarity equations. 
We have examined in some 
detail (Sect.~5.4.2) the limiting case 
$\ADh(\xi)=\uni$ which admits the explicit solution 
(\ref{hd080}) for 
$\DIh\uu$ and $\AIh\uu$. This leads to further simplifications
in the functional (Eq.~(\ref{hd150})) and in the 
self-consistent equations (\ref{hd160}-\ref{hd190}). 
Several thermodynamic consistency 
properties were verified.
As an example the projected entropy, energy and grand potential
were shown to satisfy the usual thermodynamic identities
(Eqs.~(\ref{e260}-\ref{e263})).

In Sect.~6 the theoretical framework
elaborated in Sect.~5.4 was applied to
study the effects of the 
particle-number parity in a finite superconducting system
at thermodynamic equilibrium. 
In the trial density operator (\ref{ia080}),
$\PDh$ given by (\ref{ha030}) is the projection on either even or odd
particle number.
This problem is of interest
for recent or future experiments on 
mesoscopic superconducting islands, small metallic grains
or heavy nuclei.
The explicit form of the coupled equations 
(see Eq.~(\ref{he090})) was written in Sect.~6.1. 
The parity-projected 
grand partition function (or more generally the characteristic
function for conserved single-particle 
operators $\QDh_\ggg$) was evaluated
in Sect.~6.2, where Eq.~(\ref{he230}), together with
by the definitions (\ref{he250}-\ref{he260}), was shown to replace
the standard HFB self-consistent equation. 

Sect.~6.3 deals with a system (electrons in a
superconductor or nucleons in a deformed nucleus) governed by a
BCS model Hamiltonian, where the non-interacting part 
involves twofold degenerate levels $p$ with energies
$\epsilon_p$\,, and the interaction takes place
only between the two states of each level.
The BCS gap is now replaced by the quantities $\Delta_{\eta p}$ defined
in Eq.~(\ref{hf170}), and the
usual BCS equation is replaced
by the number-parity projected self-consistent equation (\ref{hf230}),
\be\label{ia090}\displaystyle
\Delta_{\eta p}=\usd\sum_{q}\,\Gbcspq\,
\frac{\displaystyle
\Delta_{\eta q}\,\tbcsq
}{\displaystyle
\sqrt{(\epsilon_q-\mu)^2+\Delta_q^2}
}\,\frac{\displaystyle
1+\eta\,r_0\,\tbcsp^{-2}\,\tbcsq^{-2}
}{\displaystyle
1+\eta\,r_0\,\tbcsp^{-2}
}\q,
\ee
in which $\Gbcspq$ is a pairing matrix element
of the BCS Hamiltonian (\ref{hf010}). 
The notation $\tbcsp$ stands for
\be\label{ia100}
\tbcsp\equiv\tanh{\textstyle\usd}\beta e_p\q,
\ee
where the quasi-particle energy $e_p$, 
given by Eq.~(\ref{hf210}), differs
from the standard BCS expression 
$\sqrt{(\epsilon_p-\mu)^2+\Delta_p^2}$.
The number $r_0$ is defined (Eq.~(\ref{hf130})) as
\be\label{ia110}
r_0\equiv\prod_p\,\tbcsp^2\q.
\ee
The formulae (\ref{ia090}-\ref{ia110}) depend on the
parity of the particle number $N$
through $\eta=+1$ (even $N$) or
$\eta=-1$ (odd $N$). The effects of the projection
appear indirectly through the occurence of $\ebcsp$
in (\ref{ia100}), and directly
through the last fraction of Eq.~(\ref{ia090}).
As discussed in Sect.~6.3.3, this fraction is larger
than one for even-$N$ systems, smaller for odd-$N$ systems.
As a consequence, the even-$N$
projected gap is larger than the BCS gap,
while the odd-$N$ projected gap is smaller,
in agreement with the idea that pairing is inhibited in small
odd-$N$ systems.
The analysis of this fraction reveals, moreover, that
the odd-$N$ projection differs from BCS more than the
even-$N$ projection.

In Sect.~6.4 we considered some limiting situations.
At zero temperature the BCS and even-$N$ projected gaps are
equal. However, while the BCS quantities deviate from their
zero-temperature limit by terms of the form ${\rm e}^{-\beta\Delta}$,
the deviations are of the form
${\rm e}^{-2\beta\Delta}$ for the even-$N$ projection.
The differences are more striking between BCS
and the odd-$N$ projection.
At zero temperature,
the BCS gap is larger than the
odd-$N$ projected gap.
When the temperature grows,
the latter starts to increase before
reaching a maximum. The  odd-$N$ entropy
tends to $\log2$ when $T$ tends to zero, reflecting
the twofold degeneracy of the
ground state; in Sect.~6.5.3 we commented upon the fact that
this result
is not completely trivial.
At low temperature the odd-$N$ entropy
coincides with the Sakur-Tetrode
entropy of a single-particle with a mass $\Delta$
moving in a one-dimensional box, which is consistent with
gapless elementary excitations in odd-$N$ systems.
When the single-particle level density is sufficiently large, the 
critical temperatures become equal for the BCS and 
the odd or even projected solutions.
Moreover for a large system, or when the 
level spacing becomes small with
respect to the pairing gap at the Fermi surface, 
Eq.~(\ref{ia090}) reduces to the usual BCS equation. 
However, the value of the projected (odd or even) entropy is lower than 
the BCS value by $\log2$.

In order to push further the comparison between the BCS and the
projected results
we presented in Sect.~6.5 three schematic 
calculations which exemplify
three different physical situations.
The crucial parameter turns out to be the
product $w_{\rm F}\Delta$ where $\Delta$ is the zero-temperature
BCS gap and $w_{\rm F}$ 
the single-particle level density at the Fermi surface.
This parameter $w_{\rm F}\Delta$, interpreted as a 
``number of Cooper pairs'', indicates how many single-particle 
levels lie within the energy range of the gap.

When $w_{\rm F}\Delta$ is large,
the projection effects are weak. Fig.~2 illustrates the
case $w_{\rm F}\Delta= 10$.
The BCS and even-$N$ projected gaps are almost
undistinguishable. The relative difference
between the BCS and odd-$N$ gaps is $1/2w_{\rm F}\Delta$ at $T=0$.
The difference between the free energies of the odd and even systems
can be estimated perturbatively
from the BCS solution. The outcome (Eq.~(\ref{hj040})),
equal to $\Delta-1/4w_{\rm F}$ at $T=0$,
decreases almost linearly with
$T$ and displays a cross-over towards
the asymptotic value $-1/4w_{\rm F}$ at a
temperature $T_\times$ smaller than the critical temperature (see Fig.~5).
Between $T_\times$ and $T_{\rm c}$, the gap function
is the same wathever the $N$-parity.
These results are consistent with experiments
on mesoscopic superconducting islands which show
that, when $w_{\rm F}\Delta\gg1$,
the odd-even effects disappear at some 
temperature below $T_{\rm c}$\cite{LJE93}.

The value $w_{\rm F}\Delta\simeq 2$ corresponds roughly to the
case of a superdeformed heavy nucleus (Fig.~1). 
In the even system, the difference 
between the projected and BCS gaps becomes more appreciable than in
Fig.~2.
In the odd system, the projected gap
at zero temperature is reduced by about 
$30\%$ with respect to BCS. This result can also be
obtained from a BCS calculation in which the pair formation
is blocked for the level that coincides with the Fermi surface.
As the temperature grows so does the odd-$N$ projected gap.
The BCS, even and odd projected gaps merge here again at higher temperature
where they all display the same critical behaviour.

Fig.~6, where $w_{\rm F}\Delta\simeq 1.1$, corresponds to 
a situation which could be encountered in ultrasmall
Aluminium grains\cite{BRT96}. In the even-$N$ projected state,
the critical temperature is higher than for the BCS state. 
In the odd system a new phenomenon occurs: 
there is no gap at zero temperature.
However, the phase diagram ($T$, $\Delta$) obtained by projecting on
odd number exhibits a {\it reentrance effect} : 
when the temperature increases, a gap appears at a first transition
temperature, reaches a maximum and disappears at a second
transition temperature smaller than the BCS one.
This behaviour can be understood as follows : 
at $T=0$, at least one particle remains unpaired and it
fully occupies one of the two single-particle states at the
Fermi surface, forbidding 
the formation of a pair with components
on this strategic level. 
When the temperature increases, the
unpaired particle partially frees up this level, 
which then becomes available 
for the formation of Cooper pairs.
According to Fig.~9, the existence of two transitions
in odd ultrasmall metallic grains could be experimentally detected by 
specific heat measurements of odd and even systems.

Besides its relevance for finite superconducting systems,
one can imagine using the variational method introduced 
in this paper
for studying other physical problems.
In particular the formulation of Sect.~5.4, which treats situations
with unbroken $\PDh$-invariance, is
applicable in various problems.
Consider for instance a finite fermion system at equilibrium
without pairing correlations. In this case the exponents in
the trial quantities (\ref{ia020}) and (\ref{ia030}) need only
contain single-particle terms in $a_i^\dagger a_j$ which
commute with $\NDh$. The projection $\PDh$ has the form
(\ref{ha020}). A variational method to
evaluate the {\it canonical} partition function is then
furnished by Sect.~5.4.2, which constitutes
a projected extension (with a well-defined number of particles)
of the thermal Hartree-Fock approximation.
It suffices in Eqs.~(\ref{hd080}-\ref{hd090}) to take a
trial operator $\HDh_0$ of the single-particle type.
This procedure also provides, along the lines of Sect.~2.3, a consistent
variational estimate for the characteristic function
of any conserved single-particle observable (commuting with
both $\NDh$ and $\HDh$).
For other single-particle observables (commuting with $\NDh$ only)
one should resort to the method of Sect.~5.4.1 where the
trial operators $\TDhd\uu$ and $\TDha\uu$
have a more complicate $u$-dependence.
Likewise, for finite systems without odd-multipole 
deformations, the projection over a given space-parity 
can be performed by implementing the operator (\ref{ha060})
in Sects.~5.4.1 and 5.4.2. 

In case the symmetry is broken (Sects.~5.2 and 5.3),
the functional (\ref{hb050}) takes a more intricate form, due to
the occurence of two projections. One could then make use of a
further approximation by using the scheme described
in Ref.~\cite{FOn96} where an 
integral such as (\ref{ha020}) on the group elements is replaced by a
truncated sum. This would
generate approximate expressions 
for the canonical thermodynamic functions and correlations
which should be easier to
handle numerically than the general 
expressions (\ref{hc020}-\ref{hc040}) and (\ref{hc070}-\ref{hc080}).

In the variational methods 
described above,
one could imagine implementing 
trial quantities $\DIh\uu$ and $\AIh\uu$ with a more
general form than (\ref{ia020}) and (\ref{ia030}).
For instance Ref.~\cite{Flo89} has demonstrated, in the related
context of time-dependent Hartree-Fock equations,
the feasibility of extensions where $\TId\uu$ is multiplied
by some polynomial in creation and annihilation operators.
Using the methods developed here,
one could obtain extensions of the HFB formalism,
or of the projected HFB formalism, which would take into account
a wider class of correlations. 

The variational setting of Sect.~2 could
easily be adapted to problems of {\it classical statistical mechanics}.
A first possibility would be to start from the classical version
of the functional (\ref{e040}) in which traces, 
density operators
and observables are replaced by 
integrals, distribution functions and random
variables in phase space. 
The commutative nature of these quantities implies that 
a characteristic function would be obtained as a {\it maximum}.
Alternatively, one could 
start from the quantal functional (\ref{e040}) and then 
apply a semi-classical limiting process, for instance through the
use of the Wigner transform.
The latter procedure would yield in addition the first
quantum corrections to the classical limit. This could be of interest 
for heavy nuclei or for sufficiently large clusters of atoms and 
molecules.

Minor modifications are needed to 
deal with systems of {\it interacting bosons}. Besides the change 
of the anticommutation rules (\ref{f020}) into commutation ones,
linear terms in $\gbd$ and $\gbt$ should be added to 
the observables (\ref{f120}) and to the exponents of the trial
operators (\ref{ia020}) and (\ref{ia030}). 
This should result in a systematic variational approach
of Bose-Einstein condensation in finite interacting systems, 
a phenomenon which has become of direct experimental
interest with the recent achievement of condensates 
in dilute atomic gases at ultra-cold temperatures.
It is worthwhile to note that here again
the projection on the right number of particles 
is an important requirement.

Another physical problem where a projection is essential arises
in particle physics, when mean-field like
approximations violate the colour invariance.
One may imagine restoring it by means of a projection on
colour singlets along the lines of Sect.~5.

In conclusion we would like to stress 
the {\it consistent} character of the 
present method, besides its 
generality and its flexibility. We have verified the 
agreement of our approximation scheme with
thermodynamics. We have recovered naturally the RPA (the complete
series of bubble diagrams). 
Finally, although we have not touched upon the subject in 
this already long article, we point out the complementarity
of the present static approach with the variational 
treatment of dynamical problems of Ref.~\cite{BVe92}.
Indeed, the use of the Bloch equation
as a constraint for the initial state combines
coherently with the use of the (backward) Heisenberg equation as
a dynamical constraint; a variational principle comes out
which (within restricted trial spaces) optimizes both the initial
state and the evolution of the system.
This comprehensive static and dynamic variational 
method should provide a convenient tool for evaluating
time-dependent correlations, 
and hence transport properties, either in the 
grand-canonical formalism as in Sect.~4 or with  
projections as in Sects.~5 and 6.\vspace{1cm}

\noindent{\Large\bf Acknowledgements}

We are indebted to M.H. Devoret, D. Est\`eve and N. Pavloff for very
instructive discussions, to S. Creagh and N. Whelan for reading parts of our 
manuscript, and to M.T. Commault for her help in the preparation of the
figures.
\newpage

\appendix
\section[APPENDIX A:]{Geometric Features of the HFB Theory}
\setcounter{equation}{0}
\renewcommand{\theequation}{A.\arabic{equation}}

\subsection[A1]{The Reduced Liouville Space}

The Liouville representation of quantum mechanics
(see for instance
\cite{BAR86}, Sect.~3) relies on the idea that the algebra of observables,
when regarded as a vector space,
can be spanned by a basis of operators.
Any observable is then written as a
linear combination of the operators of the chosen basis, with
coefficients interpreted as ({covariant}) coordinates.
Correlatively, for any state, the
expectation values of the operators
of the basis are regarded as the ({contravariant})
coordinates of this state. In the HFB context,
it is sufficient to consider the subalgebra
of quadratic forms of creation and annihilation operators $\gb_\lambda$
(equal to either $\abd_i$ and $\abt_i$). With the
{ notation (\ref{f120})},
keeping aside the trivial c-number $\qi$,
we are led to regard the
coefficients $\QI_{\lambda\lambda'}$ as the
{\it covariant coordinates of the operator}
$\usd\sum_{\lambda\lambda'}\gb^\dagger_\lambda
\QI_{\lambda\lambda'}\gb_{\lambda'}$. The factor $\usd$ and the
constraint (\ref{f050}) imposed on $\QI$
are consistent with the duplication
$\gb^\dagger_{\lambda}\gb_{\lambda'}=
-\gb_{\lambda'}\gb^\dagger_{\lambda}+\delta_{\lambda\lambda'}=
-\sum_{\mu\mu'}\sigma_{\lambda'\mu}
\gb^\dagger_{\mu}\gb_{\mu'}
\sigma_{\mu'\lambda}+\delta_{\lambda\lambda'}
$
occuring in the operator basis.
We shall thus consider $\QI_{\lambda\lambda'}$ both as
{ a {\it matrix}
in the $2n$-dimensional $\lambda$-space of creation and
annihilation operators (Sect.~3.1) and as a {\it vector} }
in the $2n\times 2n$-dimensional reduced Liouville space
of observables.

Likewise, since in the HFB approximation we focus on
density operators  of the form (\ref{f040})
and since they
are characterized by the contraction matrices
$\RI_{\lambda\lambda'}$
defined by Eqs.~(\ref{f080})
and (\ref{f100}-\ref{f110}),
the {\it contravariant coordinates of these density operators}
reduce
to the set $\RI_{\lambda\lambda'}$, with the constraint (\ref{f090}).
Again, $\RI_{\lambda\lambda'}$ is both
{ a {\it  matrix} in the $\lambda$-space and a
{\it vector} }
in the
reduced Liouville space of states.

As in the full space of observables and states, the
{\it expectation value} of an operator
$\usd\gbt\QI\gb$,  in a state characterized by its contractions
$\RI_{\lambda\lambda'}$, is given here by the {\it scalar
product} of the vectors $\RI$ and $\QI$ in the Liouville
space. We denote this scalar product as
\be\label{aa010}
\QI : \RI = {\textstyle\usd}\trd{\QI\RI} =
{\textstyle\usd}\sum_{\lambda\lambda'}
\QI_{\lambda\lambda'}\RI_{\lambda'\lambda}
\q.
\ee
In (\ref{aa010}) the {colon symbol $\ :\ $  
indicates therefore both a}  summation on a twisted pair
of indices and { a multiplication by the} factor $\usd$
{ accounting} for the
duplication of coordinates {both in}  $\RI$ { and} $\QI$.

Among others, we shall use in this Appendix the
following two matrices in
Liouville space :
\be\label{aa020}
\UL_{(\lambda\lambda')(\mu\mu')}=
2\,\delta_{\lambda\mu'}\delta_{\mu\lambda'}\ ,\q
\SIL_{(\lambda\lambda')(\mu\mu')}=
2\,\sigma_{\lambda\mu}\sigma_{\mu'\lambda'}\q.
\ee
They are invariant under the exchange
$(\lambda\lambda') \leftrightarrow (\mu\mu')$
(symmetry in Liouville space). One has also 
$\SIL^2=\UL$. Moreover, the relation
(\ref{f090}) implies that the variation
$\delta\RI$ of a contraction matrix satisfies
\be\label{aa030}
\SIL : \delta\RI = \sigma\delta\RI^{\rm T}\sigma = -\delta\RI\q.
\ee

\subsection[A2]{Expansion of the HFB Energy,
Entropy and Grand Potential}

{\it The HFB energy}, defined by
$\EI\{\RI\}=\Tr{\DIh\KDh}/\Tr{\DIh}$ (including therefore
the term $-\mu\langle\NDh\rangle$), is given by Wick's
theorem as the function (\ref{f310}) of the Liouville
vector $\RI_{\lambda\lambda'}$. Its partial derivatives
have been defined in Eqs.~(\ref{f380}-\ref{f400})
as the $2n\times2n$ matrix
$\HI\{\RI\}$  constrained by
(\ref{f410}). In the Liouville space, 
in agreement with (\ref{f380}), $\HI_{\lambda\lambda'}$
appears as the gradient of $\EI$, that is, a covariant vector
which, with the
notation (\ref{aa010}), reads
\be\label{ab010}
\delta\EI\{\RI\}=\HI\{\RI\} : \delta\RI\q.
\ee

Using
(\ref{f070}) and (\ref{f080}),
{\it the HFB entropy} is evaluated  as
\be\label{ab020}\ba{rl}
\SI\{\RI\}&\equiv -\Tr{\DIh\log\DIh}/\Tr{\DIh}-\log\Tr{\DIh}\\
&=-\usd\trd{[\,\RI\log\RI+(1-\RI)\log(1-\RI)\,]}\q,
\ea\ee
where the two terms under the trace are equal. 
Similarly to (\ref{ab010}), its
gradient, written by means of (\ref{f090})
so as to satisfy the same identity (\ref{f410}) as $\HI$,
is
given by
\be\label{ab030}
\delta\SI\{\RI\}= \LI : \delta\RI\q,
\ee
with $\LI\equiv[\log(1-\RI)/\RI]$.

The second derivatives of $\EI\{\RI\}$ and $\SI\{\RI\}$
define {\it the twice covariant tensors}
$\ELl$ and $\SLl$ entering the expansions
\be\label{ab040}
\EI\{\RI+\delta\RI\}=\EI\{\RI\} + \HLl\{\RI\} :\delta\RI
+{\textstyle\usd}\delta\RI : \ELl : \delta\RI\q,
\ee
\be\label{ab050}
\SI\{\RI+\delta\RI\}=\SI\{\RI\} + \LI :\delta\RI
+{\textstyle\usd}\delta\RI : \SLl : \delta\RI+\ldots\q.
\ee
Both matrices $\ELl$ and $\SLl$ are symmetric
(for instance,
$\ELl_{(\lambda\lambda')(\mu\mu')}=\ELl_{(\mu\mu')(\lambda\lambda')}$).
Moreover they satisfy
\be\label{ab060}
\SIL : \ELl : \SIL = \ELl\ ,\q\SIL : \SLl : \SIL = \SLl\q,
\ee
so that
$\ELl : \delta\RI$ obeys the same symmetry relation
(\ref{aa030}) as $\delta\RI$. 
The explicit form of the second derivative
$\ELl$ of $\EI\{\RI\}$ comes out from  (\ref{f310})
and $\delta\HI = \ELl : \delta\RI$. 
When $\RI_{\lambda\lambda'}$ and $\delta\RI_{\lambda\lambda'}$
are hermitian matrices in the $\lambda$-space, the 
matrix $\SLl$ of the second variations
of $\SI\{\RI\}$
is always negative in (\ref{ab050}).
As shown in \cite{BAR86}, it can be expressed as
\be\label{ab070}
\SLl_{(\lambda\lambda')(\mu\mu')}=
-2\int_0^\infty\,\hbox{d}v\,
\crg\frac{1}{\RI+v(1-\RI)}\crd_{\lambda\mu'}
\crg\frac{1}{\RI+v(1-\RI)}\crd_{\mu\lambda'}\q,
\ee
which, in a basis where $\RI$ is diagonal
($\RI_{\lambda\lambda'}=\RI_\lambda\,\delta_{\lambda\lambda'}$),
reduces to
\be\label{ab080}
\SLl_{(\lambda\lambda')(\mu\mu')}=
2\,\delta_{\lambda\mu'}\delta_{\mu\lambda'}\,
\frac{\LI_\lambda-\LI_\mu}{\RI_\lambda-\RI_{\mu}}
\q.
\ee
The ratio is equal to $-1/\RI_\lambda(1-\RI_\lambda)$ when
$\RI_\mu=\RI_\lambda$.

The standard HFB variational principle
(Sects.~4.1) provides the grand potential
at grand canonical equilibrium as the minimum of
\be\label{ab090}
\FI_{\rm G}\{\RI\}=\EI\{\RI\}-\frac{1}{\beta}\,\SI\{\RI\}\q,
\ee
which is attained for $\RI=\RIze$.
From (\ref{ab010}) and (\ref{ab030}),
the stationarity condition yields
the self-consistent equation
\be\label{ab100}
\HI\{\RIze\}=\frac{1}{\bb}\,\LI_0\q,
\ee
which defines the HFB solution $\RIze$, in agreement with
Eq.~(\ref{ga010}). Around this equilibrium, the expansion of the grand
potential $\FI_{\rm G}\{\RI\}$ reads
\be\label{ab110}
\FI_{\rm G}\{\RI_0+\delta\RI\}=\FI_{\rm G}\{\RI_0\}+
{\textstyle\usd}\dRLl : \FLl_0 : \dRLl +\dots \q,
\ee
where $\FLl_0\equiv\ELl_0-\frac{1}{\bb}\,\SLl_0$
is a positive matrix, like $-\SLl_0$.

\subsection[A3]{The HFB Grand Potential and the RPA Equation}

The expansion of Eqs.~(\ref{f520}) around
$\RIb=\RIze$ (or around $\RIc=\RIze$) leads to Eqs.~(\ref{gg090}) 
which involve the RPA kernel $\KLl_0$ defined in 
Eq.~(\ref{gg100}) as
\be\label{ac010}
\KLl_0\,:\,\delta\RLl\equiv
\displaystyle
[\HI\{\RIze\}\,,\,\delta\RI]
+[\sum_{\mu\mu'}
\left.\frac{\partial\HI\{\RI\}}{\partial\RI_{\mu\mu'}}\right\vert_{\RI=\RIze}
\,{\delta\RI}_{\mu\mu'}\,,\,\RIze]\q.
\ee
In Liouville space the matrix $\KLl_0$ is non symmetric ;
it depends on $\RIze$ and on the parameters of $\KDh$.
We show here that $\KLl_0$ is directly
related to the matrix $\FLl_0$ which enters 
(\ref{ab110}), or (\ref{gd050}).

To this aim, we first introduce in the Liouville
space a new matrix which {\it generates the commutators
with $\RI$}.
Its action on any vector $\MI$ is defined by
\be\label{ac020}
\CLl : \MI = -\MI :\CLl =[\RI\,,\,\MI]\q.
\ee
Accordingly, its expression as a matrix is
\be\label{ac030}
\CLl_{(\lambda\lambda')(\mu\mu')}=
\delta_{\mu\lambda'}\RI_{\lambda\mu'}
-\delta_{\lambda\mu'}\RI_{\mu\lambda'}
+\sigma_{\lambda'\mu'}(\RI\sigma)_{\mu\lambda}
-\sigma_{\mu\lambda}(\sigma\RI)_{\lambda'\mu'}\q,
\ee
where the last two terms yield the same contribution as
the first two when applied to a vector $\MI$
satisfying (\ref{aa030}) (i.e.,  $\SIL : \MI = -\MI$).
The matrix $\CLl$ is antisymmetric, and it moreover satisfies
\be\label{ac040}
\CLl : \SIL = \SIL : \CLl = -\CLl\q.
\ee
In a basis where $\RI$ is diagonal, it reduces to
\be\label{ac050}
\CLl_{(\lambda\lambda')(\mu\mu')}=
2\,\delta_{\lambda\mu'}\delta_{\mu\lambda'}(\RI_\lambda-\RI_\mu)\q,
\ee
as it is obvious from (\ref{ac020})

The {\it commutators with the matrix}
$\LI$ can then be represented in Liouville space
by the product $\CLl : \SL$ which acts according to
\be\label{ac060}
\CLl : \SLl : \delta\RI =
\SLl : \CLl : \delta\RI =
-\delta\RI : \CLl :\SLl =
[\LI\,,\,\delta\RI]\q.
\ee
Indeed, in a basis where $\RI$ is diagonal,
this identity, analogous to (\ref{ac020}), is a direct
consequence of (\ref{ab080}) and (\ref{ac050}).
Note that the matrices $\CLl$ and $\SL$ commute.

The first term of (\ref{ac010}) is thus proportional
to (\ref{ac060}), in which $\CLl$ and $\SL$ have to be taken at the
stationary point where the relation
$\LI_0\equiv\log(1-\RIze)/\RIze=\bb\HI\{\RIze\}$ is satisfied.
To write the second
term, we note that
\be\label{ac070}
\sum_{\mu\mu'}
\frac{\partial\HI}{\partial\RI_{\mu\mu'}}{\delta\RI}_{\mu\mu'}=
\ELl : \delta\RI\q,
\ee
a consequence of (\ref{ab040}).
Hence, using (\ref{ac020}) and then adding (\ref{ac060}),
we find for any $\RI$ and $\delta\RI$~:
\be\label{ac080}\ba{rl}
\displaystyle[\sum_{\mu\mu'}
\frac{\partial\HI\{\RI\}}{\partial\RI_{\mu\mu'}}{\delta\RI}_{\mu\mu'}
\,,\,\RI] + \frac{1}{\bb}[\LI\,,\,\delta\RI]&=
\displaystyle
-\CLl : \ELl :\delta\RI + \frac{1}{\bb}\,\CLl : \SL :\delta\RI\\
&= -\CLl : \FLl : \delta\RI \q.
\ea\ee
The sought for expression of the kernel $(\ref{ac010})$ 
is then simply
\be\label{ac090}
\KLl_0= -\CLl_0 :\FLl_0\q,
\ee
where the matrices $\CLl_0$ and $\FLl_0$ are evaluated at the
equilibrium point $\RI=\RIze$.
The {\it RPA kernel\,} $\KLl_0$ is thus directly related to the
{\it matrix $\FLl_0$ of the second derivatives of the grand potential}.

At the minimum (attained at $\RI_0$) of the trial grand potential
$\FI_G\{\RI\}$,
the matrix $\FLl_0$ is non-negative.
Therefore {\it the eigenvalues of} $\KLl_0$, which are the same as those
of $-\FLl_0^{1/2} : \CLl_0 : \FLl_0^{1/2}$
in the space of Liouville vectors
satisfying (\ref{aa030}), {\it are real}.
We thus recover the well-known consistency between the stability of the
variational HFB state and the nature of the associated RPA dynamics 
in real time.
Moreover, due to the antisymmetry
of $\CLl_0$, the non-vanishing eigenvalues of $\KLl_0$ come
 in opposite pairs.

\subsection[A4]{Riemannian Structure of the HFB Theory}

By relying on the very existence
of the von Neumann entropy $\SI(\DDh)$,
it has been shown that the set
of density operators $\DDh$ can be
endowed with a natural Riemannian structure\cite{BAR86}.
The basic idea is to introduce a metric such that
the distance ${\rm d}s^2$ between two neigbouring states
$\DDh$ and $\DDh+\delta\DDh$ is defined by
\be\label{ad010}
{\rm d}s^2=-\delta^{2}\SI\q.
\ee

With this metric the general projection method of
Nakajima-Zwanzig appears as an orthogonal projection in the
space of states. A reduction of the exact
description through projection 
on a subset of simpler states, in our case the HFB states
of the form (\ref{f040}), induces from
(\ref{ad010}) a natural metric on this subset.
It is thus natural to define $-\SL$, the
positive symmetric matrix describing the second variations
of the entropy (\ref{ab050}), as
a {\it metric tensor}
in the reduced $2n\times 2n$ Liouville space of states.
This definition allows us to introduce distances between
neighbouring HFB states, and also to establish a correspondance between
the covariant and contravariant components of the vectors of Sect.~{A.1}.
More precisely, this {\it canonical mapping}
relates the variations of the observables
$\usd\gbt\LI\gbd$ to the variations of those
states which are their exponentials; it reads :
\be\label{ad020}
\SLl : \delta\RI = \delta\LI\q.
\ee

We can also introduce
the twice contravariant metric tensor $-\SLl^{-1}$
in the subspace of single quasi-particle observables by inverting
$\SLl$ in Liouville space
( $\SLl^{-1} : \SLl = \UL$ ).
In a basis where $\RI$ and
$\LI$ are diagonal, its matrix elements are
\be\label{ad030}
\SLl^{-1}_{(\lambda\lambda')(\mu\mu')}=
2\,\delta_{\lambda\mu'}\delta_{\mu\lambda'}\,
\frac{\RI_\lambda-\RI_\mu}{\LI_\lambda-\LI_\mu}\q,
\ee
where the ratio is $-\RI_\lambda(1-\RI_\lambda)$
when $\RI_\mu=\RI_\lambda$.  In an arbitrary basis, it
can be written as
\be\label{ad040}
\SLl^{-1}_{(\lambda\lambda')(\mu\mu')}=
-2\int_0^1\,\hbox{d}u\,
\crg\frac{e^{u\LI}}{e^\LI+1}\crd_{\lambda\mu'}
\crg\frac{e^{(1-u)\LI}}{e^\LI+1}\crd_{\mu\lambda'}\q.
\ee

Finally, we can introduce as in 
(\ref{ga070}) the reduced
thermodynamic potential
\be\label{ad050}
\log\ZI\{\LI\}=\log\Tr{\exp\crg
-\usd\gbt\LI\gbd
\crd}=\usd\trd{\log(1+e^{-\LI})}
\q.\ee
The expression of the first-order term in the expansion 
in powers of $\delta\LI$,
\be\label{ad060}
\log\ZI\{\LI+\delta\LI\}=\log\ZI\{\LI\} -\RLl : \dLLl
-\usd \dLLl :\SLl^{-1} :\dLLl + \ldots\q,
\ee
exhibits the fact that {\it the entropy}
(\ref{ab050}) is the
{\it Legendre transform of the
reduced grand potential} (\ref{ad050}),
while the second-order term is seen to describe a distance
between two neighbouring observables.
This last feature is the counterpart in the reduced HFB 
description of a structure already recognized\cite{BAR86} in the full
Liouville space.

All these properties, on which we shall not dwell here,
confer a geometric nature to the reduced HFB
description and set it into the general
projection method \cite{BAR86}.

\subsection[A5]{Lie-Poisson Structure of the Time-Dependent HFB Theory}

We shall now introduce another geometric type of structure,
of a symplectic type.
This will allow us to regard the reduced energy
$\EI\{\RI\}$ or the grand potential 
$\FI_{\rm G}\{\RI\}$ as a classical Hamiltonian for the
dynamics associated with the HFB theory.

The evolution of the variables
$\RI_{\lambda\lambda'}$ is governed in the {\it time-dependent
mean-field HFB theory} by the equation
\be\label{ae010}
i\hbar\,\frac{{\rm d}\RI}{{\rm d}t}=[\HI\{\RI\}\,,\,\RI]\q,
\ee
written in the $2n\times 2n$ matrix representation.

In order to analyze this equation, we 
first note that the basic operators $\gbd\gbt$ of the
Liouville representation are the generators
of a Lie group with the algebra
\be\label{ae020}
[\gbd_\lambda\gbt_{\lambda'}\,,\,\gbd_\mu\gbt_{\mu'}]=
\delta_{\mu\lambda'}\gbd_\lambda\gbt_{\mu'}
-\delta_{\lambda\mu'}\gbd_\mu\gbt_{\lambda'}
+\sigma_{\lambda'\mu'}\gbd_\mu\gbd_{\lambda}
-\sigma_{\mu\lambda}\gbt_{\lambda'}\gbt_{\mu'}
\q.\ee
Following the arguments of Ref.\cite{BVe89}, where a similar
formulation was set up for the Hartree-Fock case, we note that
the contractions $\RI_{\lambda\lambda'}$ are through (\ref{f080})
in one-to-one correspondence
with the operators $\gbd_\lambda\gbt_{\lambda'}$.
Regarding the $\RI_{\lambda\lambda'}$'s as a
set of {\it classical dynamical variables},
we introduce among them a Poisson structure; indeed,
the Lie-algebra relation (\ref{ae020}) 
suggests to characterize the Lie-Poisson bracket between
the dynamical variables by the {\it Poisson tensor}
\be\label{ae030}
i\hbar\,\acg\RI_{\lambda\lambda'}\,,\,\RI_{\mu\mu'}\acd=
\CLl_{(\lambda\lambda')(\mu\mu')}\q,
\ee
where each $\CLl$ is the linear function of the $\RI$'s inferred
from the right hand side of (\ref{ae020}). As a matrix $\CLl$
is readily identified with the
matrix introduced in (\ref{ac020},\ref{ac030}) which described, 
in the Liouville space,
the commutation with $\RI$. While the metric
of Sect.A.4 was characterized by the symmetric Riemann tensor $-\SL$,
the present Poisson structure is characterized by the antisymmetric
Poisson tensor $\CLl$.
For more details on Poisson structures, see for instance
Ref.\cite{Wei84}.

In contrast to the standard Poisson brackets between
canonically conjugate position and momentum variables, which are
c-numbers, the Lie-Poisson brackets (\ref{ae030})
depend on the dynamical variables $\RI_{\lambda\lambda'}$.
In the Liouville representation, the
bracket of two functions $f$ and $g$ of the
$\RI$'s is evaluated by saturating their gradients
with the tensor $\CLl$ according to
\be\label{ae040}
i\hbar\, \acg f \,,\, g \acd=
\frac{\partial f}{\partial\RI}
: \CLl :
\frac{\partial g}{\partial\RI}\q.
\ee
The compatibility of (\ref{ae040})
with the definition (\ref{ae030}) results
from the property (\ref{ac040}) of the tensor $\CLl$ and from the relation
\be\label{ae050}
\frac{\partial\RI_{\lambda\lambda'}}{\partial\RI_{\mu\mu'}} =
\usd (\UL - \SIL)_{(\lambda\lambda')(\mu\mu')}\q,
\ee
which itself is a consequence of (\ref{f080}).

Using the rules (\ref{ae040}) and (\ref{ae050}), we 
identify the r.h.s. of (\ref{ae010})
as a Lie-Poisson bracket since
the time-dependent mean-field HFB
equation reads
\be\label{ae060}\ba{rl}
\displaystyle
\frac{{\rm d}\RI}{{\rm d}t}&=\displaystyle
\frac{1}{i\hbar}[\HI\{\RI\}\,,\,\RI]\\
&=\displaystyle
\frac{1}{i\hbar}
\frac{\partial\EI}{\partial\RI} : \CLl =
\frac{1}{2i\hbar} \frac{\partial\EI}{\partial\RI} : \CLl : (\UL - \SIL)
=\frac{1}{i\hbar} \frac{\partial\EI}{\partial\RI} : \CLl :
\frac{\partial\RI}{\partial\RI}\q,
\ea\ee
that is
\be\label{ae070}
\frac{{\rm d}\RI}{{\rm d}t}=\acg\EI\{\RI\} \,,\,\RI \acd
\q.
\ee
This equation has thus the form of a classical dynamical
equation of motion~; {\it the HFB energy} $\EI\{\RI\}$
plays the r\^ole of {\it a classical Hamiltonian}, while
a Lie-Poisson structure is associated with the dynamical
variables $\RI$ by Eq.~(\ref{ae030}).

Since the gradient (\ref{ab030}) of $\SI\{\RI\}$, as a matrix
in the $\lambda$-space, commutes with $\RI$, the relation
(\ref{ac020}) implies that the Lie-Poisson bracket
$\acg\SI\{\RI\}\,,\,\RI\acd$ vanishes.
Hence, we can alternatively regard the grand potential (\ref{ab090}),
instead of $\EI\{\RI\}$, as a classical Hamiltonian and write
\be\label{ae080}
\frac{{\rm d}\RI}{{\rm d}t}=\acg\FI_{\rm G}\{\RI\} \,,\,\RI \acd
\q.
\ee

This remark is useful when the time-dependent
mean-field equation (\ref{ae010}) is used in the small amplitude
limit to describe approximately collective excitations
around the grand-canonical equilibrium state. By expanding $\RI$
as $\RI_0+\delta\RI$, Eq.~(\ref{ae010}) then reduces to the RPA
equation associated with the HFB approximation,
\be\label{ae090}
{\displaystyle\frac{{\rm d}\delta\RI}{{\rm d}t}}
=\frac{1}{i\hbar}\,\KLl_0\,:\,\delta\RLl\q,
\ee
where the RPA kernel $\KLl_0$ is
expressed by (\ref{ac010}). We can now identify
the r.h.s. of (\ref{ae090}) with the
first-order term
$\acg\delta\FI_{\rm G}\{\RI\} \,,\,\RIze \acd$
in the expansion of  (\ref{ae080})
in powers of $\delta\RI$
around $\RIze$. Using the expansion (\ref{ab110}) and the
expression (\ref{ae030}) of the Lie-Poisson bracket,
we find that (\ref{ae090}) is equivalent to
\be\label{ae100}
i\hbar\,{\rm d}\delta\RI/{\rm d}t=-\CLl_0 : \FLl_0 :\delta\RI\q,
\ee
and thus recover the expression
(\ref{ac090}) for $\KLl_0$.
This new derivation of (\ref{ac090}) provides us with a simple
interpretation of the RPA equation (\ref{ae100}) 
as an equation for small
motions in ordinary classical dynamics.
Indeed, small changes of canonically conjugate variables $q_i, p_i$
around a minimum of a Hamiltonian $H\{q,p\}$
are governed by
\be\label{ae110}
\frac{{\rm d}\ }{{\rm d}t}\left(\ba{c}\delta q\\\delta p\ea\right)=
\left(\ba{c}
-\delta \frac{\partial H}{\partial p}\\
\delta \frac{\partial H}{\partial p}
\ea\right)\simeq
-\left(\ba{cc}
\left\{q\,,\,q\right\}&\left\{q\,,\,p\right\}\\
\left\{p\,,\,q\right\}&\left\{p\,,\,p\right\}
\ea\right)
\left(\ba{cc}
\frac{\partial^2 H}{\partial q\partial q}
&\frac{\partial^2 H}{\partial q\partial p}\\
\frac{\partial^2 H}{\partial p\partial q}
&\frac{\partial^2 H}{\partial p\partial p}
\ea\right)
\left(\ba{c}\delta q\\ \delta p\ea\right)\q,
\ee
where one recognizes the same multiplicative
structure as in the r.h.s. of (\ref{ae100}),
within the replacement of the 
ordinary Poisson bracket by (\ref{ae030})
and of $H\{q,p\}$ by $\FI_{\rm G}\{\RI\}$.

For small deviations around an arbitrary TDHFB
trajectory, the time-dependent RPA kernel $\KLl$ is defined by
\be\label{ae120}
i\hbar\,{\displaystyle\frac{{\rm d}\delta\RI}{{\rm d}t}}
=\KLl : \delta\RI
\equiv[\HI\{\RI\}\,,\,\delta\RI]+
[\sum_{\mu\mu'}
\frac{\partial\HI\{\RI\}}{\partial\RI_{\mu\mu'}}{\delta\RI}_{\mu\mu'}
\,,\,\RI]\q.
\ee
Since we no longer lie at the minimum of $\FI_{\rm G}\{\RI\}$,
this variation (\ref{ae120}) involves not only the expansion of
$\FI_{\rm G}$ but also of $\RI$, which appears
both directly and through the Poisson tensor $\CLl$. 
We thus get, in the product
\be\label{ae130}\ba{rl}
\KLl : \delta\RI
&=\HI :\delta\CLl - \CLl : \ELl : \delta\RI\\
&=-\CLl :\FLl :\delta\RI + (\HI-\bb^{-1}\LI) : \delta\CLl\q,
\ea\ee
an additional term proportional to the deviation between $\bb\HI\{\RI\}$
and $\LI=\log[(1-\RI)/\RI]$.
Apart from this term, which arises from the dependence
of the structure constants $\CLl$ on the variables $\RI$,
the present classical dynamics differs from an ordinary 
Hamiltonian dynamics through the presence of vanishing 
eigenvalues in the tensor $\CLl$. The Poisson structure
generated by (\ref{ae030}) therefore differs from the
usual symplectic structure associated with canonically 
conjugate variables because some combinations of the
dynamical variables $\RI$ have a vanishing bracket with
{\it any} variable. This property reflects the existence 
of {\it structural} constants of the motion, which never vary
{\it whatever the effective Hamiltonian} in (\ref{ae070}).
In particular, the eigenvalues of $\RI$ and $\SI\{\RI\}$
always remain constant.
\newpage

\section[APPENDIX B :]{\label{B} Liouville Formulation of the Projected
Finite-Temperature HFB Equations}
\setcounter{equation}{0}
\renewcommand{\theequation}{B.\arabic{equation}}

Using the Liouville-space formalism introduced in
Appendix A, it is possible to recast
the coupled equations (\ref{hc020}) and (\ref{hc070})
governing the evolution of $\TId\uu$ and $\TIa\uu$,
defined in Eqs.~(\ref{f060}) and (\ref{f160}-\ref{f170}),
so as to make these equations formally similar
to Eqs.~(\ref{f370}) and (\ref{f430}).

To this aim, we first introduce two
operators $\TLa$ and $\TLd$
whose action on a Liouville vector
$\QI$ is given by
\be\label{ba010}
\TLa : \QI\equiv \TIa\QI\TIam\ ,\q\q \TLd : \QI\equiv \TId\QI\TIdm\q.
\ee
They satisfy the relations
\be\label{ba020}
\SIL : \TLa : \SIL = \TLa\ ,\q\q\SIL : \TLd : \SIL = \TLd\q ,
\ee
\be\label{ba30}
\TLam={\TLa}^{\rm T}\ ,\q\q\TLdm={\TLd}^{\rm T}\q,
\ee
where the symbol ${}^{\hbox{T}}$ denotes the matrix transposition
in the Liouville space.

Next, we define the matrix
\be\label{ba040}\ba{rl}
\MLcd_{(\lambda\lambda')(\mu\mu')}\equiv&
\sff\YIgg\{2([1-\RIcup]\exup)_{\lambda\mu'}(\exum\RIcup)_{\mu\lambda'}\\
&\displaystyle\q\q +
(\RIcup{}-\RB^\pcq)_{\lambda\lambda'}(\RIcdp-\RB^\pcp)_{\mu\mu'}\}\q,
\ea\ee
and the two Liouville vectors
\be\label{ba050}\ba{rl}
\GLcd{}\equiv
&\sff\YIgg\{(1-\RIcup)\HI\{\RIcup\}\RIcup\\
&\q\q\q\q\q+(\EI\{\RIcup\}-\EBc)(\RIcup-\RB^\pcq)\}\q,\\
\GLbu{}\equiv
&\sff\YIgg\{(\RIbup\HI\{\RIbup\}(1-\RIbup)\\
&\q\q\q\q\q+(\EI\{\RIbup\}-\EBb)(\RIbup-\RB^\pbp)\}
\q.\ea\ee

After multiplication to the right by $\TIa$, 
Eq.~(\ref{hc020}) can be reformulated as
\be\label{ba060}
\MLcd : (\duh{\TId}\TIdm) +
\usd(\GLcd + \TLam : \GLbu) =0\q.
\ee
By introducing in the Liouville space the matrix
\be\label{ba070}
\MLbu\equiv\TLa : \MLcd : \TLd\q,
\ee
namely
\be\label{ba080}\ba{rl}
\MLbu_{(\lambda\lambda')(\mu\mu')}\equiv&
\sff\YIgg\{2(\RIbup\exdm)_{\lambda\mu'}(\exdp[1-\RIbup])_{\mu\lambda'}\\
&\displaystyle\q\q +
(\RIbup{}-\RBbu)_{\lambda\lambda'}(\RIbdp-\RBbd)_{\mu\mu'}\}\q,
\ea\ee
and the vectors
\be\label{ba090}
\HLcd=\MLcdm : \GLcd\q,\q\HLbu=\MLbum : \GLbu\q,
\ee
we can rewrite (\ref{ba060}) as
\be\label{ba100}
\duh{\TId}=-\usd[\HLcd\TId+\TId\HLbu]\q,
\ee
a form which resembles closely (\ref{f370}).

In the same way Eq.~(\ref{hc070}) can be transformed into
\be\label{ba110}
\duh{\TIa}=\usd[\HLbd\TIa+\TIa\HLcu]\q,
\ee
where the matrices
$\HLbd$ and $\HLcu$ in the single-particle space
are obtained from Eqs.~(\ref{ba040}-\ref{ba050})
and (\ref{ba080}-\ref{ba090})
by the exchanges $a\leftrightarrow d$ and
$g\leftrightarrow g'$. One major difference, however, between
the coupled equations (\ref{ba100}-\ref{ba110}) and
Eqs.~(\ref{f370}) and (\ref{f430}) is that the former involve four
different quantities ($\HLcd$, $\HLbu$, $\HLbd$, $\HLcu$)
instead of two ($\HI\{\RIc\}$, $\HI\{\RIb\}$) for the latter.

The matrices $\MLbd$ and $\MLcu$ required for the calculation
of the quantities $\HLbd$ and $\HLcu$ appearing in (\ref{ba110})
are related to $\MLbu$ and $\MLcd$ through
\be\label{ba120}
\MLbd=\MLbu{}^{\rm T}\q,\q\MLcu=\MLcd{}^{\rm T}\q.
\ee

From the definition (\ref{ba040}-\ref{ba050}) and (\ref{ba070}) of
the matrices $\MLcd$ and $\MLbu$
and of the Liouville vectors $\GLcd$ and $\GLbu$,
one establishes the relations
\be\label{ba130}
\SIL : \MLcd : \SIL = \MLcd\ ,\q\q\SIL : \MLbu : \SIL = \MLbu\q.
\ee
\be\label{ba140}
\SIL : \GLcd =-\,\GLcd\ ,\q\q\SIL : \GLbu =-\,\GLbu\q.
\ee
One deduces that, as the reduced HFB Hamiltonian $\HI\{\RI\}$,
the vectors in Liouville space $\HLcd$ and $\HLbu$
verify the relation (\ref{f410})
which in the Liouville formulation becomes :
\be\label{ba150}
\SIL : \HLcd =-\,\HLcd\,\q\q\SIL : \HLbu =-\,\HLbu\q.
\ee
Through the exchanges $a\leftrightarrow d$ and $g\leftrightarrow g'$
one extends the results (\ref{ba130}-\ref{ba150})
to the matrices $\MLbd$, $\MLcu$ and vectors $\GLbd$, $\GLcu$,
$\HLbd$, $\HLcu$.

When the operator $\ADh$ (defined in Eq.~(\ref{I060})) is hermitian,
Eqs.~(\ref{ba100}), (\ref{ba110}), (\ref{hc020}) and (\ref{hc070}),
have hermitian solutions.
Indeed, in such a case these equations preserve the hermiticity of
$\TIa$ and $\TId$ and the reality of $\lia$ and $\lid$.
To prove it, one first notices that the relations
$\exup^\dagger=\TI^{[g{}^{-1}]}$,
$\TIa^\dagger=\TIa$, $\TId^\dagger=\TId$
result in
\be\label{ba160}
\RIcup^\dagger=\RI^{[ag'{}^{-1}dg{}^{-1}]}\q,
\ee
\be\label{ba170}
\YIgg{}^*=\YI^{g'{}^{-1}g{}^{-1}}\q.
\ee
Then, since
$(\sff)^*=\int_{g'{}^{-1}\,g{}^{-1}}$,
one sees from Eq.~(\ref{ba170}) 
that $\YI$ is real.
The use of this property along with the 
equality $\EI\{\RI^\dagger\}=\EI\{\RI\}^*$ in 
Eqs.~(\ref{hb080}) and (\ref{hb090}) implies
\be\label{ba180}
\RBcu{}^\dagger=\RBbd\ ,\q\q\RBcd{}^\dagger=\RBbu\q,
\ee
\be\label{ba190}
\EBc{}^*=\EBb\q.
\ee
Finally, on formulae (\ref{ba040}-\ref{ba050}) and
(\ref{ba080}), one readily checks the relations
\be\label{ba200}
\MLcu_{(\lambda'\lambda)(\mu'\mu)}{}^*=
\MLbd_{(\lambda\lambda')(\mu\mu')}\ ,\q\q
\MLcd_{(\lambda'\lambda)(\mu'\mu)}{}^*=
\MLbu_{(\lambda\lambda')(\mu\mu')}\q,
\ee
\be\label{ba210}
\GLcu{}^\dagger=\GLbd\ ,\q\q\GLcd{}^\dagger=\GLbu\q.
\ee
They yield the equations
\be\label{ba220}
\HLcu{}^\dagger=\HLbd\ ,\q\q\HLcd{}^\dagger=\HLbu\q,
\ee
which entail the hermiticity of $\hbox{d}\TId/\hbox{d}u$ and
$\hbox{d}\TIa/\hbox{d}u$.
\newpage

\end{document}